# A C T

## Advanced Compton Telescope

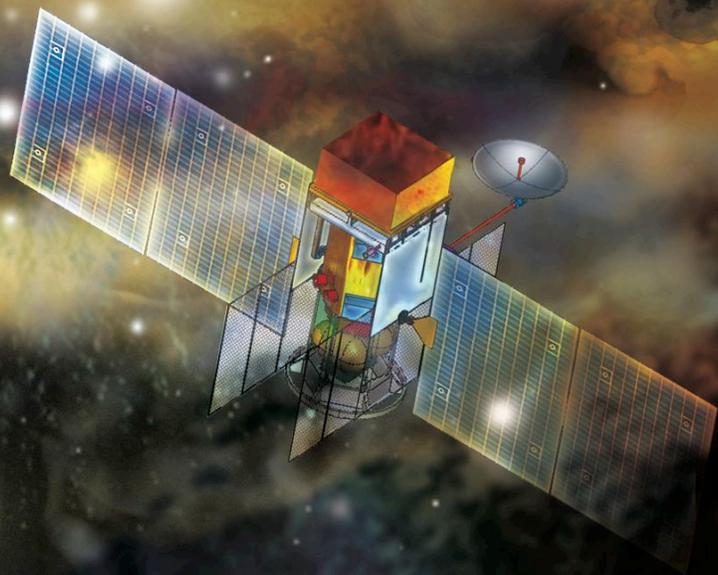

## Witness to the Fires of Creation

# TABLE OF CONTENTS

# A. ACT Study Team

## Executive Committee

**Steven Boggs (Principal Investigator)**, *University of Califonia, Berkeley*
**James Kurfess (co-Chair)**, *Naval Research Laboratory*
**James Ryan (co-Chair)**, *University of New Hampshire*
**Elena Aprile**, *Columbia University*
**Neil Gehrels**, *Goddard Space Flight Center*
**Marc Kippen**, *Los Alamos National Laboratory*
**Mark Leising**, *Clemson University*
**Uwe Oberlack**, *Rice University*
**Cornelia Wunderer**, *University of California, Berkeley*
**Allen Zych**, *University of California, Riverside*

## ACT Science Working Group

**Mark Leising (Chair)**
**Matthew Baring**, *Rice University*
**John Beacom**, *Ohio State University*
**Lars Bildsten**, *University of Califonia, Santa Barbara*
**Charles Dermer**, *Naval Research Laboratory*
**Dieter Hartmann**, *Clemson University*
**Margarida Hernanz**, *IEEC-CSIC, Spain*
**Peter Milne**, *University of Arizona, Tucson*
**David Smith**, *University of California, Santa Cruz*
**Sumner Starrfield**, *Arizona State University*

## ACT Simulations Working Group

**Marc Kippen (co-Chair)**
**Cornelia Wunderer (co-Chair)**
**Peter Bloser**, *University of New Hampshire*
**Michael Harris**, *CESR, France*
**Andrew Hoover**, *Los Alamos National Laboratory*
**Alexei Klimenko**, *Los Alamos National Laboratory*
**Dan Kocevski**, *Rice University*
**Mark McConnell**, *University of New Hampshire*
**Elena I. Novikova**, *Naval Research Laboratory*
**Uwe Oberlack**, *Rice University*
**Bernard Phlips**, *Naval Research Laboratory*
**Mark Polsen**, *University of California, Riverside*
**Steven Sturner**, *Goddard Space Flight Center*
**Derek Tournear**, *Los Alamos National Laboratory*
**Georg Weidenspointner**, *CESR, France*
**Eric Wulf**, *Naval Research Laboratory*
**Andreas Zoglauer**, *University of California, Berkeley*

## ACT Collaborators

**Morgan Burks**, *Lawrence Livermore National Laboratory*
**Wayne Coburn**, *University of California, Berkeley*
**Lynn Cominsky**, *Sonoma State University*
**Roland Diehl**, *MPE, Germany*
**Stanley Hunter**, *Goddard Space Flight Center*
**Kevin Hurley**, *University of California, Berkeley*
**Pierre Jean**, *CESR, France*
**Neil Johnson**, *Naval Research Laboratory*
**Gottfried Kanbach**, *MPE, Germany*
**Jim Matteson**, *University of California, San Diego*
**William Paciesas**, *University of Alabama, Huntsville*
**William Purcell**, *Ball Aerospace*
**Robert Savage**, *Goddard Space Flight Center*
**Peter von Ballmoos**, *CESR, France*
**Stan Woosley**, *University of California, Santa Cruz*

# B. ACT FACTS

## SCIENCE OBJECTIVES

The *Advanced Compton Telescope* (ACT) will be a powerful new survey instrument for studying supernovae, Galactic nucleosynthesis, γ-ray bursts (GRBs), compact objects, and fundamental physics. The ACT instrument design is driven by its primary science goal: spectroscopy of the $^{56}$Co (0.847 MeV) line from type Ia supernovae (SNe Ia), which we expect to be broadened to ~3%. ACT will allow hundreds of SNe Ia detections over its primary 5-year survey lifetime. In the process, ACT becomes a powerful all-sky observatory for all classes of γ-ray observations (Table B1).

**Table B1.** ACT source numbers vs. COMPTEL.

| Sources (5yr) | COMPTEL | ACT |
|---|---|---|
| Supernovae | 1 | 100–200 |
| AGN & Blazars | 15 | 200–500 |
| Galactic | 23 | 300–500 |
| GRBs | 31 | 1000–1500 |
| Novae | 0 | 25–50 |

## SCIENCE INSTRUMENT

ACT employs the Compton imaging technique pioneered in COMPTEL on CGRO, but utilizing recent advances in detector technologies and integrated readout electronics. Modern 3-D position sensitive γ-ray detectors, arranged in a compact, large volume configuration, will improve efficiency by two orders of magnitude, provide a powerful new tool for background rejection, and utilize high spectral resolution to dramatically improve sensitivity (Table B2).

**Table B2.** ACT baseline instrument performance.

| | |
|---|---|
| Energy range | 0.2–10 MeV |
| Spectral resolution | 0.2–1% |
| Field of view (FoV) | 25% sky |
| Sky coverage | 80% per orbit |
| Angular resolution | ~1° |
| Point source localization | 5' |
| Detector area, depth | 12,000 cm$^2$, 47 g/cm$^2$ |
| Effective area | ~1000 cm$^2$ |
| 3% broad line sensitivity | $1.2\times10^{-6}$ γ cm$^{-2}$ s$^{-1}$ |
| Narrow line sensitivity | $5\times10^{-7}$ γ cm$^{-2}$ s$^{-1}$ |
| Continuum sensitivity | $(1/E)\times10^{-5}$ cm$^{-2}$ s$^{-1}$ MeV$^{-1}$ |
| GRB fluence sensitivity | $3\times10^{-8}$ erg cm$^{-2}$ |
| Data mode | every photon to ground |

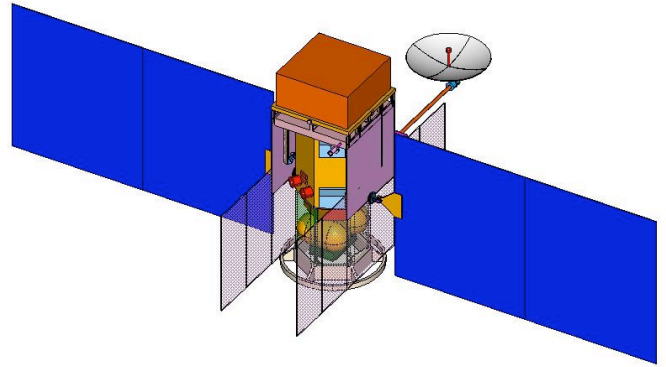

**Fig. B1.** *ACT science instrument (orange) on its spacecraft bus, with solar panels (blue), thermal radiators (transparent grey), and high-gain TDRSS antenna deployed.*

## MISSION OVERVIEW

The ACT mission involves a single instrument composed of a large array of multi-channel γ-ray detectors, surrounded by anti-coincidence (ACD) shields on all sides, mounted on a zenith-pointing S/C (Fig. B1). ACT could be launched as early as 2015 from KSC on a Delta IV 4240 vehicle into a 550 km circular orbit with 8° inclination for a 5-year minimum (10-year desired) lifetime (Table B3).

**Table B3.** ACT mission requirements.

| |
|---|
| Launch ~2015, 5–10 year lifetime |
| 550 km LEO, <10° inclination, Delta IV (4240) |
| 1° attitude, 1' aspect, zenith pointer |
| Instrument 2100 kg, S/C 1425 kg, propellant 462 kg |
| 3340 W power, 69 Mbps telemetry |
| $760M (FY04, original IMDC estimate) |

## MISSION OPERATIONS

In order to both continually survey the γ-ray sky, as well as build up statistics for sensitivity to faint nuclear line emission from distant supernovae, ACT is required to have a very large field-of-view (FoV), covering over 25% of the sky at any instant. In its primary survey mode, ACT will be continuously pointed at the zenith, sweeping out over 80% of the entire sky over the course of each orbit.

## C. Executive Summary

Since the gravitational collapse of matter into stars and galaxies a few hundred thousand years after the Big Bang, much of the visible matter in the Universe has been through the slow but spectacular lifecycle of matter: stellar formation and evolution ending in novae or supernovae, with the ejection of heavy nuclei back into the galaxy to seed a new generation of stars. The unstable balance between gravitational and nuclear forces produces a cycle in which the death of stars leads to the birth of others, maintaining a rich, dynamic story of life, death, and rebirth on the galactic stages of our Universe.

Nuclear γ-ray astrophysics is the study of emission from radioactive nuclei as tracers of this cycle of creation. Nuclear decays are "fingerprints" of the isotopes, and the γ-rays emitted characterize their quantities, speeds, and depths in their environments. The unstable nuclei provide a direct means of quantifying the underlying processes of nuclear burning in supernovae, novae, and hydrostatic stars. They carry unique information about the otherwise hidden, extreme conditions under which they were produced. These nuclei allow us to see to the very core of a supernova explosion, revealing critical information about the underlying nuclear ignition, structure, and dynamics of these events that shape our Galaxy.

Because of this potential, the *Advanced Compton Telescope* (ACT) has been identified as the next major step in γ-ray astronomy in NASA's roadmap. Its main goal is to probe the nuclear fires creating the chemical elements. For example, thermonuclear supernovae (SN Ia) are used as standard candles across the Universe, yet even those near us are poorly understood. ACT will detect and measure nuclear species produced in those explosions, providing otherwise unattainable information on the dynamics of SN nuclear burning. Supernovae, novae, and stellar winds populate our Galaxy with fresh nuclei. ACT will measure the radioactive γ-ray and positron emitters among them across the entire Milky Way, mapping our galaxy in a broad range of nuclear line emissions (Fig. E1) from radioactive decays ($^{22}$Na, $^{26}$Al, $^{44}$Ti, $^{60}$Fe), nuclear de-excitations ($^{12}$C, $^{16}$O, $^{56}$Fe) and $e^+e^-$ annihilations.

As additional objectives, γ-rays from accretion of matter onto galactic compact objects and massive black holes in AGN will test accretion disk and jet models and probe relativistic plasmas. Gamma-ray polarization will be used to study the emission processes in GRBs, pulsars, AGN, and solar flares—opening a new dimension in diagnostic phase space. The origins of the diffuse cosmic MeV background will be identified.

Previous γ-ray missions began to address this science, but the key to real advancement is a dramatic improvement in sensitivity. The optimal approach for achieving this improvement is an imaging Compton telescope using advanced detector technology. The technology enabling ACT is the development of 3-D position-sensitive detectors with excellent energy resolution. They resolve interaction sites and energies as photons Compton scatter throughout the instrument, providing a powerful new tool for background rejection, Compton imaging, and polarization studies. The excellent position resolution of these detectors facilitates compact Compton telescopes, increasing the detection efficiency by up to two orders of magnitude over COMPTEL on CGRO. The combination of improved efficiency and improved background rejection will yield 50× better sensitivity than any previous or current γ-ray spectrometer or imager. The large FoV achievable with compact designs is also a major improvement, enabling the discovery and monitoring of transient sources, and long exposures of steady-state sources. We restricted this study to instruments designed to attack the SN Ia problem, as opposed to broad MeV scientific objectives. We made this tough choice despite the demonstrated success of multi-instrument missions, e.g., CGRO.

Our primary goals of this ACT Concept Study are to (1) transform the key scientific objectives into specific instrument requirements, (2) identify the most promising technologies to meet these requirements, and (3) design a viable mission concept for this instrument. To this end, we developed technology recommendations that identify the detector and readout requirements and lay out goals for their development, demonstration, and implementation. Furthermore, we developed a baseline ACT mission concept, including mission requirements. The primary finding of our mission design studies is that all of the primary mission architecture requirements are achievable with technologies readily available for a 2015 launch.

# D. Concept Study Overview

The ACT collaboration is composed of over 50 collaborators from 25 institutions in the US and abroad. This concept study has included the broad US γ-ray astronomy community, and we anticipate this collaboration to vigorously continue this work after this concept report is finalized.

## ACT Workshops

The ACT collaboration participated in five workshops over the course of this study. These workshops were designed to solidify the goals of the study, develop detailed plans, and to finalize our report recommendations. The dates and primary goals of the workshops are listed here:
- 4/04, Berkeley: kickoff meeting, study outline
- 9/04, HEAD: science requirements review
- 2/05, Berkeley: simulations working group
- 3/05, UNH: instrument status reports
- 8/05, Berkeley: final report planning

In addition, the Executive Committee participated in regular bi-weekly telecons to discuss the progress of the study and to set the study priorities. The Simulations Working Group participated in weekly telecons to discuss simulation priorities and status of the ACT simulation tools.

## ACT Simulation Tools

A major part of the ACT concept study was the development and application of computer simulations and models for estimating realistic instrument performance parameters. Space-based instruments operating in the energy range of nuclear lines are subject to complex backgrounds generated by cosmic-ray interactions and diffuse γ-rays. The count rate from backgrounds typically far exceeds that from astrophysical γ-ray sources. Maximizing the signal and minimizing the background depends critically on complex event selection and reconstruction algorithms, which in turn depend on the exact geometry of the instrument. Detailed computer simulations allow us to efficiently explore this parameter space and are thus vital for optimizing instrument designs and predicting performance. For the ACT simulation effort, we combined previously existing tools into a complete, powerful package for γ-ray astronomy. These tools were applied to several different instrument concepts operating in different candidate environments as described throughout the report.

## ISAL & IMDC Runs

During the course of the ACT Concept Study, our team participated in two one-week studies at the NASA/Goddard Integrated Design Capability laboratories in order to develop a viable mission concept, and to identify the key technology requirements for the mission. The first study was a one-week run through the Instrument Synthesis & Analysis Laboratory (ISAL) in September 2004. The primary goals of the ISAL study were to develop a more detailed science instrument model, identify its technical requirements (mass, power, telemetry), and identify the key requirements placed on the mission architecture. The second study was a 1-week run through the Integrated Mission Design Center (IMDC) in November 2004, whose primary goal was to take the instrument requirements from the ISAL run and develop an integrated mission architecture package for ACT.

While one of the goals of this concept study was to determine an optimal science instrument design by comparing the various detector technologies, the ISAL run occurred too early in the course of this study to go in with an optimized design. Thus we selected a "baseline" instrument for the study that was a best guess of what an optimal design could look like. This baseline was designed also to encompass a variety of the challenges an ACT instrument might pose: a combination of Ge detectors (stringent cooling requirements) and Si-strip detectors (moderate cooling, high number of channels). This baseline instrument was optimized over the course of our study (Section F), and the performance parameters quoted are for this optimized design.

## Technology Scope

Given the finite time and resources of this study, we limited our comparisons among instrument designs and technologies to their performance with regards to ACT's single-highest science priority: study of $^{56}$Co emission from SNe Ia. While we have identified promising technologies to achieve the SNe Ia science, and broadly characterized overall performance, we cannot yet infer that these technologies are optimal for *all* ACT science goals.

# E. Science Objectives

*"to uncover how supernovae and other stellar explosions work to create the elements"*

### Relation to NASA Strategic Goals

NASA's May 2005 *Universe Strategic Roadmap* identifies the *Nuclear Astrophysics Compton Telescope* (NACT, of which ACT is an example) as one of the *Pathways to Life Observatories*, with a primary goal of addressing Objective 4C, "Trace the evolution of nuclei, atoms, and molecules that become life."

ACT is a mission in the *Cycles of Matter and Energy* section of the January 2003 Roadmap for the *Structure and Evolution of the Universe* theme in NASA's Office of Space Science. An *Advanced Compton Telescope* is explicitly called for in the Roadmap for future progress in γ-ray astronomy. The mission was described as the primary tool "to uncover how supernovae and other stellar explosions work to create the elements." The scientific excitement in the fields of explosive nucleosynthesis, annihilation radiation and AGN was presented as an illustration of the discovery potential available to a future γ-ray spectrometer and imager.

In the γ-ray astronomy community ACT remains a top priority for the future in MeV astrophysics. NASA's Gamma Ray Astrophysics Program Working Group named ACT as its highest priority major mission in reports published in 1997 and 1999.

### Scientific Objectives

The unique and substantial astrophysical information carried by γ-ray photons with energies near one MeV has long been clear. Nuclear γ-ray line transition energies are practically unique, identifying individual isotopes. In the case of nuclear levels populated by radioactive decay, we also infer the ages of these isotopes, to within a few half-lives. At the electron rest-mass energy (0.511MeV) the annihilation of electron-positron pairs can be studied in numerous high-energy sources, and the onset of relativistic effects in photon production and interaction processes produces distinctive spectral shapes. Gamma-ray photons are highly penetrating, easily traversing entire galaxies and escaping from supernova ejecta on timescales of weeks. Gamma rays provide probing diagnostics of many astrophysical sources, in some cases the only accessible means of understanding the physics of those sources. The greatest challenge for γ-ray astronomy has always been small photon fluxes and large backgrounds, so instrument sensitivities are paramount.

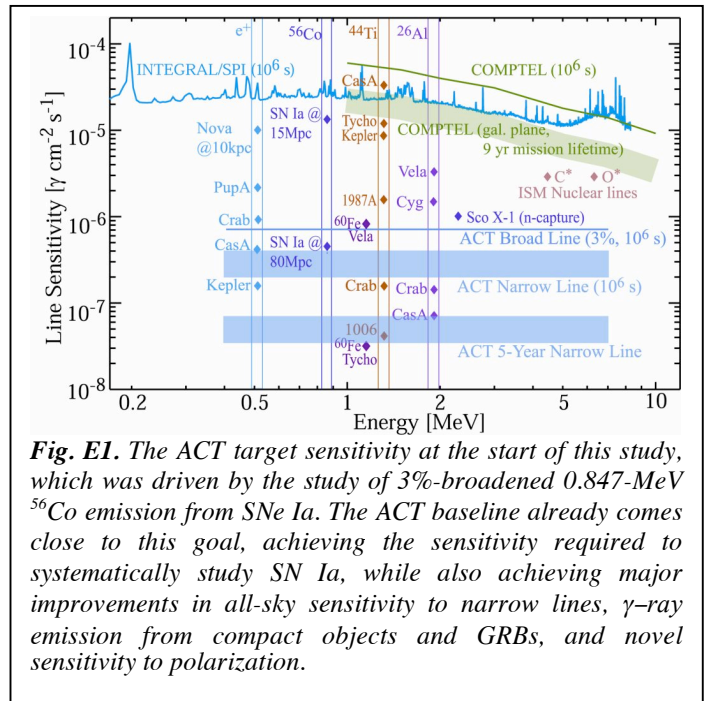

**Fig. E1.** *The ACT target sensitivity at the start of this study, which was driven by the study of 3%-broadened 0.847-MeV $^{56}Co$ emission from SNe Ia. The ACT baseline already comes close to this goal, achieving the sensitivity required to systematically study SN Ia, while also achieving major improvements in all-sky sensitivity to narrow lines, γ–ray emission from compact objects and GRBs, and novel sensitivity to polarization.*

**ACT is driven in its instrumental performance by some of the most pressing astrophysical questions of our age, the observations of distant Type Ia supernovae and their role in the production of the most abundant heavy element, Iron.** The understanding of SNe Ia explosion physics is of critical importance for cosmology, astrophysics, and nucleosynthesis, and places the most stringent sensitivity requirements on ACT (Fig. E1). Therefore, we spent substantial effort studying this requirement. We believe this mission *must* achieve the Type Ia supernovae objectives, thus the discussion and recommendations that follow are couched in the context of this overarching objective. While most of the other science objectives will follow naturally from the ACT improvements in line and continuum sensitivity, some secondary objectives could potentially suffer from this focus.

The formation of stars and planets, and the development of the chemistry of life, can be understood only in the context of the creation and evolution of the elements. The nucleosynthesis products are interesting as signposts of our origins, as well as diagnostics of the poorly understood supernova explosion mechanisms. ACT's direct observations of nucleosynthesis products in diverse environments, including individual new supernovae as well as supernova remnants, will yield deep new insights.

## SUPERNOVAE AND NUCLEOSYNTHESIS

Supernovae have synthesized most of the elements heavier than He and supplied much of the energy input to the interstellar medium. Most of the newly created nuclei are indistinguishable from older nuclei that predate the star or were produced in hydrostatic phases; however, radioactive species among the ejecta serve as definitive tracers of recent nuclear processing. Some nuclei, such as the atomic mass A=56 chain, provide much of the power for the optical displays. The $\gamma$-ray lines from the decay of these nuclei reveal the location of the radioactivity within the ejecta through the time-dependence resulting from the unfolding of the attenuating material, and the ejection velocities of various layers from the line Doppler shifts. Spectroscopy and lightcurve measurements of these $\gamma$-ray lines allow direct measurement of the underlying explosion physics and dynamics.

## SNE IA: COSMIC YARDSTICKS, ALCHEMISTS

*ACT will provide unprecedented broad-line sensitivity, high spectral resolution (<1%) for resolving lines, and a wide FoV (25% sky) for continuous all-sky monitoring.*

SNe Ia, the thermonuclear explosions of degenerate white dwarfs, are profoundly radioactive events. As much as one-half of the white dwarf mass is fused to $^{56}$Ni ($t_{half}$ = 6.1d). After a short time, it is the decay of this nucleus and its daughter, $^{56}$Co ($t_{half}$ = 77d), which power the entire visible display of the supernova. Most of this power, however, is emitted in the form of $\gamma$-ray lines, some of which begin to escape after several days (Fig. E2). These $\gamma$-rays are the most direct diagnostic of the dominant processes in the nuclear burning and explosion.

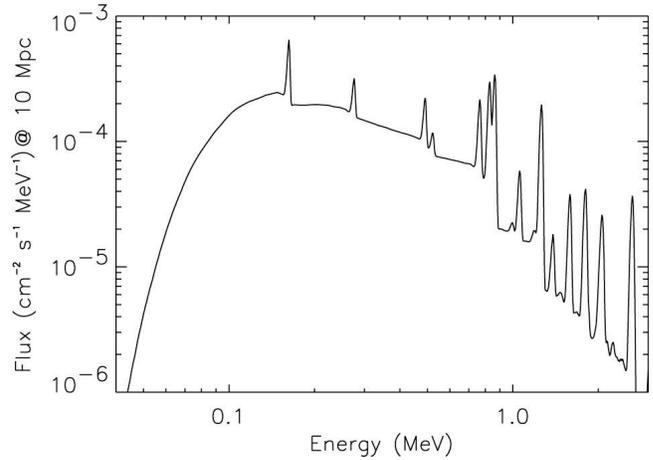

**Fig. E2.** *The $\gamma$-ray spectrum of delayed detonation model DD202c (see below) at 25 days post-explosion. At this time lines of both $^{56}$Ni (158, 749, 812 keV) and $^{56}$Co (847, 1238, 2599, 511 keV) are prominent, as well as the Compton-scattered continuum.*

Despite extensive study of SNe Ia, and their use in cosmology, major fundamental questions about these explosions remain unanswered. For example, we do not know their progenitor systems [13]. Almost certainly they occur in binary systems, but the nature of the companions, whether normal stars, white dwarfs, or some of each, is unknown. Thermonuclear supernovae are also grand experiments in reactive hydrodynamical flows. We do not understand how the nuclear flame propagates, how it proceeds as fast as it does, or how and when it turns into a shock (or even whether, though there is some evidence that it does) and propagates supersonically [42]. We do not know to what extent instabilities break spherical symmetry, or whether their effects are wiped out by subsequent burning if they do. We do not fully understand the empirical correction to their magnitudes that allows them to be used as standard candles for measuring the geometry of our Universe.

**Nuclear $\gamma$-ray lines from SNe Ia hold the key to solving these mysteries. ACT has three primary science goals for studying SNe Ia, and these goals are the primary drivers of ACT's instrumental performance requirements (broad-line sensitivity, spectral resolution, FoV).**

**ACT's goals are aggressive:**

**1. Standard Candles.** Characterize the $^{56}$Ni production distribution for SNe Ia, and correlate with the optical lightcurves to determine the relationship between empirical absolute magnitude corrections and $^{56}$Ni production.

*Requirements: measurement of $^{56}$Ni production in >100 SNe Ia at >5σ levels.*

**2. Explosion Physics.** Clarify the nuclear flame propagation and ejecta mass and kinematics for a key handful of SNe Ia, to uniquely distinguish among (or reject) current models of SNe Ia explosions.

*Requirements: high sensitivity (>15σ) lightcurves and high-resolution spectra (ΔE/E <1%) of several SNe Ia events of each subclass over the primary 5-year survey.*

**3. SN Ia Rate—Local and Cosmic.** Measure the SN Ia rates in the local universe, unbiased by extinction and solar constraints, and the cosmic SNe Ia history.

*Requirements: all-sky survey of SNe Ia, sensitive to distances beyond the Virgo cluster and measurement of the cosmic γ-ray background spectrum with sufficient energy resolution to separate the contributions from SNe Ia and AGN*

OSSE and COMPTEL detection of $^{56}$Co emission, and discrimination between SN explosion scenarios were two of the primary goals of the CGRO mission. However, the two SNe Ia that were observed with CGRO, SNe 1991T and 1998bu, were too distant to conclusively discriminate among dominant explosion scenarios [78,64,63,31].

**Standard Candles: Physics & Evolution**

Although many fundamental questions remain about SNe Ia, they are still being used as standard candles to measure cosmological distances with dramatic implications [84, 92]. Though intrinsically variable in their absolute magnitudes, the Phillips relation can be used to correct the measured luminosities, $\Delta m_{15}$(B). This empirical relation, deduced from the width in time of their visible light curves, assumes there is a one to one mapping between duration and luminosity [85]. Whether there is truly one physical parameter of the explosion that determines both is not certain. Whether the same relationship should hold exactly to redshifts approaching unity and beyond, when, for example, the metallicity of the systems was lower, is also not clear [28]. Complete confidence in the accuracy of using SNe Ia to measure such distances awaits better understanding of the explosions themselves.

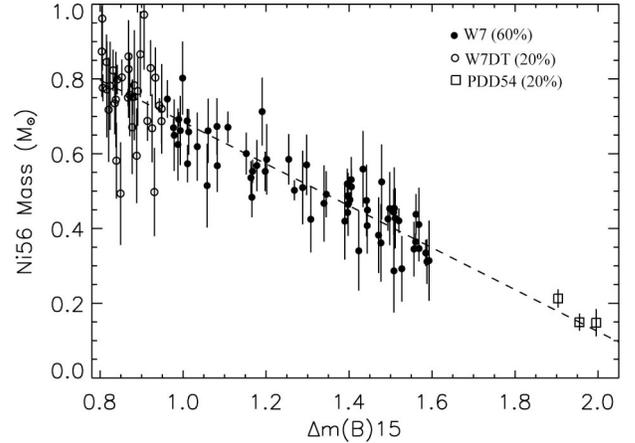

**Fig. E3.** *This figure shows an example of how ACT studies can quantify the relationship between SNe Ia $^{56}$Ni production and the optical lightcurve variations. Shown is a 5-year cumulative measurement of $^{56}$Ni from all SNe Ia detected above 5σ, assuming a ratio of SN explosions of 60:20:20 between normal (W7), superluminous (W7DT), and subluminous (PDD54) SNe Ia. It is assumed for this simulation that the $^{56}$Ni mass is correlated with the optical magnitude correction by the relation $M_{56}$=1.24–0.56*$\Delta m_{15}$(B) (dashed line).*

ACT will allow direct correlation for over 100 SNe Ia between the optical properties and the $^{56}$Ni production, which is likely to be the underlying factor in the optical lightcurve variations. For example, Fig. E3 shows a simulated distribution of measured $^{56}$Ni production versus the optical magnitude correction, assuming a linear correlation between the two. ACT is also sensitive to other possible causes of the light curve variation, such as $^{56}$Ni distribution in velocity (through spectroscopy), and total ejecta mass (through light curve monitoring), which can be distinguished from the total $^{56}$Ni mass explanation. Such measurements will allow us to directly probe the underlying physical mechanisms driving the variations in the optical curves. Coupled with ACT's ability to uncover the explosion mechanism and dynamics, a much better understanding of how SNe Ia evolve with redshift will be possible, as well as confidence in their use as standard candles to high z.

## Uncovering the Explosion Physics

Although a variety of SN Ia progenitors have been suggested, three dominant scenarios have emerged. (1) A single CO white dwarf accretes mass from a donor star until it (nearly) reaches the Chandrasekhar mass and suffers a thermonuclear runaway. (2) Two smaller CO white dwarfs merge to form a single CO white dwarf of the Chandrasekhar mass surrounded by an envelope, perhaps leading to an off-center thermonuclear runaway. (3) A single, low mass CO white dwarf accretes and burns hydrogen to form a thick helium shell that suffers helium ignition at its base, triggering a thermonuclear runaway in the CO white dwarf (sub-Chandrasekhar). Within the first scenario, it has been suggested that the burning front proceeds entirely sub-sonically as a deflagration, or alternatively, that it later accelerates to produce a detonation in the outer, near-surface layers of the exploding white dwarf (referred to as a delayed detonation).

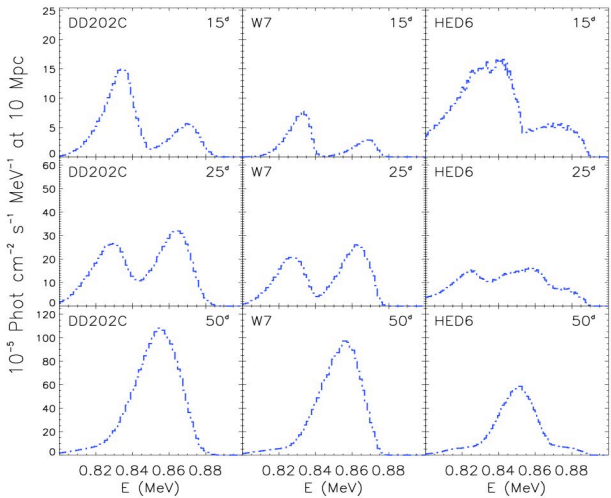

**Fig. E4.** *The spectrum of the three models described in the text at three epochs (15, 25, 50 days since explosion) in the energy range where bright $^{56}Ni$ and $^{56}Co$ lines are found. With sufficient signal-to-noise even moderate resolution instruments (~1% FWHM) can distinguish these models with single observations, and somewhat more easily from the evolution of the fluxes over longer times (the light curves).*

During the 1990's many SNe Ia were well studied optically, and the dominant explosion scenarios were simulated with radiation transport codes. It is the consensus (although not universally accepted), that the sub-Chandrasekhar mass explosions produce optical spectra that do not well match the observations. Generally, sub-Chandrasekhar mass models appear to be inconsistent with normal SNe Ia, but perhaps are able to explain some fraction of the peculiar events. This stance is perhaps too charitable, as the spectra predicted for these events have never been clearly detected from an observed SN Ia. Recent studies of the optical emission from individual SNe Ia find features that tend to favor delayed detonations over deflagrations (e.g., [94, 7]).

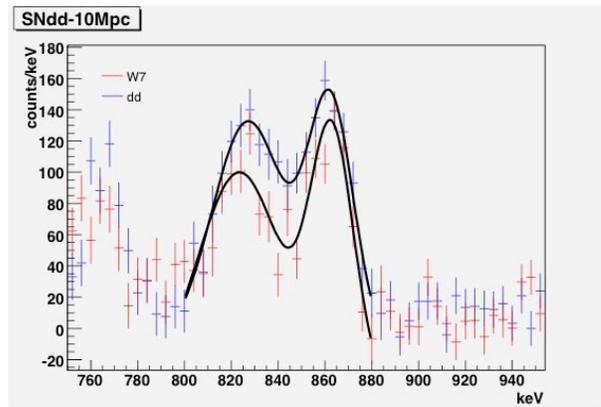

**Fig. E5.** *Simulated observations of models W7 and DD202c around 25 days after initial explosion and a distance of 10 Mpc, using the baseline Si/Ge instrument described below. All known background components are included in the simulation, as well as instrument characteristics and realistic event reconstructions. The solid lines show simple two-Gaussian fits to each spectrum. These two models are easily distinguished in a single observation of $10^6$ seconds at this distance.*

It has long been recognized that the amount and location of the $^{56}Ni$ in supernova explosions could be deduced from γ-ray line observations [17], and specifically that the differences between the Chandrasekhar mass explosions (that produce more total $^{56}Ni$, but bury it under a significant amount of more slowly expanding material) and sub-Chandrasekhar mass explosions (that burn a large fraction of their total mass to $^{56}Ni$ but produce lower total $^{56}Ni$ yields than the equivalent CM explosion) could lead to significant differences in the γ-ray line flux light curves and line spectra from SNe Ia (Fig. E4). Moreover, within the Chandrasekhar mass explosion scenarios the near-surface $^{56}Ni$ produced in the delayed detonation would also substantially alter the γ-ray properties from a deflagration explosion [44]. There has not been as much effort simulating merger scenarios as there has been in the other explosion scenarios, but indications are that their γ-ray spectra are intermediate between sub-Chandrasekhar mass models and Chandrasekhar mass models.

In this study, we concentrate on deflagration versus delayed detonation flame propagation, and, to a lesser extent, Chandrasekhar-mass versus low-mass models.

The most stringent performance constraint on ACT is set by its goal to definitively discriminate between deflagrations and delayed detonations for a handful of events during the ACT mission. To achieve this, we must discriminate between these two scenarios to 20 Mpc, i.e., roughly the distance within which there is one SN Ia per year. (We adopt that distance for our comparisons.). With peak line fluxes of ~$10^{-5}$ cm$^{-2}$ s$^{-1}$ for even near SNe Ia (~15 Mpc), which surpasses current detection sensitivities, it is clear that a significant improvement in sensitivity over previous and current missions is required (Fig. E1). We select three models for this investigation: a Chandrasekhar mass deflagration (W7 [80]), a delayed detonation (DD202C [44]), and a sub-Chandrasekhar mass explosion (HED8 [43]). We use the sub-Chandrasekhar mass HED8 in our study rather than the more commonly referenced HED6, because of the larger $^{56}$Ni production in HED8. By doing this, we compare three models that are more likely to represent normally-luminous SNe Ia.

Our requirement is that we distinguish the deflagration model from the delayed detonation model, even if the distance (and therefore absolute line luminosity) is unknown. Clearly, the combined information from optical and γ-ray studies will be utilized, but we set the goal: ACT must be able to detect the differences in the line profiles and/or light curve shapes alone. This is a difficult problem we have set in front of ACT: DD202C and W7 have similar γ-ray properties (Fig. E5). For the two models we calculate the spectra versus time, do a least squares normalization of the two sets, subtract them, and determine whether we can detect the difference for a given instrument spectral resolution, sensitivity, and observing strategy. Note that these lines are Doppler broadened to 3–5%, so that broad line sensitivity is (for this objective) the relevant instrumental requirement.

For distinguishing explosion models it is not adequate to set a single sensitivity requirement: instruments with higher spectral resolution can distinguish explosion models better than instruments with poorer spectral resolution, given the same sensitivity. Therefore, we developed curves that show our ability to distinguish explosion models as a function of both the instrumental spectral resolution and broad-line sensitivity (Fig. G8, Section G). The differences between the models manifest themselves in two primary ways in the γ-ray data: lightcurves of the emission lines, and the Doppler-broadened profiles of the lines. Therefore, both sensitivity and spectral resolution are required to distinguish the models. These requirements are discussed further in Section G, but typical requirements to clearly distinguish these two models (5σ) at 20 Mpc are: (1) a 3%-broadened line sensitivity at 847 keV below $1\times10^{-6}$ cm$^{-2}$ s$^{-1}$ (3σ in $10^6$ seconds), and (2) a corresponding spectral resolution better than 1%. The poorer the spectral resolution for an instrument, the stricter is the sensitivity requirement to distinguish these models.

### SNe Ia Rates—Local

ACT will detect a large number of SNe Ia, but it will be the closest few that will most illuminate thermonuclear supernova physics (Fig. E6). It is difficult to extrapolate the rates on larger scales to the very local, nonuniform universe, but actual SN Ia discoveries suggest that the rate within 20 Mpc distance is adequate for our goals. Simply taking confirmed SNe Ia in the past ten years, and taking distances of their host galaxies from Tully's (1988) Nearby Galaxies Catalog, gives the cumulative distribution of SNe Ia vs. distance shown in Fig. E6. Clearly this sample still suffers from incompleteness (even the brightest SN display a zone of avoidance along the galactic plane), but at least one SN Ia per year closer than 20 Mpc is assured.

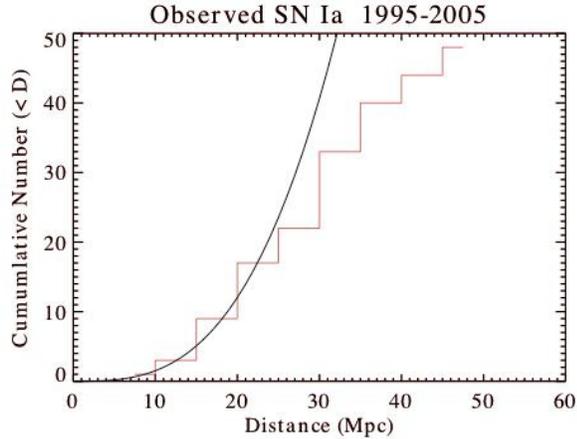

**Fig. E6.** *The cumulative distribution of nearby SNe Ia discovered in the past 10 years vs. distance. The smooth line shows the number expected if 150 per year are homogeneously distributed within 100 Mpc. The sample is by no means complete, but suggests that one per year closer than 20 Mpc is reasonable. ACT will detect $^{56}Ni$ emission from SNe Ia as far as 100 Mpc.*

ACT will survey the entire sky continuously, uniformly detecting SNe Ia to distances well beyond the Virgo cluster. Its wide FoV and all-sky survey strategy will discover over a hundred supernovae over its 5-year survey, and provide an unbiased, direct determination of the SN Ia rate in this volume—an independent check of SN Ia rates based on optical studies. In many cases ACT will discover relatively nearby SNe Ia before ground-based searches, including some that would be otherwise missed completely. Scanning much of the sky frequently and sensitive to nearby SNe 10-12 days pre-maximum, ACT will detect these events with high efficiency.

**SNe Ia Rate—Cosmic Gamma-Ray Background**

As noted originally by Clayton and Silk [18], it should be possible to measure the diffuse cosmic background of MeV γ-rays from the integrated SNe Ia in the universe, in addition to individual nearby SNe Ia. The accumulated and redshifted γ-ray line spectra form a continuum, with its normalization and spectral features probing the rate and redshift evolution of the contributing SN Ia. While both COMPTEL and SMM measured a γ-ray continuum in the MeV region [116, 115], it was unclear if it could be associated with the expected SN Ia signal [114, 95]. The normalization of the SN Ia prediction was quite uncertain, and the data were not precise enough to conclusively test for the expected spectral features.

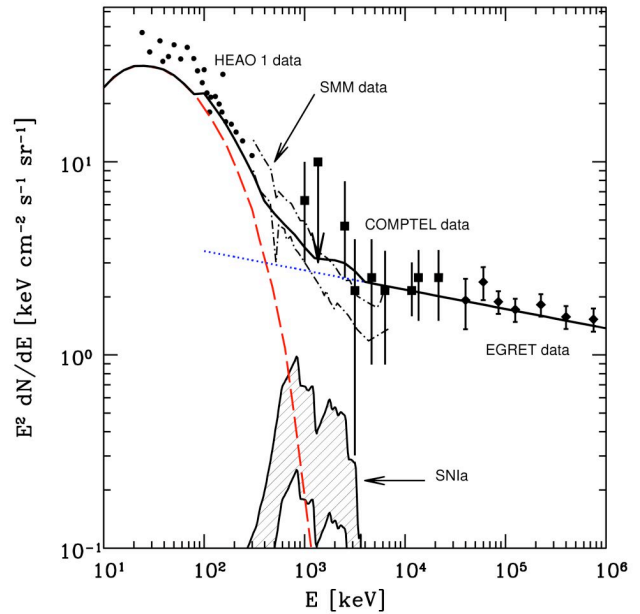

**Fig. E7.** *The cosmic γ-ray background, as measured by the HEAO-1 (circles), SMM (dot-dashed band), COMPTEL (squares), and EGRET (diamonds) missions [106]. The expected contributions from active galaxies, Seyferts and blazars respectively, are shown by the curves to reasonably reproduce the HEAO-1 and EGRET data. The shaded band is the expected contribution from SN Ia, using modern results for the star formation and supernova rate histories; surprisingly, it falls well below the measured data, leaving the origin of the signal unknown. A speculative extrapolation of the blazar background to lower energies is shown by the blue dotted line, and the sum of this with the supernova and Seyfert backgrounds by the black solid line.*

The SN Ia contribution to the MeV cosmic γ-ray background depends on three basic ingredients: (i) the cosmic star formation rate history, (ii) the progenitor models of SNe Ia, which determine the efficiency and time delay with which newly formed stars produce SN Ia, and (iii) the γ-ray emission per SN Ia. Recent improvements in the measurements of the star formation rate and cosmological parameters have helped constrain (i). Assuming ~0.6 $M_{sol}$ of $^{56}Fe$ per SN Ia, and the limited SN Ia rate data available now, the SN Ia contribution to the cosmic γ-ray background falls short of the measured data [106, 1].

This means that there is presently *no accepted explanation* for the origin of the MeV cosmic γ-ray background, as illustrated in Fig. E7. Speculative extrapolations of the backgrounds from active galaxies may account for some of the data, as shown. As discussed in Ref. [106], the MeV region is of special importance for constraining models of exotic physics, e.g., dark matter decay or annihilation. In order to make further progress on resolving this mystery, better characterization of the rates and emission from SNe Ia and active galaxies are needed—ACT will provide this information.

The other essential item that ACT will provide is a new and precise measurement of the MeV cosmic γ-ray background. The science requirements focus attention on the importance of very good energy resolution (to test for possible spectral features) and control of detector backgrounds (to allow the full-sky diffuse measurement). In addition, good angular resolution (~1°) would enable angular correlations of the data with themselves and other tracers of cosmic structure, much as has been done for the cosmic microwave background data. As shown in Ref. [119], cosmic γ-ray background from sources that follow the observed clustering of galaxies should produce detectable angular correlation signals; in contrast, the signals expected in some exotic physics models will be purely isotropic. With very good energy and angular resolution, it may be possible to isolate the SN Ia contribution, even if it is small, which would provide an important test of the predictions of SN Ia progenitor and explosion models.

## Gamma Ray Bursts

*ACT will provide rapid localizations (5') over a wide FoV (25% sky), broad spectral coverage (0.2–10 MeV Compton imaging, 10 keV – 10 MeV spectra), a minimal detectible fluence ~3×10⁻⁸ erg cm⁻², accurate timing (1 μs), and novel capabilities for measuring polarization.*

The early successes of the Swift mission are pointing toward key questions that ACT will address in the post-Swift era. GRBs are proving a powerful tool for probing the transition of the Universe from the "dark ages" after hydrogen recombination to the era of the first luminous objects (e.g., stars and accreting black holes) which re-ionized the matter at z~10. To probe the Universe to such early epochs with individual objects one must rely on objects that are bright (detection), understood (evolution corrections), and present at this epoch. Galaxies are bright today, but were much less so during their early assembly. Supernova properties are well established, but the theoretical models of SN II, Ibc and SN Ia are far from complete.

GRBs emerge as a solution to some of these problems. The observed bimodal GRB distribution (in duration and spectral properties) has led to two classes of models. The long-duration GRBs are believed to be associated with the final stages of rapidly rotating, massive stars. The leading collapsar scenario, e.g. [68], predicts that a supernova should follow GRBs. This association has been established directly through optical spectroscopy in a few cases, e.g. the nearby (z = 0.168) pair GRB030329-SN2003dh. But even when the sources are so far away that spectroscopy is not possible, extra light above the power-law afterglow lightcurve (see Fig. E8) has been detected in essentially all cases [118]. It is now believed that all long duration GRBs are associated with supernovae, providing a very bright (due to beaming) source that is easy to spot at even the highest redshifts, tracing star formation to a few million years after it started in the early Universe. The current redshift record is z(GRB050904) = 6.29 [48], which is at par with the most distant quasars and galaxies. Conceivably, GRB progenitors may even prefer a low metallicity environment (Woosley, private communications), which would make them an even better probe of the first generation of stars. The sensitivity and FoV of ACT promises a significant sample of high-z GRBs, continuing to enable the utilization of GRBs as a probe of the reionization epoch.

GRB afterglows provide the necessary illumination to sample the chemical composition of material along the line of sight—probing "cosmic chemical evolution"; i.e., metallicity as a function of redshift, Z(z). Metallicity measurements (Zn is commonly used as a substitute for Fe) through spectroscopy are a well-established method for QSO damped Lyman-alpha (DLA) systems. QSOs provide a few hundred lines of sight, and sample a wide dynamic range in distance and host column densities. New surveys will increase the sample size in the future, but GRBs are a potentially larger source of new directions and distances. The CCE Z(z)-relation established with QSO-DLA systems can be augmented with GRB-DLAs (e.g., [98]). Fig. E9 shows an example of a GRB-DLA, obtained with FORS at the VLT [112]. The afterglow (at z = 3.32) propagates through an effective neutral hydrogen column density of about $10^{22}$ cm$^{-2}$, larger than any (GRB- or QSO-) DLA HI column density inferred directly from Ly-alpha absorption. The host galaxy that underlies this GRB is very faint, AB ~ 28. Metallicity, as derived from the detected sulfur lines, is [S/H] = -1.26, indicating a low Z host/circum-burst environment. However, this metallicity is not low considering the high redshift of this burst. In fact, the GRB-DLAs seem to exhibit enhanced metallicity in comparison to the QSO-DLAs!

**Fig. E9.** *Spectrum of the afterglow of GRB 030323 at z = 3.372, showing the DLA and metal absorption lines [112]. The hydrogen column density measured in this GRB-DLA is very large, N(H) ~ $10^{22}$.*

Short, hard bursts are now known from Swift measurements to come from events at lower redshift (typically 0.2) and appear to have all the hallmarks of neutron-star coalescence or black-hole neutron star coalescence. Like supernovae they require a delayed power source to explain their later light curves [57], which might also be radioactivity. Those models feature very small $^{56}$Ni masses, for example, which would be detectable by ACT only at very close distances. Both the prompt and delayed emissions might be of great interest for ACT studies.

In addition to probing the first epochs of star formation, there is recent evidence that prompt GRB emission could be usable as calibrated standard candles themselves [33, 32] to study the expansion of the Universe to redshifts larger than currently reached by SNe Ia (Fig. E10). To use them as such requires good measurements of: (1) the energy of the spectral ($vF_v$) peak, (2) the burst fluence, (3) the time of the break in the afterglow slope due to the jet geometry, and (4) the redshift. ACT, with its large sensitive area from 10 keV – 10 MeV can measure the first two directly, and positions can be determined rapidly onboard to about 5'. This is adequate for follow-up measurements of the last two quantities. As many bursts peak at >400 keV, ACT is an ideal instrument for measurement of the required spectral properties for a wide range of bursts.

**Fig. E8.** *Essentially all GRB afterglows (z< 0.7) show late-time Supernova light components [118].*

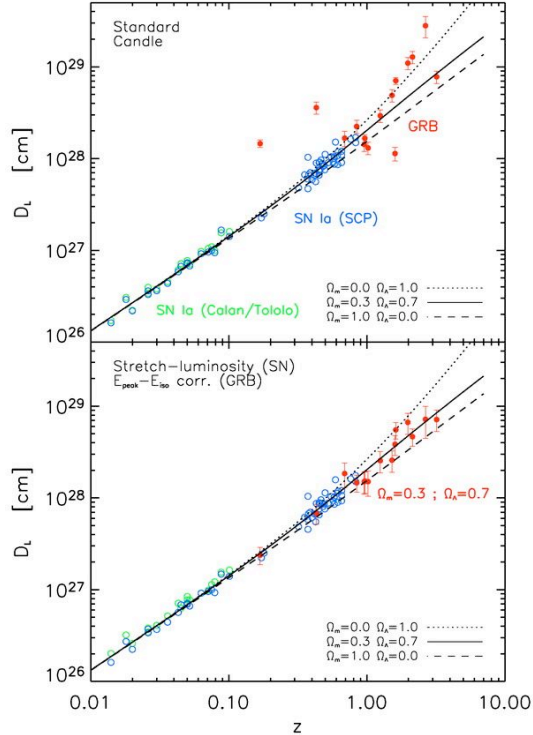

**Fig. E10.** *Hubble diagrams (luminosity distance versus redshift) for SN Ia and GRBs [32, and references therein]. Top: assuming the SNe Ia and GRBs are perfect standard candles. ($E_{GRB} = 10^{51}$ ergs) Bottom: SNe Ia are corrected using the empirical stretch-luminosity relationship, GRBs are corrected using the empirical $E_{peak}$-$E_{iso}$ relationship. With their young stellar progenitors, GRBs have the potential to probe the cosmic geometry to much higher redshifts.*

Given its broad spectral band (10 keV – 10 MeV), excellent timing (µs), high spectral resolution, and sensitivity to polarization, ACT is also an ideal instrument for studies of prompt emission processes in GRBs. The non-thermal spectra of bursts are commonly interpreted as synchrotron and inverse Compton radiation from electrons accelerated to ultra-relativistic energies in internal shocks (e.g., [73]). The required magnetic fields are substantial, and assumed to be generated in the forward/reverse shock zones of the jetted ejecta. Recent microscopic PIC simulations of the Weibel instability, a mechanism that can generate these fields [79], indicate that the field structure is highly tangled on length scales much shorter than the coherence length for the formation of synchrotron radiation. The resulting radiation spectrum must thus be derived in greater sophistication, and recent theoretical developments

(e.g., [29, 72]) suggest that observational challenges (from BATSE data) in the MeV regime [89] can be overcome with these new approaches. To address the open issue of whether classical synchrotron radiation or modified processes operate during the prompt GRB phase, ACT will provide high S/N hard X-ray spectra with good time resolution, so that we can study the temporal hard-to-soft evolution (e.g., [24]), determine the detailed spectral shape below the peak in the so-called Band function, and follow its correlation with the photon flux.

Polarimetric information will probe the GRB geometry and emission mechanism. Detections of high polarization (>40%) would strongly argue for the synchrotron mechanism (e.g., [23]), and suggest that field tangling can only arise on scales larger than the synchrotron coherence length. Polarization evolution through the burst can provide diagnostic probes of the character of the underlying electron distribution, such as the relative thermal and non-thermal content, and perhaps also weigh in on whether or not self-absorption is present.

Afterglow observations and the associated studies of their host galaxies dominate the current emphasis of GRB studies, but investigations of the prompt GRB γ-ray signal may hold the key to a better understanding of the central engine and the jets they create. ACT can address questions about magnetic field generation in relativistic lepton/baryon outflows that are not accessible with UVOIR observations. ACT will serve as a GRB trigger and locator to drive ground- and space-based follow-up, as well as provide the opportunity to advance basic understanding of radiation processes in complex, relativistic plasmas.

### DIFFUSE GALACTIC NUCLEAR LINES

*ACT will provide a deep all-sky exposure over its 5-year survey, with two orders of magnitude improvement in narrow-line sensitivity and high spectral resolution.*

Diffuse line emissions from interstellar radionuclides, electron-positron annihilations, and nuclear excitations by accelerated particles afford us the opportunity to study stellar evolution, the ongoing production of the elements, and the most energetic processes throughout the Milky Way

Galaxy. The decay of $^{26}$Al shows directly (Fig. E11) a million years of massive star and supernova activity [88]. Given its wide FoV, ACT will accumulate deep exposures on persistent sources over its 5-year survey (>2×10$^7$ s), reaching narrow-line sensitivities below $10^{-7}$ cm$^{-2}$ s$^{-1}$, enabling a detailed study of the production of $^{44}$Ti, $^{26}$Al, and $^{60}$Fe in various types of supernovae. With greatly improved sensitivity and angular resolution, we expect these apparently diffuse emissions to be resolved, at least in part, into hundreds of distinct regions, which can understood in terms of individual massive-star groups visible at other wavelengths. At least 16 known massive-star remnants in our Galaxy and Local Group are expected to be detectable in at least one line assuming a five-year mission lifetime (Fig. E1).

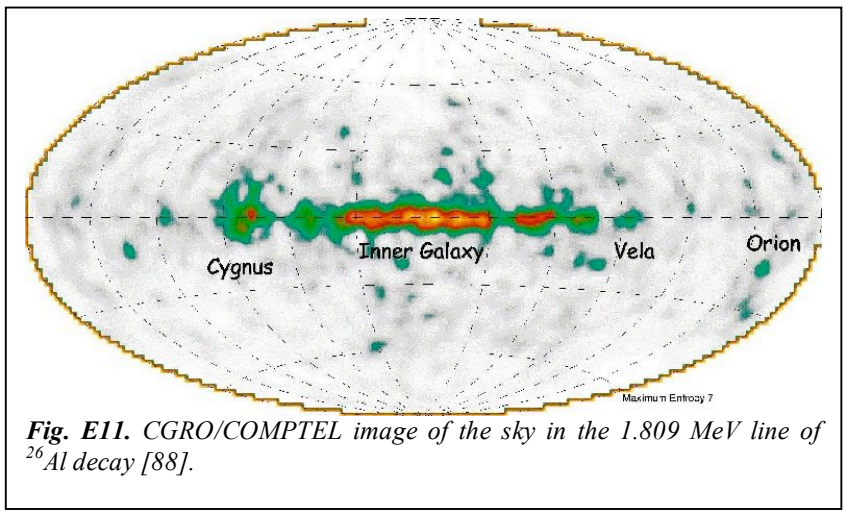

**Fig. E11.** *CGRO/COMPTEL image of the sky in the 1.809 MeV line of* $^{26}$*Al decay [88].*

supernova remnants, as well as from stellar ejecta already merged into the interstellar medium. Any minor contributions from longer-lived progenitors, AGB stars and classical novae might be identifiable from their smooth spatial distributions or sites far from any recent star formation. Through this nuclear decay we can study global galactic nucleosynthesis, the massive star content and evolution of associations, the nucleosynthesis yields of individual objects, and the potentially even the dynamics of a handful of supernova remnants. For example, precise measurements of the $^{26}$Al yield from nearby Wolf-Rayet star $\gamma^2$ Vel, and from the Vela and Cygnus Loop supernova remnants, will constrain models of the individual events and allow us to use the global galaxy observations to understand the rates and distributions of these sources. As the Vela SNR will be resolved by ACT, we can also hope to study the dynamics of the mass ejection, such as, for example, whether the fast fragments seen in X-rays also contain $^{26}$Al. The nearest AGB stars can be studied to finally assess directly their contributions to the global $^{26}$Al production. For nearby star forming regions, such as the Orion molecular clouds, where the present stellar inventory is well documented (and where there is some hint of $^{26}$Al line emission in COMPTEL data), we have the chance to study the massive-star inventory of the recent past.

## $^{26}$Al Decay

The proton-rich isotope $^{26}$Al decays to the first excited state of $^{26}$Mg at 1.809 MeV with mean lifetime of 1.04×10$^6$ yr. This is the most apparent radioactivity in the sky. In the *Compton Observatory* COMPTEL map we see a line flux of ~3×10$^{-4}$ cm$^{-2}$ s$^{-1}$ from the central Galaxy (longitudes ±30°), and smaller fluxes from a handful of other star-forming regions [88]. A source related to population I stars, most likely massive star winds and explosions, best explains this distribution. With ACT's wide FoV, greatly improved sensitivity, and extensive exposure, this map will separate into hundreds of regions of high significance. We expect to see $^{26}$Al 1.809 MeV emission from nearby clusters, OB associations, and individual

## $^{60}$Fe Decay

The neutron-rich isotope, $^{60}$Fe ($t_{half}$ = 1.5 My) is exclusively ejected from core-collapse supernovae, not by any other potential $^{26}$Al source. RHESSI and INTEGRAL have possibly detected [39] line emission from $^{60}$Co, the shorter-lived daughter of $^{60}$Fe. The total flux is an order of magnitude smaller than $^{26}$Al, but ACT will measure it in a number of distinct regions and individual sources, as well as from the central galactic plane. The total galactic production as well as individual source yields of $^{60}$Fe and $^{26}$Al will provide unprecedented constraints on core collapse supernova nucleosynthesis calculations. If the nuclear flame proceeds slowly in the initial burning in thermonuclear supernovae, a significant older stellar population could also be represented in $^{60}$Fe emission. The spatial differences between these two million-year radionuclides will teach us about the nucleosynthesis of both, among several classes of sources.

**Core Collapse Mass Cut and Jets**

In core collapses, $^{44}$Ti ($t_{half} = 59$ y) is in the deepest material ejected, providing an excellent probe of the explosion mechanism, specifically how the large neutrino energy is transferred to the inner ejecta. In the Cas A supernova remnant, the ejected mass of $^{44}$Ti will be determined to ~1%. If the late light curve of SN 1987A is, as often suggested, powered by $^{44}$Ti decay positrons, ACT will measure directly the ejected $^{44}$Ti mass to ~10%. If the solar abundance of $^{44}$Ca, which is almost certainly produced as $^{44}$Ti, is made in normal supernovae, ACT should discover of order ten more $^{44}$Ti supernova remnants in the inner Galaxy, which are below current sensitivities. If axial jets occur in core-collapses, as suggested by current understanding of GRB sources, some $^{44}$Ti (and other radioactivity) might be ejected with relatively high velocity [69]. The spectral resolution of ACT will allow this to be distinguished from the slowly ejected core material. Moreover, spatial resolution of the nearer remnants, possibly in several isotopes and via positron annihilation, (e.g., the Vela SNR in 26Al), will permit a complementing study of the dynamics of the explosion and expansion.

**POSITRON ASTROPHYSICS**

*ACT will provide a deep all-sky exposure to positron annihilation, with two orders of magnitude improvement in narrow-line sensitivity, high spectral resolution, and continuous all-sky monitoring for transient and variable sources.*

The bright positron annihilation line and triplet-state positronium continuum delineate the escape of positrons from extreme environments over millions of years. The bright bulge positron emission component remains a mystery. Interstellar positrons entrained in galactic magnetic fields will provide part of the complex 511 keV map, which will also feature individual supernova remnants and stellar and compact object wind nebulae, and possibly the galactic center. Given an energy resolution of ~0.5% at 511 keV, the conditions in the various annihilation media will be revealed through the line profile and the annihilation physics.

Despite the fact that the 511 keV line produced by the annihilation of positrons with electrons is the brightest γ-ray line in the sky, the dominant source of the Galaxy's positrons remains a puzzle. Efforts to solve this puzzle investigate the spatial distribution, the time-variability, the line profile, and the fraction of positron-electron annihilations that occur after the temporary formation of the hydrogen-like lepton atom, positronium. ACT will make great leaps forward in all aspects of these investigations.

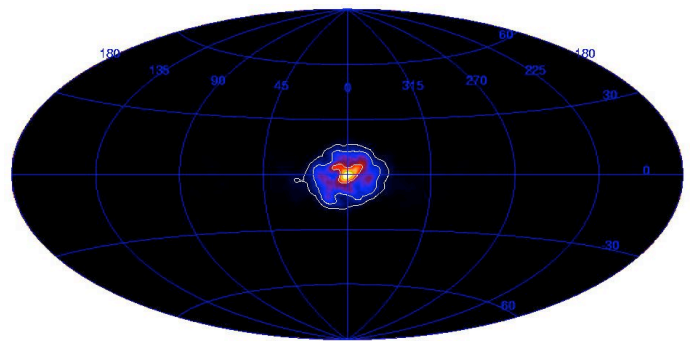

**Fig. E12.** *The intensity of the sum of 511 keV as mapped by INTEGRAL/SPI [51]. The apparent flux is dominated by the bulge component, whose positrons are of unknown origin.*

Observations made by the NASA CGRO/OSSE and ESA INTEGRAL/SPI (Figs. E12, E13) instruments suggest a time-invariant emission that features an anomalously large contribution from the galactic bulge/halo, relative to emission at other wavelengths [90, 49, 51]. As it becomes increasingly clear that the emission is brightest in the direction of the galactic center, but that the emission appears truly diffuse (rather than from a point source near the galactic center), new theories are being explored to explain the bulge/halo positrons. Mapping the spatial distribution of this emission, as well as constraining the time-variability, is limited by the low S/N of the OSSE and SPI detections. Over the course of the ACT mission, annihilation radiation will be mapped with very high significance and spectral resolution.

The 511 keV line-to-positronium continuum ratio suggests that upwards of 93% of positron-electron annihilations occur after first forming positronium [49]. A positronium fraction that high helps to exclude some media as the annihilation site for the bulk of the positron-electron annihilations. In addition, the profile of the 511 keV line appears to be composed of two components, a narrow line (~1.5 keV wide), and a broader line (~6 keV) [45]. Collectively, the positronium fraction, line widths and narrow-to-broad line ratio are consistent with the expected signature of positron-electron annihilation occurring in a warm medium [35, 45, 16]. The intense bulge emission dominates the observations of these instruments. ACT will enable the first detailed investigations of the differences in positron annihilation in the galactic disk and the galactic halo.

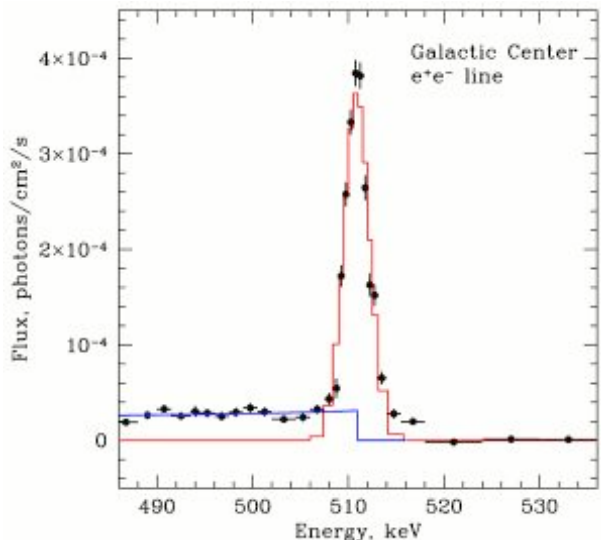

**Fig. E13**. *INTEGRAL/SPI spectrum of the $e^+e^-$ annihilation radiation from the galactic bulge. The red line shows the positron annihilation line, while the blue line shows the three photon continuum spectrum associated with the ortho-positronium decay [16].*

The mystery of the source of these galactic positrons underscores the infancy of positron astrophysics. Potential sources include hypernovae from an episode of starburst activity in the bulge [14], SNe Ia [19, 75], pulsar winds [113], and annihilations or decays of (as yet undiscovered) light dark matter particles [93, 9] (see [6] for constraints on this mechanism, [51] for a review of potential sources). For each of these potential sources, the contribution to the population of galactic positrons is dictated by the yield from the mechanism, the survival/escape of positrons from the source, the fraction of surviving/escaping positrons retained by the Galaxy, and the population of the source in the Galaxy.

The mechanisms for each of these potential sources reflect the myriad of ways in which positrons are generated in nature. Nucleosynthesis can produce positrons via $\beta^+$ decays, such as decays of $^3$He, $^{13}$N, $^{18}$F, $^{22}$Na, $^{26}$Al, $^{44}$Sc, $^{56}$Co (see [53] for a census of various positron-emitting isotopes). This mechanism occurs in explosive nucleosynthesis (SNe, novae, hypernovae, stellar winds), as well as in solar flares [54]. Pair production is another viable mechanism for positron production, occurring due to collisions of photons or due to the motion of electrons through strong magnetic or electric fields. This mechanism is expected to occur in all high-energy phenomena, including AGN, pulsars, accreting black holes, etc. An exciting mechanism, reflecting the increasing appreciation for the importance of dark matter in galactic studies, is the suggestion that galactic halo positrons are the by-product of light matter annihilations or decays, somewhat similar to earlier suggestions of positron production as the by-product of $\pi$ decays. It is through observations of candidates of each mechanism that the positron yields and transport from the source(s) will be understood.

ACT will make great strides in positron astrophysics by producing maps of annihilation radiation far superior to the OSSE and SPI maps. Importantly, it will also dramatically improve the sensitivity to potential point sources (Fig. E14), which will enable the detection of annihilation radiation from individual compact objects and SNRs. As an example, ACT will be capable of studying nearby SNe Ia remnants (e.g. Tycho's SNR, SN 1006, Lupus Loop, etc.). From the collective study of the annihilation radiation from these remnants, ACT will quantify the contribution of type Ia SNe to galactic positrons. The strength and alignment of the magnetic field in the SN ejecta determines whether positrons produced in the decay of $^{56}$Co to $^{56}$Fe escape the ejecta. Thus, studying annihilation radiation from SN Ia remnants incorporates a study of the magnetic fields of these remnants. Along the same vein, if a SN occurs within roughly 10 Mpc during the ACT mission, positron escape from a SN Ia could be observed directly via the 511 keV line to 847 keV line ratio [76]. For each of these potential sources, ACT observations of individual objects will offer new insight into the physics of that source, and into its contribution to the global galactic positron budget.

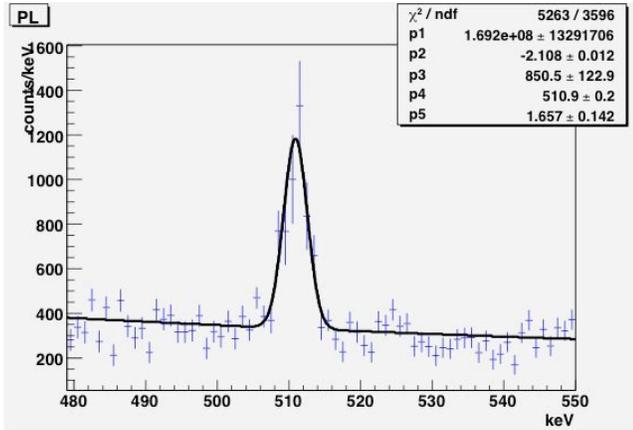

**Fig. E14.** *The simulated spectrum of a source with a power-law spectrum and 511-keV electron-positron annihilation line with flux $1\times10^{-5}$ ph $cm^{-2}$ $s^{-1}$ (i.e., 1% of the galactic center flux).*

## COMPACT OBJECTS

*ACT will provide broad spectral coverage, continuous all-sky monitoring for transient and variable sources, high spectral resolution to search for spectral features, fast photon timing, and novel sensitivity to polarization.*

### Active Galactic Nuclei

The *Compton Gamma-Ray Observatory* showed that nuclear activity in galaxies produces extraordinarily powerful and rapidly variable fluxes of γ-rays. The only compelling explanation is that a supermassive black-hole engine at the center of a galaxy generates this emission. The active galactic nuclei (AGNs) that have been detected at γ-ray energies fall into two distinct classes (see Fig. E15). The first class includes radio galaxies and Seyfert AGNs with redshifts z < 0.1 and >50 keV luminosities between $10^{40}$–$10^{45}$ ergs $s^{-1}$. The OSSE and COMPTEL instruments on *CGRO* provide strong evidence for a spectral softening or cutoffs above 100 keV for these AGNs. Blazars comprise the second class of γ-ray emitting AGNs. These include BL Lac objects, core-dominated flat-spectrum radio quasars, and highly polarized optically violently variable quasars. Blazars emit strong fluxes of radiation at 0.1–10 MeV energies (ν ~ $10^{19}$–$10^{21}$ Hz), and the bulk of their power output is often generated at these energies or higher. Blazars detected at γ-ray energies have a wide range of redshifts peaking near z ~ 1 and reaching to z ~ 5. The inferred isotropic γ-ray luminosity of these sources can exceed $10^{49}$ergs $s^{-1}$, as in the case of PKS 0528+134 at z = 2.07, making them the most luminous persistent sources in nature (see Fig. E16).

A precise characterization of the shape of the high-energy continuum radiation in Seyfert galaxies is important for testing the unification scenario for AGNs. In this scheme, Seyfert 1 and 2 galaxies differ primarily by the orientation of the observer with respect to the axis of the accretion disk. OSSE and *INTEGRAL* observations suggest that Seyfert 2 AGNs have lower high-energy cutoffs than Seyfert 1 AGNs, and X-ray observations indicate that the X-ray luminosity of Seyfert 2 galaxies is much less than that of Seyfert 1s. One interesting suggestion is that the high-energy emission of Seyfert 2 galaxies is Seyfert 1 emission that has been scattered by an ionized high-latitude gas. More sensitive observations with ACT can test this idea by comparing the measured spectra of Seyfert 2 galaxies with calculations of Compton-scattered Seyfert 1 radiation spectra. Knowledge of the high-energy continuum radiation of AGNs will also help establish the contribution of blazar jet radiation and the hard tails of Seyfert galaxies to the cosmic diffuse γ-ray background.

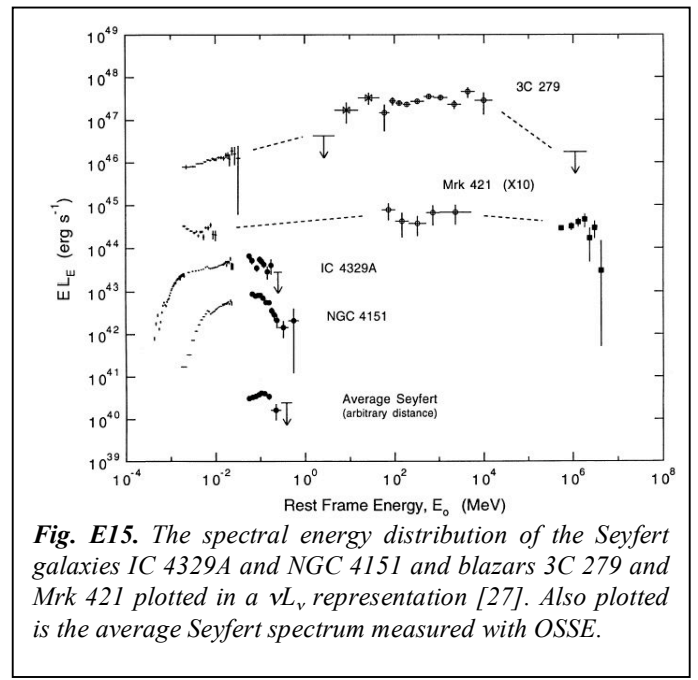

**Fig. E15.** *The spectral energy distribution of the Seyfert galaxies IC 4329A and NGC 4151 and blazars 3C 279 and Mrk 421 plotted in a $\nu L_\nu$ representation [27]. Also plotted is the average Seyfert spectrum measured with OSSE.*

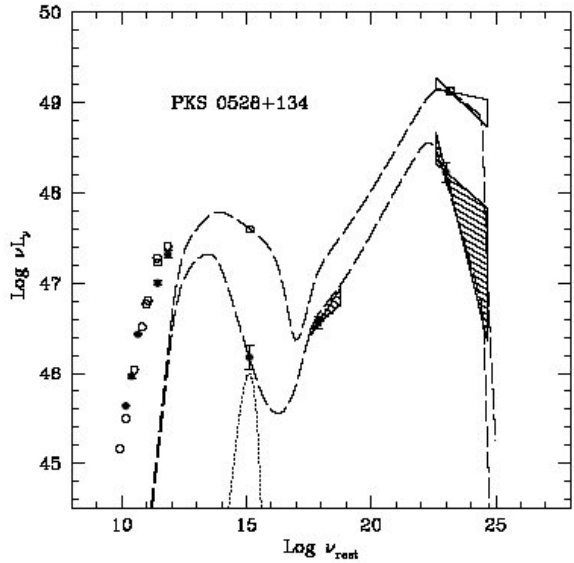

**Fig. E16.** *Spectral energy distribution of PKS 0528+134 from radio to γ-rays in the rest frame of the source [97].*

The enormous luminosities of blazars are so energetically demanding that they must result from relativistically beamed emission, as suggested by superluminal observations and the low fluxes of synchrotron self-Compton X-rays in compact core radio quasars. Observations of blazars at soft γ-ray energies provide some of the most compelling tests for relativistic beaming from γ-γ transparency arguments and the Elliot-Shapiro relation (Fig. E17). In the former test, γ-rays would be absorbed by lower energy radiation and unable to escape if the size of the emitting region, inferred from the variability time scale and light-travel time arguments, is too small. The detection of γ-rays from rapidly varying sources therefore implies that the emitting region is in motion. This test is best performed at soft γ-ray energies, where the threshold for γ-γ attenuation is largest and the cospatial assumption between target and detected photons can be avoided. The Elliot-Shapiro relation is based on the argument that if a source is assumed to be isotropically emitting and Eddington-limited, then the Eddington limit yields a lower limit on the black hole mass. This cannot be larger than the mass inferred from the variability time scale, which is limited by a size scale corresponding to the Schwarzschild radius. Violations of the γ-γ transparency test have already been established at soft γ-ray energies with the OSSE instrument. ACT will yield lower values to the speed of the plasma outflow that can be compared with superluminal observations to give the observing angle with respect to the jet axis, thus providing a completely new diagnostic for bulk plasma outflow and studies of blazars.

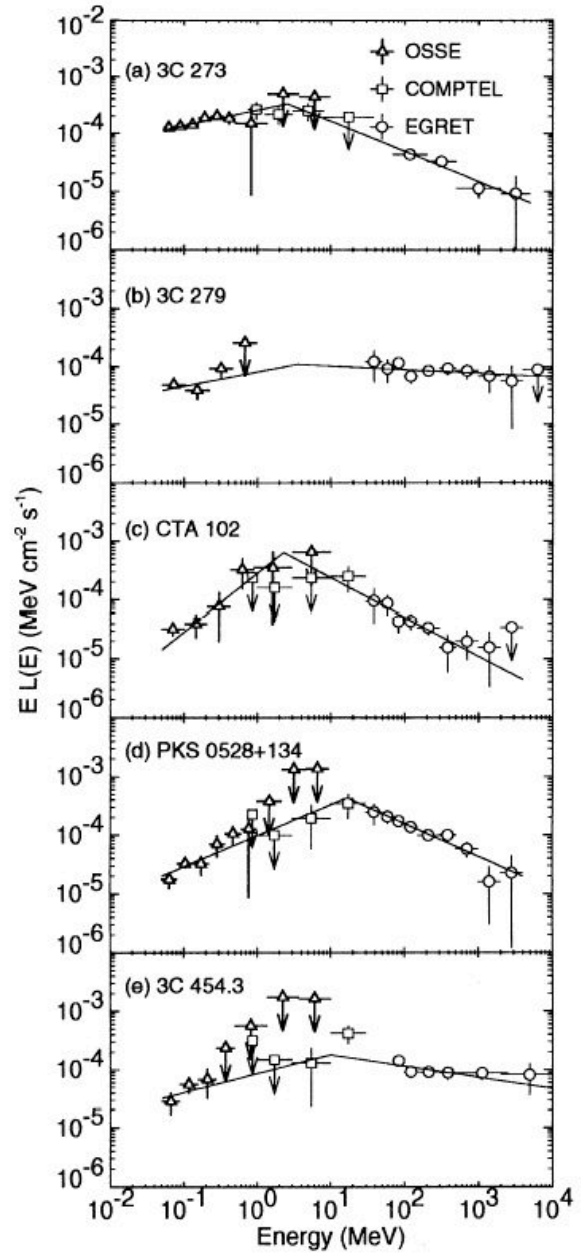

**Fig. E17.** *CGRO spectra of blazars, illustrating the peaking of their power output at γ-ray energies [71]. Spectral and variability information provide strong evidence that supermassive black hole jets ejecting relativistic plasma produce this radiation.*

High sensitivity studies of blazar jets will permit derivation of the bulk Lorentz factors for statistically complete samples of blazars of various types, including the flat spectrum radio quasars also known as "red blazars," the low-peaked BL Lac objects, and the high-peaked BL Lac "blue blazars." This will permit us to identify their parent radio galaxy populations, and to perform a statistical census of blazars through cosmic time in order to determine how supermassive black holes form and evolve.

Orientation effects from Doppler boosting are also thought to relate properties of radio galaxies, quasars, and blazars. This scheme can be tested by measuring weak hard MeV tails on radio galaxies from direct or scattered jet radiation and by discovering spectral hardenings at hard X-ray and soft γ-ray energies in sources with both disk and jet radiation. Observations of sources such as Cen A and 3C 273—which, according to the unification scenario, are misaligned and nearly aligned blazars, respectively—will clarify the relative contributions of jet and disk emission when blazar sources are observed at various angles to the jet axis. Another prospect for ACT is to discover "extreme blazars" with $\nu F_\nu$ peak synchrotron output in the soft γ-ray range. These sources—though with low luminosity in comparison to flat spectrum radio quasars and BL Lac objects—are hypothesized to exist as the most extreme type of blazar in the blazar sequence (Fig. E18). The blazar sequence has been challenged by the appearance of outliers, and ACT will provide a more secure observational basis.

Soft γ-ray observations of infrared luminous and starburst galaxies are also crucial for addressing the nature of AGN evolution. According to the standard scenario, galaxy mergers and interactions lead to infrared luminous sources that harbor active nuclei, which are fueled and uncovered as the dust settles. Searches for nuclear activity in infrared luminous galaxies will test the viability of this scheme, and such tests are most feasible at γ-ray energies because of the unique penetrating power of high-energy radiation. Indeed ACT, with substantial improvements of the continuum sensitivity over the OSSE and INTEGRAL instruments, may be the best way to find massive black holes. Such a black-hole survey capability would provide compelling evidence for or against our present ideas about quasar genealogy and the nature of the blazar sequence, which has been hypothesized to represent an evolutionary progression of black hole fueling and growth.

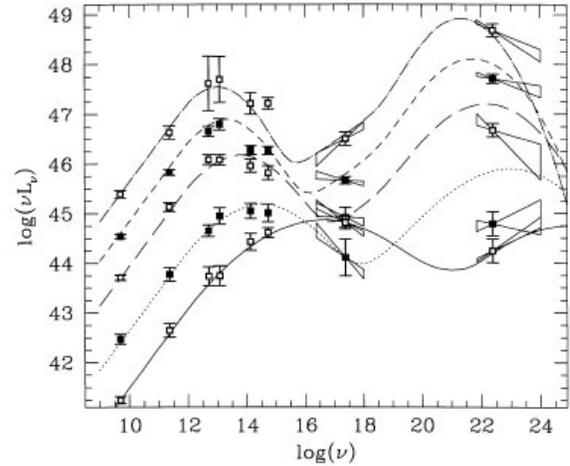

**Fig. E18.** *The blazar sequence, showing the progression of high luminosity flat spectrum radio quasars with synchrotron $\nu L_\nu$ peak frequencies $\nu_{pk}$ at radio-IR energies, to low-luminosity BL Lac objects with $\nu_{pk}$ at optical to X-ray energies* [30].

The detection of $e^+$-$e^-$ annihilation line radiation from AGNs would markedly change our understanding of the conditions in the innermost regions near a black hole. The most likely origin of electron-positron pairs in AGNs is through pair production from γ-γ attenuation, which requires energetic electrons and photons in a compact environment. Although the detection of a broad pair annihilation line from Seyfert galaxies is most feasible if a nonthermal electron spectrum is present, OSSE and *INTEGRAL* observations indicate that the Seyfert X-ray and γ-ray emission is mainly due to thermal electrons scattering soft photons. Thus a broadened annihilation line is not expected, though a narrow annihilation line from Seyfert galaxies could be formed if pairs escape from the central nucleus. For blazars, the composition of the jet plasma is not known, and the detection of transient pair annihilation lines would conclusively measure the amount of antimatter present in the jet. ACT could determine whether the MeV blazars discovered with COMPTEL on *CGRO* are due to Doppler-boosted annihilation radiation, as has been proposed. Searches for narrow pair annihilation lines and broadened features will also be conducted in luminous radio galaxies.

ACT could also help resolve the origin of the ultra-high energy cosmic rays (UHECRs). Blazars and GRBs are the most likely astrophysical acceleration sites for UHECR production because of the relativistic shocks that are formed in their jets. Hadronic cascade processes will lead to anomalous γ-ray components in the spectra of blazars that lack associated synchrotron components at lower energies. Neutral beams formed by the UHECRs will produce γ-ray halos around nearby luminous radio galaxies. Detection of such signatures, especially if accompanied by the detection of ultra-high energy neutrinos, would profoundly alter our view of how supermassive black-hole jets are formed and are powered, and the nature of the sources of UHECRs.

## Galactic Black Holes

Results from CGRO show that galactic black hole candidate sources and AGN are bright at MeV γ-ray energies, and display a wide range of spectral states both individually and among classes. Black hole X-ray binaries such as GX 339-4 and Cygnus X-1 undergo a series of transitions from the bright, soft X-ray state to a dimmer, hard γ-ray state where the photon energy of peak power output ($E_{pk}$) is at hundreds of keV.

The improved sensitivity of ACT will map these spectral states to test disk models for black holes accreting near the Eddington limit, and at low Eddington ratios where advection effects are thought to play an important role. The ACT large FoV will ensure the necessary coverage to catch many of these sources in and around transitions. ACT will measure weak, hard nonthermal components in the high soft state, and will chart the variation of the γ-ray cutoff energy in the low hard states.

Any observation of polarized soft γ-rays from binaries can probe the orientation of the accretion disk since Compton scattering models (e.g., [107]) predict that the achromatic polarization is a function of observer orientation to the direction normal to the disk. This can be a particularly potent geometrical diagnostic if transient annihilation line reflection features are also present, for which the kinematic coupling between energy and direction in Compton scattering can be used to distinct advantage.

Because black-hole X-ray binaries reach flux levels of $10^{-2}$ ph cm$^{-2}$ s$^{-1}$ MeV$^{-1}$ at 511 keV, ACT will be able to search for broadened annihilation line features from nonthermal compact plasmas that are predicted to surround accreting black holes. ACT will search for both broad and narrow annihilation lines in the galactic "microblazars" such as 1915+105 and 1655-40. Detection would finally resolve the speculation that black holes are truly the "great annihilators."

Also, neutrons produced by nuclear reactions in the accretion disks can be ejected in all directions. A fraction of these neutrons captured in the atmosphere of the secondary star will produce 2.2 MeV line emission. Fluxes for the closest XRB's (Cen X4, Ginga 2000+25, GRO J0422+32, XTE J1118+480 and A0620-00) can be detected by an ACT if the accretion rates are large enough [34].

## Neutron Stars & Pulsars

Neutron stars, both accreting neutron stars and radio pulsars, are known to be strong continuum sources above 100 keV. The phase-resolved continuum spectra from radio pulsars are indicators of the emission mechanism, as most of the emitted power is in γ-ray photons. Combining ACT data with that which will have already been collected by GLAST will help unravel the particle acceleration and photon production mechanisms in pulsar magnetospheres. ACT will be particularly suited to studying high-field pulsars such as PSR 1509-58, and perhaps magnetars with their newly-discovered hard components [56], where spectral structure and turnover information falls precisely in the ACT band: ACT can elucidate the magnetospheric locale of the emission region.

This niche will be amplified by ACT's polarimetric capability, since such spectral features can be strongly dependent on the photon polarization (e.g. [38]); observations of polarized signals in the 1–10 MeV band may discern whether or not the exotic QED process of photon splitting is occurs.

Even more exciting is the discovery space for γ-ray line emission from these same objects. There have been numerous unconfirmed reports of redshifted electron-positron annihilation lines from radio pulsars, at fluxes up to 1000 times the ACT sensitivity; given ACT's sensitivity and wide FoV, it will definitively resolve this issue. Pulsar winds might include substantial numbers of pairs as well, with estimates of positron production ranging from $10^{40}$–$10^{42}$ per second for some models. If the positrons, or a significant fraction of them, slow in the vicinity of the pulsar, as opposed to being accelerated to cosmic ray energies, pulsars could be a reasonably bright source of narrow annihilation line emission.

Accreting neutron stars can be bright continuum sources for ACT, especially at lower accretion rates when the inner disk appears hotter than normal, thus revealing emission from the innermost plunging region. A real gem would be the observation of a gravitationally redshifted γ-ray line from a neutron star. Any line originating from the radially-thin atmosphere would immediately yield the gravitational redshift (∝M/R), thereby constraining the nuclear equation of state at matter densities currently unavailable in the terrestrial laboratory. The first gravitationally redshifted atomic spectral line was reported from summed spectra of many Type I burst events, yielding R ~ 4.4 GM/c$^2$. A promising γ-ray line emission mechanism was first discussed by Shvartsman, who noted that matter accreting onto a neutron star has large enough kinetic energies to excite or destroy nuclei. The subsequent γ-ray line emission from nuclear de-excitation or neutron recombination (i.e. p(n,γ)d resulting in a 2.223 MeV photon) might just provide the needed probe of the gravitational redshift, with little uncertainty as to the height of the emission region. Later studies found that the flux of redshifted, rotationally broadened 2.2 MeV photons from Sco X-1 could be nearly 10$^{-6}$ cm$^{-2}$ s$^{-1}$. This estimate assumed that the accreting material has just a solar abundance of helium (the main neutron source). Just like during solar flares, neutrons liberated by the deceleration of the incident helium nuclei either recombine with atmospheric protons and emit 2.2 MeV γ-rays or suffer charge exchange on atmospheric $^3$He (i.e. $^3$He(n,p)T). Only a fraction of the liberated neutrons produce an unscattered 2.2 MeV γ-ray since the emission occurs at Compton scattering optical depths of order unity. This reduces the expected line emission so that only the brightest persistent X-ray sources (e.g., Sco X-1 and the GX sources) are detectable by ACT.

Other sources that would be excellent ACT targets would be bright accreting pulsars in outburst and any of the newly discovered ultracompact binaries in outburst, as these systems are accreting pure helium or a C/O mix that is neutron-rich and a likely source of emission lines.

## Classical Novae

Novae are explosions occurring on the surface of accreting white dwarfs in close binary systems as a consequence of explosive H-burning. γ-rays are expected from positron-electron annihilation (positrons are emitted by the β$^+$-unstable short-lived isotopes $^{13}$N and $^{18}$F, as well as by $^{22}$Na) and from nuclear lines emitted during the decay of the medium-lived isotopes $^7$Be and $^{22}$Na. $^{13}$N and $^{18}$F are synthesized in all nova types, whereas $^7$Be ($^{22}$Na) is mainly expected in novae from CO (ONe) white dwarfs, i.e. low-mass (high-mass) white dwarfs. Therefore, prompt emission consisting of a 511 keV line plus a continuum below this energy (with a cut-off at 20–30 keV) is expected [62, 40]. In addition, γ-ray lines at 478 keV ($^7$Be decay in CO novae) and 1275 keV ($^{22}$Na decay in ONe novae), are expected to last a couple of months and years, respectively [20, 21, 41]. It is known that around a third to half of novae occur on ONe white dwarfs (the global galactic nova rate is 35 per year).

The detection of the prompt γ-ray emission from novae is an important challenge for ACT, which would strongly constrain theoretical models. The duration of the emission, its intensity and the width of the line are clear indicators of the dynamical properties of the envelope, the efficiency of convective mixing, and the relative amount of $^{13}$N, $^{18}$F and $^{22}$Na. These relevant properties are impossible to measure through observations at other energies. Fundamental gaps in our understanding of novae remain after decades of intense study: the mechanism of the envelope enrichment, the nature of the envelope convection during the runaway, and the discrepancy between observed and calculated ejected masses [41]. It is important to keep in mind that a very large FoV as envisioned for ACT is crucial for this type of observation, since the positron annihilation transient γ-ray emission happens before the nova is discovered optically. In addition, ACT could detect essentially all galactic novae, whereas in the optical the vast majority of novae are never discovered because they are obscured by interstellar dust (the galactic nova rate can thus not be deduced from direct nova detections in our galaxy, but from extrapolations of observations in other nearby galaxies). Detectability distances for the 511 keV line with ACT are between 5 and 16 kpc ($^{13}$N peak lasting 1 hour, for all nova types) and around 1–2 kpc (for the $^{22}$Na positron tail, lasting around 4 years, for ONe novae), meaning that at least one nova per year should be detectable.

The detection of the long-duration γ-ray lines from novae is another interesting goal of the mission. The $^{22}$Na line has been extensively searched for, but remains undetected, in agreement with the predictions of the models of the last 10 years. ACT would enable this long anticipated detection, since the detectability distance for the 1275 keV line (relatively broad, $\Delta E \sim 20$ keV; $\Delta E/E \sim 1.5\%$) would be 4–5 kpc, and one ONe nova per year at this distance is guaranteed with high confidence. The detection of the $^7$Be line is more challenging, since it needs a distance around 1 kpc, and just about one nova of any type every 5 years is expected at this very short distance. Detection of $^7$Be would confirm that novae produce $^7$Li, which would help to understand the origin of galactic $^7$Li. $^7$Li still remains undetected in the optical, except marginally in Nova Velorum 1999. For $^{22}$Na, the direct measurement of this isotope from the NeNa thermonuclear cycle would constrain the temperatures attained in the explosion and the degree of core-envelope mixing. The distances mentioned here are somewhat conservative since the effective exposures for a wide-field ACT would be larger than $10^6$ sec.

## COSMIC ACCELERATORS

*ACT will have a wide FoV, large areas, high spectral resolution, and unprecedented sensitivity to narrow lines. ACT will provide a deep all-sky exposure over its 5-year survey, and continuous monitoring for flares.*

### Cosmic Rays

The 10–100 MeV/nucleon cosmic rays are an important ingredient in the interstellar energy budget, but remain undetected. Their interaction with the interstellar medium should produce signature nuclear γ-ray lines from spallation and direct nuclear excitation (most prominently from $^{12}$C and $^{16}$O) at diffuse levels that will be detectable by ACT. Enhancements could be significant where young supernova remnants are accelerating nuclei near cloud targets. Some of the radioactivity inferred in the presolar nebula (e.g., $^{10}$Be) was probably produced by particles accelerated by the active young Sun. Nearby active protostars are thus good candidates for ACT detection of nuclear lines. Continuum emission from cosmic-ray electrons interacting with the interstellar medium via bremsstrahlung and inverse Compton scattering of low energy photons will also be studied in detail. ACT will also better identify the contributions of numerous point sources, which have heretofore been included in nominally diffuse measurements, to the observed continuum.

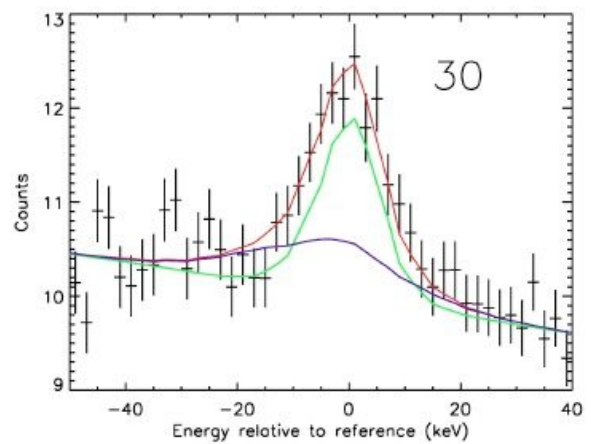

**Fig. E19.** *Expected nuclear de-excitation line profiles for lines stimulated by protons (green) and α's (purple), with their sum (red) fitted to RHESSI data [104]. For this plot, RHESSI lines from Mg, Ne, and Si were shifted and added, and data from three flares were summed. With ACT, far better spectroscopy will be possible on each of these lines in each individual flare.*

### Solar Physics

The Sun is a particle accelerator of astounding efficiency: on the order of half of the magnetic energy which fuels a flare goes into energizing coronal electrons and ions in nonthermal distributions, sometimes reaching GeV energies. The acceleration mechanism(s) remain mysterious, but when understood they may shed light on particle acceleration in other astrophysical contexts. Gamma-ray line diagnostics in flares reveal the energy spectrum, composition, and angular distribution of the accelerated particles, placing constraints on models of acceleration.

Accelerated electrons are studied by the bremsstrahlung X-rays they produce in the denser chromosphere and the radio emissions they produce in the high corona. Accelerated ions are studied via three kinds of line emission in the ACT energy range: nuclear de-excitation, neutron capture (2.2 MeV), and positron annihilation (the positrons are produced by radioactive nuclei created via spallation and by $\pi$ decay). The relative strengths of the de-excitation lines constrain the solar atmospheric abundances of the elements (which can vary from place to place and with height in the atmosphere). But more importantly, since each excited nuclear state has a different threshold energy for the ion stimulating it, the energy spectrum of the accelerated particles is also constrained. The lines resulting from spallation (annihilation and neutron capture) represent the effects of higher-energy ions.

The de-excitation occurs on picosecond timescales, and thus the lines are Doppler-broadened and redshifted due to the nuclear recoil from the exciting collision. Spectroscopy of individual de-excitation lines thus constrains the angular distribution of the ions as they reach the chromosphere from the corona—a function of the amount of scattering that occurs in the loop—and even the tilt of the loop with respect to the solar surface [103]. Since accelerated $\alpha$ particles create more recoil, they produce a characteristic red tail to the de-excitation lines, which constrains the accelerated $\alpha/p$ ratio (Fig. E19). The broadening of the de-excitation lines is typically on the order of 1%, so energy resolution at approximately that level or better is necessary. Characteristic lines that are stimulated only by accelerated $\alpha$'s provide another constraint on the accelerated $\alpha/p$ [101].

The shape of the positron annihilation line and the amount of continuum at lower energies (produced by the annihilation of orthopositronium) together constrain the density, temperature, and ionization state of the medium in which the positrons annihilate. This is a unique probe of the flaring atmosphere, a medium that is difficult to study in any other way. Recent high-resolution measurements of the annihilation line with RHESSI (Fig. E20) suggest an unexpected hot but dense medium [7, 9]. The broadening, although due to a very different process than that of the de-excitation lines, is also on the order of 1%.

ACT will also be capable of making a measurement that has never been made before: the production and decay of radioactive $^{56}$Co in the wake of a very large flare. The 847 keV line after a major X-class flare is expected to be on the order of $1 \times 10^{-5}$ cm$^{-2}$ s$^{-1}$ [91], easily detectable by ACT. Its intensity will provide another measurement of the accelerated ion fluence above the threshold for production of the isotope. More importantly, its pattern of decay (during periods when the active region is on the front of the Sun) should show the effects of mixing of material from the chromosphere and photosphere to deeper layers where the line will not escape. There is no way to measure this mixing without this radioactive tracer.

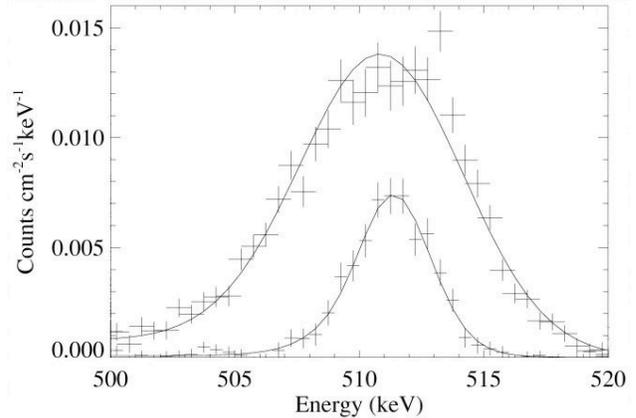

**Fig. E20.** *Positron-annihilation line shapes during the mammoth flare of October 28, 2003 from RHESSI [102]. During the middle of the flare, there is a fairly abrupt transition between the broad line (larger) and narrow line (smaller), indicating that the annihilating medium either cools quickly or shifts to a different region that is already cooler.*

The detection of both the flare and post-flare lines requires, of course, that ACT be designed to allow the Sun in its field of view. The front layers of any detector design will experience enormous fluxes of hard X-rays during any flare big enough to produce $\gamma$-ray lines. Simply ignoring the first few g/cm$^2$ of the detector volume should suffice to take care of this problem, but it is important to keep flare conditions in mind when designing the electronics and data system.

## F. COMPTON TELESCOPES

The principle of Compton imaging of γ-ray photons is illustrated in Fig. F1 [111]. An incoming photon of energy $E_\gamma$ undergoes a Compton scatter at a polar angle θ with respect to its initial direction at the position $r_1$, creating a recoil electron of energy $E_1$ that induces the signal(s) measured in the detector(s). The scattered photon then undergoes a series of one or more interactions ($E_i, r_i$), which are also measured. The initial photon direction is related to the scatter direction (measured direction from $r_1$ to $r_2$) and the energies of the incident and scattered γ-rays by the Compton formula:

$$\cos\theta = 1 + \frac{m_e c^2}{E_\gamma} - \frac{m_e c^2}{E_\gamma - E_1}$$

The total energy $E_\gamma$ of the incident γ-ray is determined by summing the measured interaction energies in the case of total absorption (which can be uniquely verified for 3 or more interaction sites [10, 81]), or calculated for partially absorbed photons that undergo three or more interactions [58]. By measuring the position and energy of the photon interactions in the instrument, the event can be *reconstructed* through use of the Compton formula at each interaction site to determine both the interaction ordering in the detector [10, 81] and the initial photon direction to within an annulus on the sky, generally known as the "event circle."

Very fast timing (sub-nanosecond) of these interactions can simplify reconstruction of the interaction order. Measuring the direction of the electron recoil in the first scatter can further restrict the initial photon direction to an arc segment on the event circle. Neither the fast timing nor electron tracking capabilities are assumed for the baseline ACT design, but were studied in a number of the alternate designs.

Two instrumental uncertainties contribute to the finite width of the event circle: the uncertainty in θ due to the finite energy resolution, and the uncertainty in the direction between $r_1$ and $r_2$ due to the finite spatial resolution. Both of these uncertainties contribute to the uncertainty (effective width) of the event circle δθ. There is also a fundamental limit on the width of the event circle set by Doppler broadening due to Compton scattering on bound electrons, which is higher for high-Z materials (see [120] and references therein).

Linearly polarized γ-rays have a higher probability of Compton scattering perpendicular than parallel to their polarization vector (along the electric field). This scattering property can be used to measure the intrinsic polarization of radiation from astrophysical sources by measuring a modulation in the distribution of azimuthal scatter angles in the instrument [61]. Compact designs maximize the efficiency for photons scattered at θ~90°—which are the most highly modulated, resulting in high sensitivity to polarized emission.

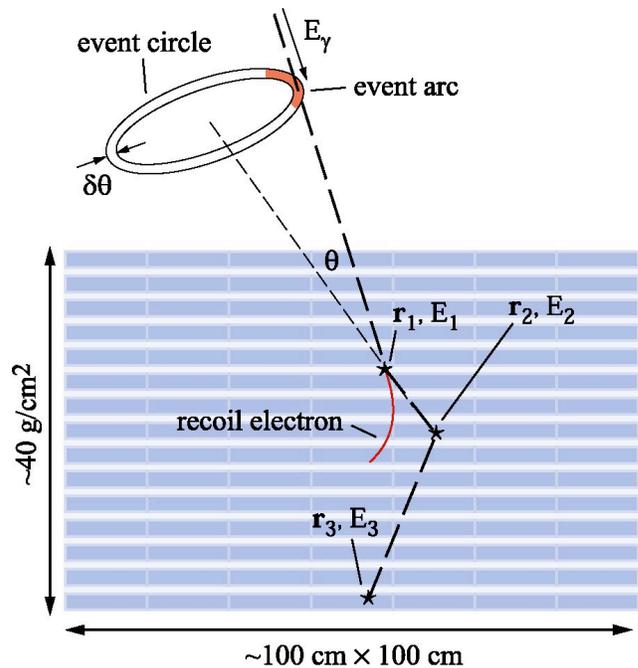

**Fig. F1.** *The principle of Compton imaging. By measuring the position and energy of each photon interaction, the initial photon direction can be determined through the Compton scatter formula to within an annulus (event circle) on the sky. Measuring the recoil electron direction could further constrain the event to an arc.*

Given the complexity of γ-ray photon interactions—including Compton scattering, photo-absorption, and pair production—as well as complex backgrounds produced by diffuse photons, cosmic rays and secondary neutrons, we must rely on Monte Carlo simulations to estimate the performance of Compton telescopes (Appendix P).

# G. Science Instrumentation

## Scientific Approach

The most promising instrument design for fulfilling the science requirements set forth in Section E is a Compton telescope utilizing recent advances in detector technologies to achieve significant improvements over the Compton Telescope (COMPTEL) on CGRO (Fig. G1). ACT utilizes advances in the Compton imaging technique pioneered by COMPTEL [99]. COMPTEL employed a design with two widely separated (1.6 m) scintillator detector arrays, with moderate 2-D position resolution (~1 cm$^2$) and energy resolution ($\Delta E/E \sim 10\%$). The large separation of the detectors enabled time-of-flight (ToF) measurement, and thus helped reduce background, but resulted in a small efficiency (<0.3%) and relatively small FoV (9% of the sky). COMPTEL achieved its success not with large effective areas, but with ToF measurements and other event-acceptance criteria which enabled suppression of the intense background in this energy range.

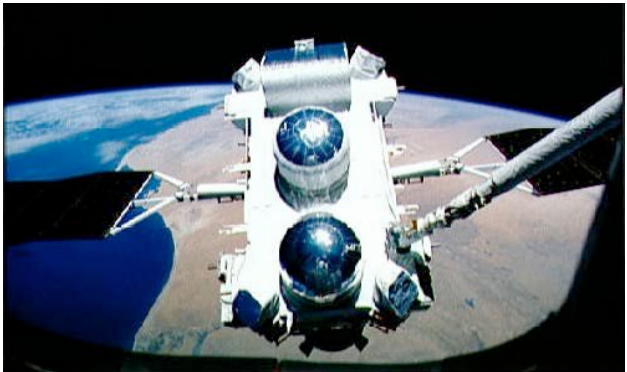

**Fig. G1.** *The Compton Gamma-Ray Observatory (CGRO) as it was being deployed from the space shuttle Atlantis (STS-37). COMPTEL is the instrument located in the center of this S/C, between the solar panel mounts.*

## Modern Compton Telescope Approaches

Modern 3-D position sensitive $\gamma$-ray detectors, arranged in a compact, large-volume configuration, will improve efficiency by two orders of magnitude, provide a powerful new tool for background rejection, and utilize high spectral resolution to dramatically improve sensitivity. [See Section F for a tutorial on Compton imaging.]

There are two required and two potential advances in detector technologies that will dramatically improve ACT sensitivity over COMPTEL: fine 3-D position resolution (1 mm$^3$), high spectral resolution (<1%), potentially tracking of the initial recoil electron for photon energies throughout the nuclear line range (0.5–2 MeV), and potentially very fast timing (sub-nanosecond) of interactions. There are a number of promising detector technologies for achieving these performance advancements. The final ACT instrument could be composed of a single detector type, or include combinations of these technologies to optimize performance.

*Fine 3-D position resolution* (<1 mm$^3$) enables Compton imaging within a uniform (or nearly uniform) detector volume, instead of requiring a scatter between two widely separated detector planes as in COMPTEL. ACT's design, utilizing a compact geometry, increases the instrumental efficiency by nearly two orders of magnitude, while also increasing the FoV to over 25% of the sky (Figs. G2, H4). All technologies considered can achieve these spatial resolutions (except LaBr).

*High spectral resolution* (<1%) combined with the fine position resolution provides a powerful new tool for background rejection (extremely important since $\gamma$-ray observations are background dominated). By resolving photon interaction sites and measuring the interaction energies with high precision, we can distinguish between external photon events that deposit their full energy in the instrument or undergo three or more interactions, and a variety of backgrounds that dominate. These backgrounds include: cosmic diffuse and earth albedo photons, as well as internal spallation products that result in multiple-site energy deposits, cosmic-ray induced $\beta$-decays, and pair production events—all of which have different spectral and spatial signatures in the instrument when resolved [10]. High spectral resolution also plays a more direct role in improving sensitivity to narrow lines. By limiting the instrumental energy bandpass covered by the line, the background under the line is significantly reduced. The most promising technologies in terms of spectral performance are Ge and thick Si.

*Tracking of the initial recoil electron* for photon energies throughout the nuclear-line range (0.5–2 MeV) could potentally improve the sensitivity even further by decreasing the fraction of background events imaged to the source location. Electron tracking has only recently been demonstrated in the critical nuclear line range (0.5–2 MeV) using silicon Controlled Drift Detectors [CDD] [15] and gas time projection chambers with micro-well readout [108]. Another promising technology for achieving this goal are thin Si strip detectors. The main challenges facing these technologies are (1) maintaining sufficient stopping power and spectral resolution such that the overall sensitivity is improved, and (2) additional complexity resulting from the large number of channels required for the high spatial resolution.

*Fast timing of interactions for ToF* (sub-nanosecond) could significantly aid in the proper reconstruction of the Compton scatter sequence and in fast (high-level trigger) rejection of background components such as upward scattered photons, γ cascades, or neutron scattering. For most detector technologies, the most likely sequence ordering is determined from Compton kinematics based on the energy and position measurements. Fast timing could enable the use of two-site interactions and remove some remaining ambiguities in multiple-site interactions. The most promising technologies for achieving very fast timing are Xe-based detectors and novel scintillators. The main challenge facing these technologies is achieving sufficient spectral resolution such that the overall sensitivity is improved by adding the fast timing.

In the course of this study we investigated a number of different detector technologies that each combine two or more of the above capabilities. **It is clear from our investigations that the first two capabilities (fine 3-D position resolution, high spectral resolution) are critical for ACT performance. The other detector capabilities (electron tracking, fast timing) are potentially very useful, but only if they can be implemented without significantly degrading the position resolution, energy resolution, or overall stopping power of the instrument.**

While we investigated in detail a number of different instrument concepts for ACT during this study (Appendices Q–W), there are several design parameters that are nearly universal (Fig. F1). The optimal ACT design consists of at least 2m² arrays of position-sensitive spectroscopy detectors in order to achieve effective areas of ~1000 cm². The depth of the sensitive volume for stopping MeV γ-rays must be 35–50 g/cm², with a weak dependence on the detector materials. Detectors must be reliable, uniform, and radiation hard. All designs require a surrounding charged-particle anticoincidence detector (ACD), composed of thin scintillator plastic, surrounding the detector array to veto cosmic-ray and trapped particle backgrounds. In addition, we have determined that in LEO, the dominant background component for all designs studied is Earth albedo γ-rays. A massive bismuth germanate (BGO) scintillator shield beneath the detector array in the baseline model helps suppress this background.

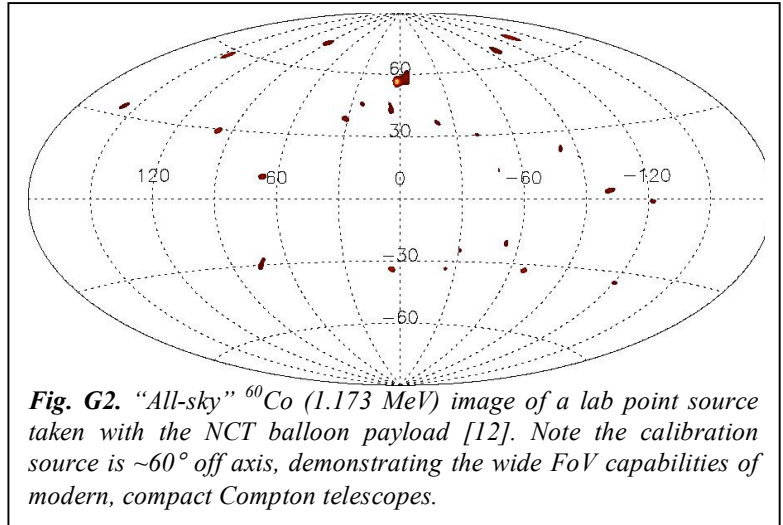

**Fig. G2.** *"All-sky" $^{60}$Co (1.173 MeV) image of a lab point source taken with the NCT balloon payload [12]. Note the calibration source is ~60° off axis, demonstrating the wide FoV capabilities of modern, compact Compton telescopes.*

## BASELINE SCIENCE INSTRUMENT

Due to the one-week duration of the ISAL study it was not possible to study a variety of instrument designs based on different detector choices. The ISAL study in particular happened too early in the study for the comparative detector concept studies to influence the choice of instrument studied. Thus we selected a "baseline" instrument for the study that was designed both to be a promising option scientifically, and to encompass a variety of the challenges an ACT instrument might pose. This *initial* baseline concept was chosen as a hybrid Si-Ge array, consisting of a 32-layer "D1" array of 2-mm thick silicon (low Z) detectors, situated immediately above a 3-layer "D2" array of 16-mm thick germanium (high Z) detectors (Fig. G3). Scientifically, this hybrid design represented a promising choice because it combines the higher intrinsic angular resolution achievable by having a first scatter in the low-Z silicon (less Doppler broadening) with the better stopping power of the high-Z germanium for higher efficiency. Both of these technologies assumed the excellent spatial resolution (<1 mm³) and excellent spectral performance (~0.2% at 1 MeV) that has been achieved in the laboratory. *Neither detector array assumed any electron-tracking or fast-timing capabilities.* This hybrid design seemed especially appropriate for the ISAL run because it combined the technical challenges of cooling a large array of Ge detectors to 80 K with that of powering a large number of readout channels with moderate cooling for the Si detectors (~300,000 channels, -30° C).

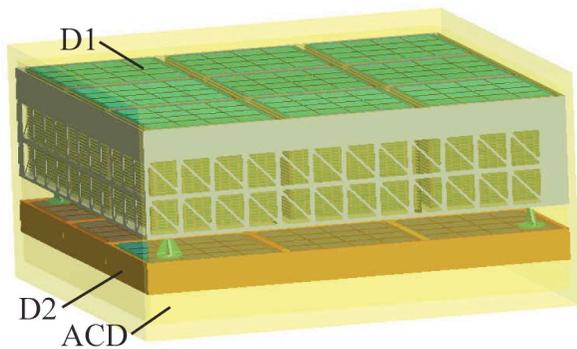

**Fig. G3.** *Baseline ACT design studied in the ISAL and IMDC runs. An optimized version of this basic design was studied in the detailed performance simulations. D1: 32 layers of 2-mm thick Si detectors. D2: 3 layers of 16-mm thick Ge detectors. (Optimized: 27 layers D1, 4 layers D2.)*

The Si-Ge instrument studied during ISAL constituted a first reasonable guess at an optimal configuration. Later tradeoff studies revealed that the instrument's sensitivity improves if four instead of three Ge layers are used, with a corresponding reduction of 32 to 27 Si layers to remain roughly within the same mass/power/size envelope, and the instrument's bottom and sides are surrounded by a 4-cm thick BGO shield. Below, we describe this improved Si-Ge baseline instrument and discuss its performance.

## Science Instrument Subsystems

### D1 Silicon Detector Array

The low-Z "scattering" array, D1, consists of 27 layers of 2-mm thick double-sided Si cross-strip detectors (Fig. G4). Each layer has 144 detectors, corresponding to 3888 detectors total. Each detector is 10×10 cm² in area, instrumented with 64×64 strips (1.5-mm pitch) and equipped with a 2-mm wide guard ring around the edges. Within one layer, 2×2 of these individual detectors' strips are connected to a "daisy-chained" readout with 248,832 channels, designed to reduce power consumption and passive material. We assumed a 0.5% failure rate of these individual electronics channels in our simulation of instrument performance. The Si layers are spaced 10 mm apart. Each Si detector is assumed to have ~0.2% energy resolution at 1 MeV, comparable to currently measured laboratory performance [59, 86]. The Si detectors must be cooled to -30° C in order to achieve their optimal spectral performance. It was determined in the ISAL run that mechanical cooling was the most efficient method for cooling this array (as opposed to passive cooling), and this approach was adopted for the IMDC run. The total active Si mass is 291 kg, with a thickness of 13 g/cm².

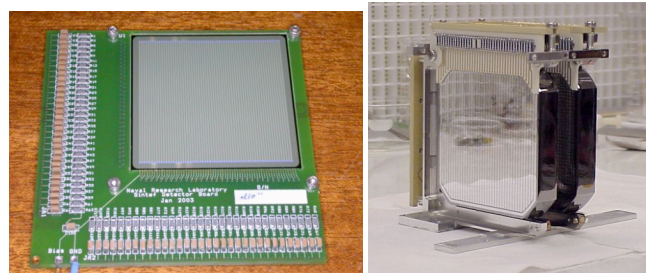

**Fig. G4.** *Detector technologies assumed for the ACT baseline instrument. Left: 2-mm thick Si cross-strip detector [59, 86], Right: 16-mm thick Ge cross-strip detectors [11, 22, 55].*

### D2 Germanium Detector Array

The high-Z "absorbing" array, D2, consists of four layers of 16-mm thick Ge cross-strip detectors (Fig. G4). Each layer has 144 detectors, corresponding to 576 detectors total. Each detector is 9.2×9.2 cm² in area, instrumented with 90×90 strips (1.0-mm pitch) and equipped with a 1-mm wide guard ring around the edges. The guard ring is instrumented to provide veto signals. Each strip is read out individually, for a total of 103,680 channels. We assumed 0.5% failure rate of these individual electronics channels in our simulation of instrument performance. The Ge layers are spaced 25 mm apart, center-to-center. Each Ge detector is assumed to have ~0.2% energy resolution at 1 MeV, and 0.4 mm depth resolution—comparable to currently measured laboratory performance [11, 22, 55]. The Ge detectors must be cooled to 80K in order to achieve their optimal spectral performance. It was determined in the ISAL run that a Turbo-Brayton cryocooler would be the most efficient method for cooling this array, which was the assumed method for the IMDC run. The total active Ge mass is 415 kg, with a thickness of 34 g/cm$^2$.

### Readout Electronics

ACT will require the design and development of low-power application specific integrated circuits (ASICs) to read out the hundreds of thousands of channels in the instrument. These ASICs must have low power, large dynamic range, low-noise to achieve the required energy resolution associated with these semiconductor detectors, and ~10 ns time resolution to provide the depth resolution within the Ge detectors. The ISAL assumed 1 mW/chn power consumption for these ASICs, based on optimistic performance and power developments. Development of low-power ASICs was identified as one of the primary technology development priorities for ACT.

Given the large volume detectors, and high event rates, of ACT, a coincidence time resolution of 150 ns or better is required for discriminating Compton scatter events from chance coincidences (Appendix P). This coincidence resolution will keep random coincidences from dominating the telemetry stream, while even further discrimination based on Compton kinematics can be performed to reject remaining random coincidences through ground-based processing

### Cryogenics

The ACT baseline instrument requires two separate cryogenic systems for the two detector arrays: a system that goes to 80K for the Ge (D2) detectors, and a system at -30° C for the Si (D1) detectors. Both the ISAL and the IMDC teams recommended that these be separate cryogenic systems given the very different cooling requirements of the two detector types (Fig. G5).

For the Ge detector array at 80K, assuming an internal power dissipation of 1 mW/chn within the cryostat at an intermediate temperature of 125 K, the heat lift requirements for the entire array are 61 W. The ISAL team identified the Turbo Brayton mechanical cooler used on HST NICMOS, combined with cooling loops, as the best candidate technology to fit these requirements. This crycooler offers the advantage of essentially zero vibrations, and it is staightforward to scale to larger sizes.

For the Si detector array at -30° C, assuming an internal power dissipation of 1 mW/chn at the detector temperature, the total heat lift requirements for the entire array are 253 W. The ISAL and IMDC teams identified the commercial Sunpower cryocoolers currently flying on the RHESSI mission as a good match for this array.

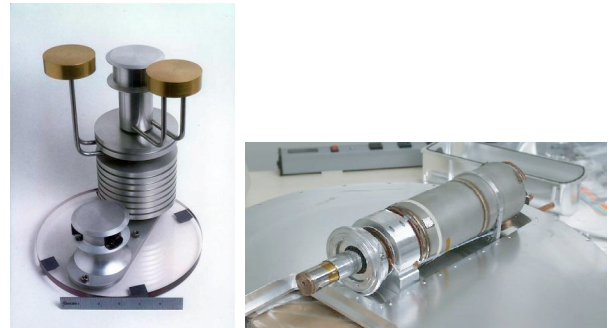

**Fig. G5.** *Cryocooler technologies for ACT. Left: compact Turbo-Brayton cooler used on NICMOS/ HST [123], recommended for cooling the Ge array to 80K. Right: Sunpower M77 Stirling cycle cooler which is flying on the RHESSI mission [65], recommended for cooling the Si array to -30° C.*

### BGO Shield

Two of the largest sources of background in an unshielded soft γ-ray detector operating in low-earth orbit are albedo photons and secondary photons from interactions of ambient photons, leptons and hadrons in the spacecraft (S/C). For a zenith-pointing instrument, both of these background components can be significantly reduced by covering the bottom and sides of the detector with an active shield that provides enough stopping power for most γ-rays to interact.

For this purpose, high-density, high-Z scintillators such as BGO are the material of choice. The optimum thickness for this shield has to be determined in a tradeoff study between (photon) stopping power on one hand and weight and neutron activation on the other hand; 3–4 cm appear a reasonable guess based on simulations and prior experience with INTEGRAL/SPI [110].

A 4-cm BGO shield for ACT comprises ~ 1000 kg of BGO, a 3-cm shield ~ 750 kg. This is on the same order as the BGO shield of INTEGRAL/SPI (517 kg, Fig. G6), with much less complexity in the required shapes of BGO crystals and consequently higher crystal light yields and less passive structural mass. We envision a redundant readout of each of 100–200 pieces of BGO with two separate PMTs or photodiodes with individually adjustable gains and thresholds for maximum adaptability of the system. The BGO must provide a prompt (< 0.5 μs) veto signal to the ACT trigger electronics whenever an interaction occurs above the trigger threshold of ~75 keV (Appendix P).

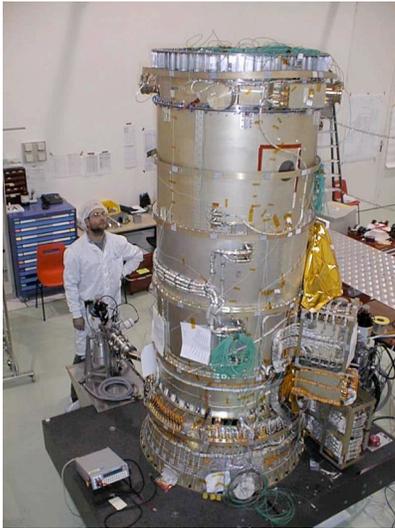

**Fig. G6.** *INTEGRAL/SPI BGO shield [110], comparable in size to that of ACT.*

### Anti-Coincidence Detector

In low earth orbit, the intensity of particle radiation is many orders of magnitude larger than the astrophysical γ-ray. Thus, the anti-coincidence detector (ACD), which provides an electronic shield, is an integral part of the ACT telescope design. The ACD must provide a prompt (< 0.5 μs) veto signal to the ACT trigger electronics whenever a charged particle crosses the active volume of the telescope, and have a probability of missing a charged particle below < 10⁻⁶. These requirements are similar to those of the GLAST ACD (Fig. G7), and a similar, modular ACD design is envisioned for ACT.

Light from each of ~100 scintillator tiles will be collected by two independent sets of wavelength shifting fibers and viewed by two separate PMTs with individually adjustable gains and thresholds, providing two-fold readout redundancy as well as a high level of adaptability.

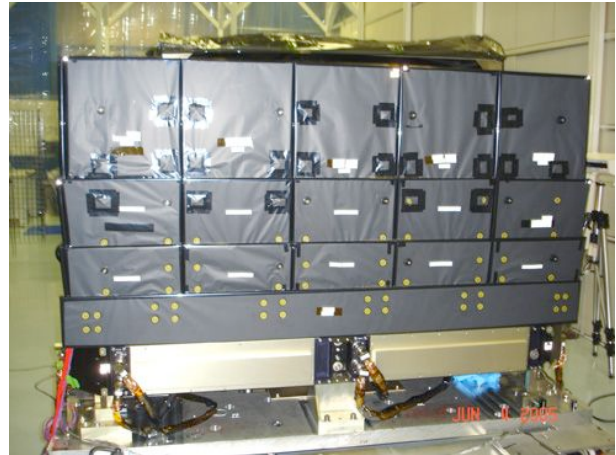

**Fig. G7.** *GLAST ACD shield [74], comparable in size and design to that required by ACT.*

### On-board Processing

In the baseline approach developed at the ISAL and IMDC runs, the ACT science instrument will require an internally-redundant science processor to perform data filtering before passing on the data to the main S/C computer, as well as two (redundant) digital signal processing (DSP) boards to perform real-time on-board analysis for transient events.

For data handling, the science processor will be performing simple event-level filtering to identify likely Compton events and package them for transfer to the main S/C computer. This filtering removes the large number of lower-energy events in the instrument, though current estimates suggest that these photons would not overwhelm the telemetry stream. The ISAL and IMDC runs identified a PowerPC 750 processor as suitable to fulfill the ACT requirements.

The DSP will be performing a real-time analysis of transient events, with the goal of localizing γ-ray bursts within a few seconds for rapid telemetry of coordinates to the ground. The IMDC study recommended that this processing should be implemented as a hardware or firmware function by addition of the DSPs to the science processor.

## Instrument Parameters—Mass, Power, Size

We designed the optimized Si-Ge baseline instrument to conform to the ISAL and IMDC findings in terms of available lifting power, size constraints, and power requirements.

The Si-Ge instrument itself, including support structure and co-located electronics as well as the plastic anticoincidence, weighs 831 kg; the lower-level electronics boxes and cryocooler another 142 kg. The 4-cm thick BGO shield consists of 1081 kg BGO and 29 kg support structure. The main instrument plus all electronics and support have a total mass of 1002 kg; including the 30% margin assumed at IMDC brings this to 1303 kg. For the BGO itself, no mass margin need be assumed since the material is well known. Thus the *total instrument mass with contingency is 2384 kg*. (The instrument assumed during the IMDC study had a mass of 2210 kg including margin.)

Scaling from the power requirements determined during ISAL for a very similar Si-Ge instrument, this ACT will require 1830 W for instrument readout and cooling, plus high-level electronics, heaters, and LV power supplies (total 270 W according to ISAL) and BGO veto readout ($\sim$ 50 W) for an *instrument total of 2150 W*—well within the 2500 W envelope (including margins) set at the IMDC.

The instrument—including BGO and ACD shields—fits within a $150\times150\times35$ cm$^3$ envelope, fitting the S/C bus and Delta IV shroud as studied at the IMDC.

The ISAL baselined ASICs with 1 mW/channel, feeling this was reasonable but pointing out these needed to be developed. ISAL recommended active over passive cooling to -30° C for the Si detectors. Additional areas of concern were the demands on telemetry (rough assumptions yielded 25000 evts/s resulting in 60 Mbit/s continuous telemetry stream) and power (with power driven by both cooling and the large number of channels), as well as the desire for a low-inclination orbit carrying a large lift penalty. All the ISAL areas of concern were addressed in the IMDC run, as presented in the Mission Implementation section.

## Instrument Calibration

The calibration of ACT will be an extensive and multi-stepped exercise. Because the instrument will be so large, a complete instrument calibration may be impractical. However, coupling detailed instrument simulations to laboratory measurements of both a prototype and the full instrument is possible and important. Additionally, exposing an ACT prototype to a balloon-flight radiation environment will test its susceptibility to prompt background radiation.

A calibration and test (CT) model should be constructed that would be smaller than the full flight ACT, but it must be instrumentally identical in terms of charged particle shield protection, passive mass fraction, and instrument electronics. This CT model, after environmental testing, would be exposed to monochromatic beams of $\gamma$-rays from (1) radioactive sources, (2) free-electron laser-induced photons and (3) perhaps nuclear-reaction induced $\gamma$-rays from a low-energy accelerator. The free-electron laser produces monochromatic and polarized $\gamma$-rays that can be used to evaluate and characterize the instrument response in the energy and polarization domains. Simulations should correctly model the measurements. Radioactive sources in the lab can be used to confirm the instrument response of the full ACT to that extrapolated from the CT model calibrations and simulations, as well as the stability of ACT's performance.

A balloon flight of the CT model exposes the instrument to intense fluxes of $\gamma$-rays and neutrons, typical of what will be encountered in orbit. The neutron flux will produce background events that must be reproduced in simulations. Furthermore, the charged cosmic-ray intensity will test the tightness and efficiency of the anticoincidence system.

The full complement of simulations of the CT and Flight models and the corresponding calibrations, including the balloon flight will provide a piece-wise, but complete, picture of the instrument response and its susceptibility to background.

## Instrument Response

Knowledge of the response of ACT to known inputs provides the fundamental means of analyzing data. The fidelity of reduced end-product data (fluxes, locations, times, etc.) depends directly upon the accuracy of the instrument response. For instruments like ACT that measure many quantities, the instrument response is a complex function of many variables. Experimental calibration measurements provide a key element in determining the response function, but it is completely impractical to use calibration data as the sole means to populate the entire data space needed for scientific operations. The ACT mission must rely on extensive computer simulations to define the detailed instrument response function required to deconvolve its data. For reliable results, these simulations must be extensively validated against, and incorporate results from, experimental calibrations.

The simulation tools required for instrument response modeling should ideally be the same ones used for design optimization and performance studies. Each step of the experimental calibration should be accompanied by a careful and through program of computer simulations, and analysis of calibration data must always include comparison with results of simulations. While this approach is resource consuming, it is imperative to identify and understand all inconsistencies between simulations and experiment, since simulations will be required as the ultimate source of instrument response data.

## ALTERNATIVE SCIENCE INSTRUMENT DESIGNS

The ultimate incarnation of ACT will be the result detailed scientific and technology trades that we have only begun in this concept study. Toward this end, however, we have studied competing technologies that could either augment or supplant those chosen for the baseline design. These include thin Si [82, 47], liquid Xe (LXe) [3–5], CdZnTe [96, 70, 83], LaBr scintillator [100], and gaseous Xe (GXe) detectors [8, 25, 26]. Each of these detectors possesses capabilities that could potentially improve the performance over the baseline instrument—in terms of electron tracking (thin Si, GXe), fast timing (LXe, LaBr, GXe), or room-temperature operation (CdZnTe, LaBr). However, these benefits currently come with a performance hit in other areas including efficiency (thin Si, GXe), spectral resolution (LXe, GXe, CdZnTe, LaBr), and overall power (thin Si, GXe).

One of the goals of the ACT concept study was to investigate different configurations of these potential detector technologies to determine the most promising technologies. We were limited, of course, to performing detailed performance simulations for only a handful of alternate designs, and primarily in context of SNe Ia $^{56}$Co sensitivity.

The configurations were chosen to take advantage of the specific capabilities of each detector type.

In order to fairly compare the different designs, we identified a science instrument envelope based on the overall *mass, power, and dimensions* as constrained by the Delta IV launch vehicle and the IMDC study results (Appendix Q). All the alternate designs were required to fit within the same science instrument envelope and power system design, so that ostensibly any of these designs could fly on the mission platform studied in the ISAL and IMDC runs. Constraints used were 2000-W power consumption and 1850-kg instrument mass without margins, sized to fit within the Delta IV shroud and on the S/C bus studied in the IMDC. Details of each design can be found in Appendices Q–W.

The designs separate themselves in several ways, but a convenient classification depends on whether the detector technologies allow either electron tracking or fast timing for time-of-flight (ToF) measurement. The three broad classes of designs are: (1) non-tracking, non-ToF, (2) ToF, non-tracking, and (3) tracking, non-ToF. Technologies adaptable to non-tracking, non-ToF instruments include thick Si, Ge, and CZT strip detectors. Potential technologies for ToF, nontracking instruments include liquid Xe and LaBr scintillators. Potential technologies for tracking, non-ToF instruments include thin Si strip detectors and gaseous Xe or Ar. For the tracking designs, one of the above non-tracking technologies is used for a calorimeter.

All technologies considered are currently in use, either in space or in the laboratory, and so all are worthy candidates for consideration in an ACT instrument.

The designs that were studied using the above technologies include:

1. Our baseline design employing thick Si strip detectors and segmented Ge detectors. It is a non-tracking, non-ToF design. We studied several versions of this baseline, including with and without the BGO shield. (Appendix P, Section H)

2. Thin Si strip detectors with CZT calorimeters. This is a tracking instrument, non-ToF with more modest cooling requirements, but large numbers of channels because there are more Si-strip detectors with finer pitch required to scatter the γ-ray. Similar designs are being prepared for a balloon flight. (Appendix R)

3.   A full Ge instrument, similar to one launched on a recent balloon flight, was also studied. This is a non-tracking, non-ToF instrument requiring low-temperature cooling (80K) but possessing the best energy resolution. (Appendix S)

4.   A full thick Si instrument, which relies on reconstructing the photon energy (3-and-out) instead of stopping the photon. This is a non-tracking, non-ToF instrument requiring minimal cooling requirements. (Appendix T)

5.   A ToF, non-tracking (except for high energies) liquid Xe instrument, requiring cooling to 170 K and pressurization to 1.3 atm. Similar detectors have flown on balloons. (Appendix U)

6.   A gaseous Xe tracking (potentially with ToF) telescope coupled to calorimeters. This technology has not yet been fully demonstrated in the lab. This instrument would require pressurization to 3 atm. (Appendix V)

7.   A ToF, non-tracking Compton telescope similar to COMPTEL, but using plastic scintillators and state-of-the-art LaBr scintillators, to achieve fast timing in a compact design. (Appendix W)

## PERFORMANCE COMPARISONS

Details and simulation results of each design can be found in the Appendices Q–W. All the designs were studied with the same set of detailed source and background simulation tools (Appendix P), with inputs tailored specifically to the baseline ACT low-earth orbit (8°, 550 km). Given the time constraints, instrument geometry revisions to improve performance were possible for only a few of the alternative instrument designs. Nonetheless, some configurations that have been studied here may ultimately prove incapable of achieving a satisfactory performance in regards to the SNe Ia science, so that the associated detector technologies need not be developed any further for this primary science goal of ACT.

In addition, we had to develop a single benchmark for comparing the alternate designs. Since all of these designs would achieve different values of sensitivity, FoV, angular resolution and spectral resolution, our scientific benchmark was based on the primary goal of the mission: distinguishing models of SN Ia explosions. Therefore, the on-axis point-source sensitivity to 3% broadened 847-keV ($^{56}$Co) emission was set as the primary benchmark. However, for distinguishing explosion models it is not adequate to set a single sensitivity requirement: instruments with higher spectral resolution can distinguish explosion models better than instruments with poorer spectral resolution, given the same sensitivity. Therefore, we developed curves that show our ability to distinguish explosion models as a function of both the instrumental spectral resolution and broad-line sensitivity (Fig. G8). Comparison of the alternate instrument configurations must be made relative to these curves.

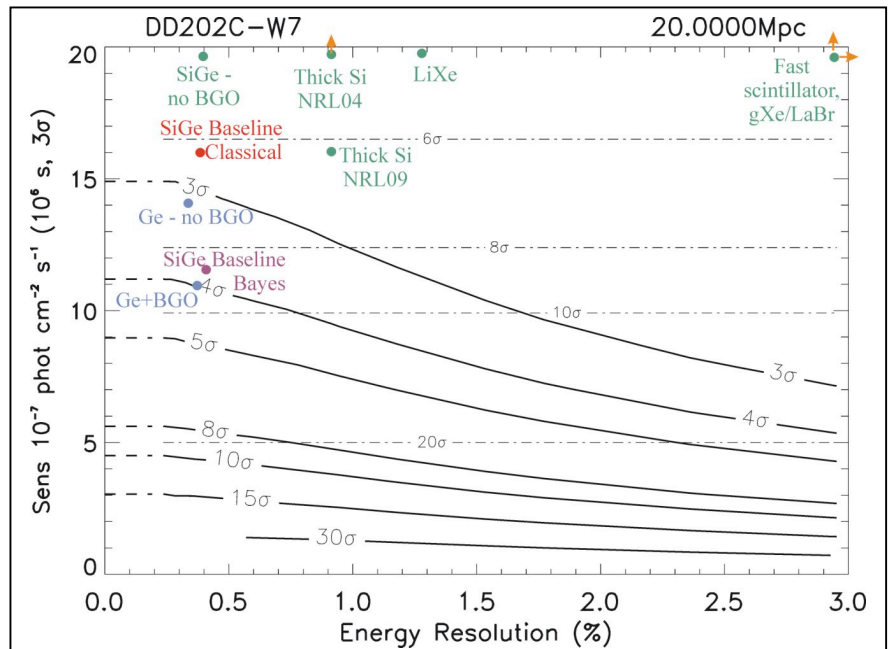

**Fig. G8.** *Curves for determining an instrument's ability to distinguish a deflagration SN Ia (model W7) from a delayed detonation (DD202C), by looking at the 3%-broadened line sensitivity and the instrumental spectral resolution (FWHM) at 0.847 MeV ($^{56}$Co). Curves show the significance of differentiating the models, assuming the SN Ia distance is well known (dot-dash), or ≥10% uncertain (solid curve) [see text]. Note, the significance for detecting the SN Ia is much higher. Details of the instrument simulations are given in Appedices P-W. For the SiGe Baseline instrument two different event reconstruction approaches have been tested, one based on a generalized $\chi^2$ approach (Classical), and one based on Bayesian statistics (Bayes) which gives better results (Appendix P). Color-coding shows which analysis method has been applied to the simulated data: Bayes (blue) or a classical approach (green). Sensitivities have typically improved by factors of 1.2-1.7 by using the Bayes method for the various Si and Ge instruments where this has been implemented.*

ACT has an extremely ambitious goal: to not only make the first measurements of $\gamma$-ray emission from SNe Ia, but to unambiguously distinguish among the various models of SN Ia explosion mechanisms and dynamics. It is not enough to detect the $^{56}$Co emission—it has to be measured with enough sensitivity to distinguish the physical mechanism underlying the explosion. The differences in the mechanisms manifest themselves in two primary ways in the $\gamma$-ray data: lightcurves of the emission lines, and the Doppler-broadened profiles of the lines. Therefore, both sensitivity and spectral resolution (<1%) are required to distinguish the models.

In order to estimate the capabilities of ACT to perform SN Ia science, we have simulated the significance with which ACT could discriminate between two models with very different underlying physical processes, but with relatively similar $\gamma$-ray emission properties: a standard deflagration model, W7 [80], and a standard delayed-detonation model, DD202C [44]. A time-series of spectra were generated for both models, consistent with the detailed comparisons of models for $\gamma$-ray emissions from these supernovae models [77].

One major complication comes from the SN distance. In our nearby Universe (<100 Mpc), galactic redshifts are still strongly affected by local motions, which limit distance measurements to typically ~10% uncertainties. A 10% distance uncertainty corresponds to a 20% uncertainty in the overall $^{56}$Co production for a given SN Ia. This uncertainty significantly affects our ability to distinguish SN Ia models based on the $\gamma$-ray lightcurves alone. So, whereas DD202C and W7 differ significantly in the overall level of their respective lightcurves for a source at a given distance, this difference cannot be strongly distinguished for a source with a 10% distance uncertainty, and we must primarily rely instead on the Doppler profile differences between $^{56}$Co emission in the models to differentiate them.

For a SN Ia at ~20 Mpc distance, Fig. G8 shows the significance of discrimination between the two models for any instrument as a function of its 3%-broadened line sensitivity and spectral resolution. The two sets of curves on this plot correspond to the case where the SN Ia distance can be measured to less than a few percent accuracy (dot-dash curves), and the case where the SN Ia distance is only measured to ~10% accuracy (solid curves). In the case where the distance is measured to high accuracy (dot-dash), it is easier to distinguish the models for a given sensitivity, and this ability is almost independent of the spectral resolution of the instrument because it relies most heavily on the overall line luminosity. In the case where the distance is more uncertain (solid), it is more difficult to distinguish the models for a given sensitivity, and this ability is strongly dependent on the spectral resolution of the instrument because it relies primarily on resolving the Doppler profile of the $^{56}$Co line, which requires good spectral resolution.

Clearly, within the same mission envelope, some of the instrument designs as modeled are excellent at discriminating the SN Ia models (Si-Ge baseline, Ge with BGO), some are marginally good (Ge no BGO, Si-Ge no BGO, thick Si NRL09), some are not adequate as modeled, but perhaps have potential if moderate improvements to sensitivity or spectral resolution are achievable (liquid Xe, thick Si NRL04), and some have little hope achieving this primary ACT goal (fast scintillator, gXe-LaBr). With the caveat of the limited resources available for this study, and the limited time for optimizing the individual designs, it is still clear that the baseline Si-Ge geometry remains a promising design to pursue for ACT, and most of the alternate designs will have to strive to reach its performance. Comparisons of the various designs in terms of ACT's secondary science goals need to be performed in further detail.

# H. ACT BASELINE PERFORMANCE

Simulations of the optimized Si-Ge baseline instrument with BGO shield in a 550-km, 8° inclination orbit indicate sensitivities for a 3% broadened 847 keV line of $1.2 \times 10^{-6}$ ph cm$^{-2}$ s$^{-1}$ ($10^6$ s, on-axis)—enabling ACT to achieve its primary science goal of systematically studying $^{56}$Co emission from SN Ia. In the process, ACT becomes a powerful instrument for all classes of γ-ray observations (Figs. H1–H4), including diffuse nuclear lines, compact objects, and GRBs.

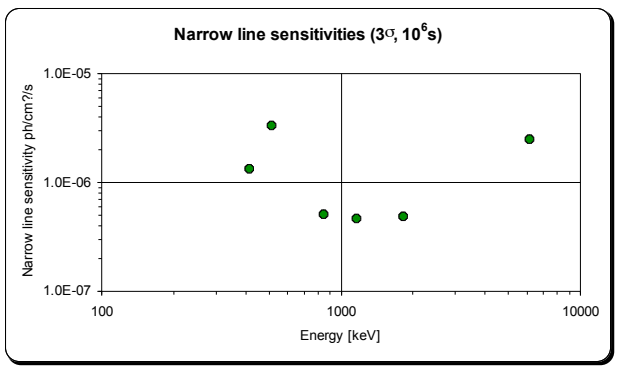

**Fig. H1.** *Narrow-line, on-axis point-source sensitivity as function of energy. Sensitivities over the main nuclear range, 0.8–2 MeV, are 100× better than COMPTEL, 50× INTEGRAL/SPI. The instrument is sensitive at least down to the r-process-induced line at 415 keV and up to cosmic-ray induced lines at 6 MeV.*

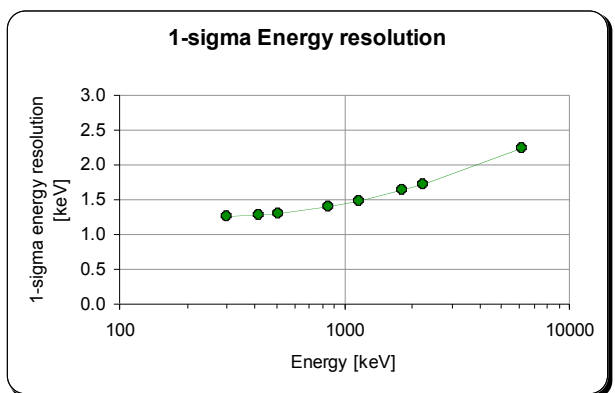

**Fig. H2**. *Instrumental spectral resolution after event reconstruction and selections. ACT will achieve comparable spectral resolution to INTEGRAL/SPI, but with much higher sensitivity to nuclear line emission.*

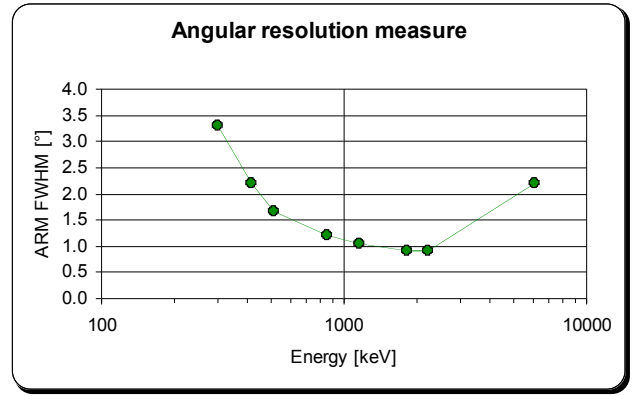

**Fig. H3.** *Instrument on-axis angular resolution. Combined with its high sensitivity, ACT will start resolving regions of Galactic nucleosynthesis. GRBs and other strong point sources can be localized to 5'.*

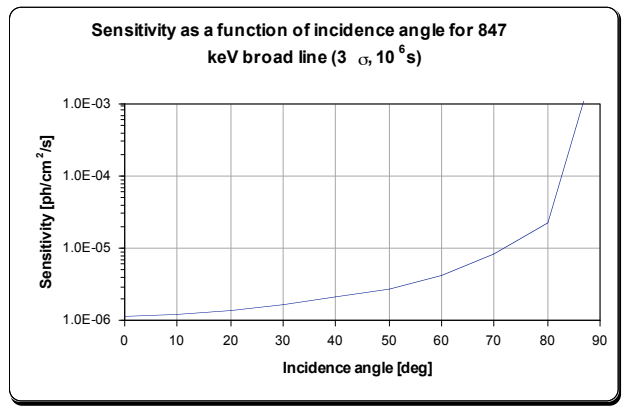

**Fig. H4.** *Baseline Si-Ge instrument field of view—expressed in terms of broad-line sensitivity for different source zenith angles. The half-width-at-half-sensitivity for the 847 keV line is reached at ~ 45°, with sensitivities below $1 \times 10^{-5}$ ph cm$^{-2}$ s$^{-1}$ achieved at zenith angles below 70°.*

A description of the numerical simulations behind these estimates is given in Appendix P, including photon, charged particle, and neutron backgrounds. While we believe these results to be robust, due to time constraints there are a couple of aspects of the background that have not been investigated in detail. Multi-photon nuclear gamma cascades are included in the simulations, but have not been verified in as much detail as other components. In addition, the simulations do not include backgrounds due to chance coincidences, though we estimate this component to be small (details in Appendix P).

# I. MISSION IMPLEMENTATION

The ACT mission consists of a single instrument composed of a large array of position-sensitive detectors, surrounded by anti-coincidence shields and mounted on a zenith-pointing spacecraft (S/C) (Fig. I1). From the mission perspective (Table I1), ACT could be launched as early as 2015 from Kennedy Space Center (KSC) on a Delta IV 4240 vehicle. The Delta IV 4240 can deploy ACT into its baseline 550-km, 8° inclination circular orbit. A 5-year minimum (10-year desired) lifetime is required to meet the primary science goals of the mission. Launch later than 2015 may require a slightly higher orbit, depending on the phase of the solar cycle.

During launch, selected elements of the S/C will be powered ON while the instrument is powered OFF with all detectors at room temperature. Any detector cryostats will be vented during launch. Upon reaching the planned orbit, the S/C will separate from the second-stage propulsion vehicle; deploy the two solar arrays, TDRSS antenna, and four thermal panels. The S/C will establish a zenith-pointed attitude (for its normal survey mode) using reaction wheels. The reaction wheels will maintain the zenith attitude with periodic angular momentum dumping via magnetic torquers. (ACT will maintain the ability to point at individual objects after its primary 5-year survey, or for once-in-a-lifetime opportunities such as a galactic SN.)

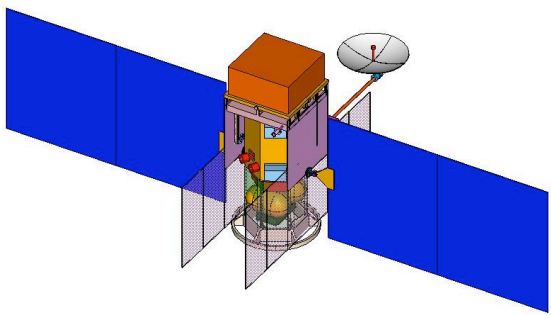

***Fig. I1.*** *ACT science instrument (orange) on its S/C bus, with solar panels (blue), thermal radiators (transparent grey), and high-gain TDRSS antenna deployed.*

Once the operational attitude and orientation have been achieved, the detector cryostats will continue to be vented to space vacuum to ensure the good vacuum required for efficient cooling and operation. The ACD and BGO shield systems will be powered and tested. Once vented, the detector electronics and cryocooler systems will be powered. Using a programmed thermal profile, the cryocooler will cool the detectors to their operational temperatures within a few days. During this time, the aspect sensors will be turned on and checked out. A detector verification period will follow, during which Si-Ge high voltages are turned on, before normal operations begin.

**Table I1.** ACT mission requirements.

| Launch ~2015, 5–10 year lifetime |
| --- |
| 550 km LEO, <10° inclination, Delta IV (4240) |
| 1° attitude, 1′ aspect, zenith pointer |
| Instrument 2100 kg, S/C 1425 kg, propellant 462 kg |
| 3800 W power, 69 Mbps telemetry |
| $760M (FY04, original IMDC estimate) |

ACT is a wide FoV instrument (25% of sky), surveying the entire sky by maintaining a zenith-pointed orientation and sweeping out the sky over the course of its orbit. Pointing attitude (±1°) and aspect (±1′) are fairly relaxed.

In normal operational mode, the detectors must be cooled and high voltages maintained on while observations are conducted. A low-inclination orbit will allow the instrument to operate with a nearly 100% duty cycle while remaining out of the South Atlantic Anomaly (SAA) with its intense radiation. All photon data will be passed through a simple filtering routine in the science instrument processor to identify candidate events. For the selected events, energy, position, and timing data, together with instrument aspect data, will be packaged and stored in the S/C 100-Gbyte solid-state data recorder for telemetry to the ground.

Semi-weekly, the instrument would be operated for a few hours in an engineering mode where on-board event selections are relaxed in order to acquire background for calibration purposes. In this mode, detector thresholds would be monitored, and cosmic-ray charged-particle events would be used to calibrate and monitor detector spatial alignment.

Data will be telemetered through TDRSS to White Sands and then transferred to the ACT Mission Operations and Distribution Center at the Goddard Space Flight Center for zero-level processing. The instrument data and required S/C data will be transferred to the Science Operations

Center (SOC) located at the PI institution or GSFC. The SOC will make these data and analysis software available on-line to the investigator community after initial processing. Final archiving of the data and analysis software will be at the NASA High Energy Astrophysics Science Archive Research Center (HEASARC).

In addition, data are processed in real-time in the on-board science processor to detect and localize γ-ray bursts and other transient events. Within tens of seconds, the transient's coordinates will be sent through a dedicated TDRSS S-Band transmitter to the ground to enable prompt multi-wavelength observations.

The mass constraints on the mission come from the baseline launch vehicle, a Delta 4240 (4772 kg to 550km/8° orbit). An instrument of 2100 kg and 2500 W power requirement (contingency included) could be launched as part of a S/C with total (dry) mass 4062 kg, total power 3.34kW (both incl. margins), and TDRSS-supported Ku-band communications to achieve the ACT objectives. All S/C bus components were found to be TRL 7–9 for a launch in 2015. (The mass and power envelopes for the Delta 4240 launcher and IMDC S/C were used to provide a design envelope for the different alternative ACT instrument designs considered as part of the ACT study; for details see Appendix Q.)

## SPACE SYSTEMS ARCHITECTURE

### Instrument Accommodation

The ACT S/C design was derived from the GLAST S/C with modifications to support the more demanding ACT power and thermal requirements. The primary structure supports the scientific instrument well above the S/C bus, for an unobstructed FoV and reduced background from the S/C itself (Fig. I1). The cryocooler system, the propulsion tank, and the Li-ion battery box dominate the central interior of the S/C bus. Two 16 m$^2$ GaAs solar arrays, deployed after launch, with a single-axis drive for Sun tracking, power the S/C. Thermal radiator panels, body mounted orthogonal to the solar arrays with one-time deployable wings, provide the required thermal dissipation. A 1.5-m deployable, gimbaled high-gain antenna provides the Ku-band link to TDRSS.

### Mass

The ACT estimated masses are 2100 kg for the science instrument, 1425 kg for the S/C bus, and 462 kg for the propellant. Significant structural components of carbon fiber composites are envisioned to reduce the instrumental background associated with cosmic-ray interactions in the S/C and ACT instrument. Including a 30% contingency, the total weight (minus propellant) was allocated 4062 kg total dry mass. The propellant is necessary for periodic altitude maintenance and for controlled re-entry at the end of the mission.

### Power

The ACT power system consists of two 16m$^2$ GaAs solar cell arrays with a single-axis drive for Sun tracking (Fig. I1), and two 275 A-hr Li-ion batteries for power storage. This is a scaled-up version of the PSE on WMAP, and the IMDC identified this as TRL-9 in time for a 2015 launch technology. This system is designed to deliver 3.34 kW of power, which includes the science instrument requirement (2.5 kW with contingency), plus the S/C power and contingency. The science instrument power is dominated by the large number of electronics channels and the cryogenic cooling. While possibly the most challenging technical aspect of the ACT mission, the power system requirements are compatible with available technology.

### Launch Vehicle & Orbit

The preferred orbit for ACT is a near-equatorial, low-altitude orbit. This is driven by the necessity to avoid the South Atlantic Anomaly (SAA) in order to minimize the instrumental background, provide continuous instrument operation, and thus optimize sensitivity. The preferred orbit will minimize the exposure to the protons in the SAA, which produce delayed radioactive nuclei in the S/C and instrument, and also minimize the exposure to cosmic rays. The IMDC study showed that a Delta IV 4240 (Fig. K2) could get ACT to a <10° inclination orbit from Kennedy Space Center (KSC). This is the highest-inclination orbit that will keep the SAA protons below the preferred levels. The sensitivity for low-inclination orbits has been extensively studied as part of the ACT Mission Concept Study and the instrument and mission requirements for this orbit were baselined for the ISAL and IMDC studies.

A high-altitude orbit, essentially beyond the Earth's radiation belts, is also a candidate, but has not been as extensively studied. (Preliminary studies suggest that this orbit hurts the sensitivity. See also Section K.) A high orbit has the disadvantage of a considerably enhanced cosmic-ray fluence that would result in a larger instrumental background. But this would be offset by elimination of the atmospheric $\gamma$-ray and neutron albedo, and might also enable ACT to have a larger FoV (Appendix T). Additional simulations are required to reliably compare the relative sensitivities of the two options. Also, if a high orbit is preferred, more extensive ISAL and IMDC studies for this option will be needed.

## Attitude Control System

ACT has a large FoV (25% sky), and is designed as a three-axis stabilizied, zenith pointer instrument, to sweep out over 80% of the entire sky for each orbit. The advantage of such a large FoV is that the ACT pointing requirements are modest (~1°). Furthermore, absolute aspect knowledge requirements, driven by our point source localization capabilities, are also modest (1′). The fully-redundant attitude control system (ACS) will consist of four reaction wheels, three magnetic torquers plus two 3-axis magnetometers for momentum unloading, eight coarse Sun sensors for safehold conditions, and two star trackers for absolute aspect knowledge. One of the star trackers will be positioned along the zenith-pointed direction (center of the ACT FoV). The reaction wheels were sized to enable a 30° re-orientation of the S/C within 30 minutes if required for optimum observation of selected targets. This time could be significantly reduced by use of larger reaction wheels, at the cost of additional mass and power.

## Command & Data Handling

The ACT C&DH system consists of a single internally-redundant S/C computer, an internally-redundant solid-state data recorder (690 Gbit), and two ultra-stable oscillators. The S/C computer will include embedded drive electronics for the thruster valves, the solar-array gimbal, and the high gain antenna gimbal. The ultra-stable oscillators are necessary to achieve the ACT timing specifications of 1 ms absolute and 150 ns relative. Fast timing is required to support the science objectives of $\gamma$-ray transients (e.g. GRBs) and pulsar timing studies. Since these transient data are processed in the instrument processor, the timing precision is needed on-board in real time. Spacecraft position is derived from on-board GPS receivers. The 690 Gbit solid-state data recorder can store two orbits of data to satisfy our original estimates for the telemetry requirement. However, the baseline instrument with the BGO shield may require a considerably smaller capability. Total power for the C&DH system is 160 W. The required technology is currently available, with the exception of the 625 Mbps Ku-band downlink, which will be available by the ACT mission timeframe.

## Communications

Two separate concerns drive the design of the communications system, though the requirements assumed for the IMDC run turn out to be very conservative for the Si-Ge baseline instrument: (1) the high continuous data rate expected from a Compton telescope the size of ACT that must be downlinked within hours of acquisition, and (2) the desire for a fast, lower-bandwidth downlink capability to alert ground observers to GRBs and other transients. Using a Ku-band link at 625 Mbps to TDRSS from a high-gain antenna (available in the next few years), data from one orbit (science data rate 60 Mbps, plus overhead) would take 11 minutes to downlink. The same link could provide 100 kbps command uplink capability. Rapid downlink of aperiodic events, such as GRB information, as well as limited continuous command capabilities, will be available through S-band TDRSS (15 kbps downlink, 2 kbps uplink using TDRSS SSA).

During the ISAL run we identified our telemetry requirement as ~60 Mbps, which the ISAL team regarded as one of the larger challenges of the ACT mission. Subsequently, the IMDC team met this requirement with a Ku-band to TDRSS using a gimbaled 1.5-m high gain antenna, capable of 69 Mbps. (Data dumps required on every orbit.) In addition, the IMDC identified an S-band omni to TDRSS to achieve rapid downlink of $\gamma$-ray transient processed data, and S-band omni to ground for commanding, state-of-health and safety data.

Subsequent to the IMDC run, our detailed Si-Ge baseline intrument simulations demonstrated that our initial telemetry requirements (assumed during ISAL and IMDC runs) were overly conservative by nearly an order of magnitude, due in part to the addition of the BGO shield. Other ACT instrument options may still require larger telemetry rates. Therefore, given our newly relaxed requirements (<10 Mbps, see Section L for details), the telemetry for the baseline ACT will be straightforward, with options to either relax the antenna requirements or the requirement to dump every orbit.

Furthermore, all our calculations assumed the telemetry of nearly all triggered events. Implementation of more elaborate on-board event selections to reduce background could dramatically reduce telemetry requirements, and thus provide additional margins. Such measures are being successfully employed for the GLAST mission, which must cope with somewhat similar background issues.

**Thermal Control**

The ACT thermal control is semi-passive, consisting of 20-m$^2$ AgTef radiators, body-mounted with deployable wings (Fig. I1), which satisfy the ACT thermal dissipation requirements. This design takes advantage of the ACT primary zenith-pointing strategy—two sides of the S/C with radiators stay continuously pointed away from the solar direction and perpendicular to the Earth, simplifying the radiator geometry. The IMDC team baselined a Central Thermal Bus (CTB) with the instrument thermally isolated from the S/C, and the instrument's mechanical cryocoolers located in the S/C bus. The CTB requires no new technologies, but the TRL level must be advanced. An IMDC concern is some difficulty with the CTB supporting all ground performance demonstrations.

**Flight Software**

The IMDC team identified no flight software issues for ACT beyond the risks that typically apply to all space missions. The primary recommendations were consideration of using an Ethernet as the prime data bus onboard the S/C, and that high CPU resources margins (memory and speed) be allocated at time of launch to maintain flexibility during the long mission life.

**Table B.** Mission TRL levels for 2015 launch.

| System | Current | Heritage | ACT | TRL |
|--------|---------|----------|-----|-----|
| DC Power | 2 kW | AQUA | 3.3 kW | 9 |
| Data Bus (Spacewire) | 32 Mbps | Swift | 60 Mbps | 7 |
| TDRSS Ku-band | 1 Gbps | GLAST | 625 Mbps | 8/9 |
| Cryocooler (80 K) | 300 W | NICMOS | 600 W | 9 |
| Cryocooler (-30° C) | 100 W | RHESSI | 300 W | 9 |

## MISSION INFRASTRUCTURE

ACT does not appear to place significant new demands on the space infrastructure (Table B). The capability to launch ACT into a near equatorial orbit from KSC currently exists with the Delta IV 4240 launch vehicle. The size and weight of the ACT S/C and instrument are limited by the necessary change in orbit inclination. If, by the time ACT is launched, capabilities are approved for a launch from a near equatorial launch site (e.g. Sea-Launch, Falcon-V), this could prove advantageous for a larger, more sensitive ACT, or alternatively provide a lower-cost launch.

## CONTROLLED RE-ENTRY

The ACT mission is a moderate-size S/C (~4772 kg). For the baseline instrument, we anticipate that controlled re-entry will be required with an implementation similar to that planned for the GLAST mission. A propellant mass of 462 kg was included in the IMDC baseline to support orbital maintenance and a controlled re-entry. Depending on the final configuration of the baseline instrument, a controlled re-entry may not be required.

## ROLE OF HUMANS

No human role is currently envisioned in the launch, deployment, or maintenance of ACT in space.

## J. Technology

During the course of the IMDC studies, it became clear to the team members that ACT, while a challenging mission, is not constrained by, or placing new requirements on, space mission architecture. All the mission technologies are either flight ready or soon forthcoming (TRL 7–9 for 2015 launch). The primary hurdles in the flight of ACT are in developing the enabling technologies for the science instrument itself. Even here, the prospects for ACT are encouraging because most of the technologies have been demonstrated either in flight or in the laboratory. The primary challenges arise from the scale of ACT. Technologies that have been demonstrated and/or flown for smaller instruments create challenges for ACT in terms of dead time, power, complexity, reliability, telemetry, backgrounds, and ultimately sensitivity.

The primary recommendations for ACT enabling technologies, in support of the primary SNe Ia science goals, are:

- **enabling development for Ge detectors**
- **continued development of thick Si detectors**
- **lab performance demonstration for LXe**
- **development of low-power readout ASICs**
- **implementation of scaled-up cryogenics**
- **minimization of passive structural materials**
- **improved simulation toolset**

In addition, to validate the performance of the ACT baseline instrument, we recommend that a subscale prototype, incorporating the primary flight technologies in the flight configuration, be flown on a high altitude balloon flight for verification of both the instrument technologies and the simulation tools in a space environment.

### Germanium Detectors

**Requirements.** For the Si-Ge baseline D2 germanium detector assembly we have baselined 576 cross-strip GeDs, each 16-mm thick with an active area of 81 cm$^2$. These detectors must be cooled to 80 K for optimal spectral performance and lifetime. Orthogonal electrode strips on the opposite faces, combined with signal timing, provide full 3D position resolution to < 0.4 mm$^3$.

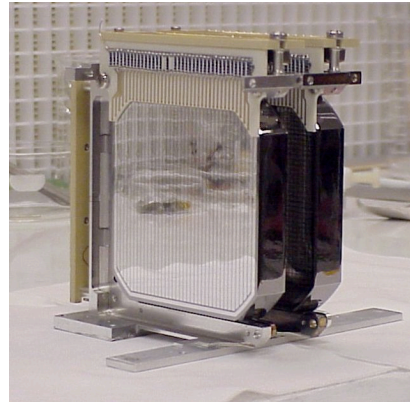

**Fig. J1.** *A two-detector prototype GeD array, consisting of 16-mm thick detectors of active area ~54 cm$^2$ each. This array flew on the NCT prototype balloon flight in June 2005 [12].*

**Status.** These detectors are nearly identical to the detectors (Figs. J1, J2) developed by the Lawrence Berkeley National Laboratory (LBNL) for the Nuclear Compton Telescope (NCT) balloon payload [12], utilizing their pioneering amorphous Ge contact technologies [66]. These GeDs operate as fully-depleted p-i-n junctions, with the blocking electrode made from a ~0.1-μm thick layer of amorphous Ge which is deposited on the entire detector surface. The strip electrodes are defined by evaporating a layer of metal through a shadow mask on top of the amorphous Ge film. The amorphous Ge film serves as the blocking contact, and fully passivates every detector surface not used for contact connection. The bipolar blocking behavior of the amorphous contacts allows them to be used on both sides of the detector, replacing the conventional n-type lithium and p-type ion-implanted contacts. The amorphous contact fabrication has proven reliable, providing high yields of successful detectors. Only a few simple steps are involved in the manufacturing process. The amorphous contacts are also robust. First-generation prototype strip detectors operated successfully for over 5 years with no measurable degradation of the performance. In addition, the contacts have been shown to be stable with temperature up to 100° C for more than 12 hours [67], which could permit detector annealing for radiation damage repair in ACT if required.

**Challenges.** The GeDs baselined in this concept study are nearly identical in geometry and performance to those used in laboratory and balloon experiments today. There is significant development required, however, to enable their use in ACT. The primary technological challenges for these detectors are (1) optimization of the electrode and guard-ring geometries, which minimizes the guard-ring widths and allows the guard rings to be used as active veto shields for minimizing passive material near the detector, and exploration of inter-strip interpolation to optimize position resolution, (2) detailed radiation damage testing and verification of ruggedness, and (3) transfer of these processing techniques to a commercial manufacturer to provide the large number of detectors required for ACT.

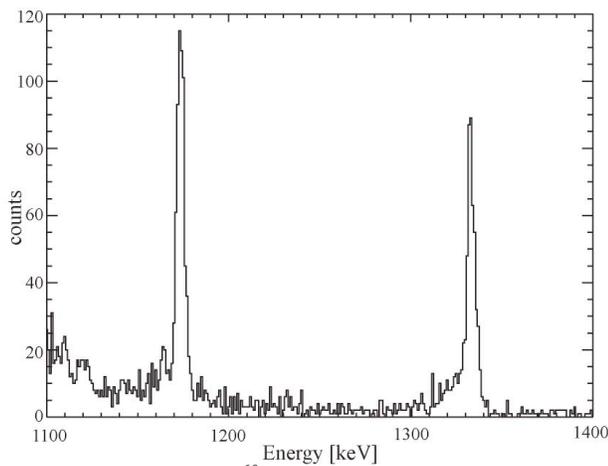

**Fig. J2.** *Preliminary $^{60}$Co (1173, 1333-keV) spectra of the Compton-scatter events for the two-GeD prototype array shown in Fig. J1. Resolutions of ~4 keV FWHM are currently achieved, with 2–3 keV expected as cross-calibration between the strips improves [22].*

**Recommendation.** We recommend support of the enabling technology development for cross-strip GeDs on ACT including electrode optimization studies, environmental testing, and manufacturing of large number of detectors.

**Demonstration.** These detectors will be extensively tested in the laboratory for their spectral and positioning performance, as well as uniformity. Tests will be performed at accelerator facilities for radiation hardness. In addition, these detectors should be tested independently on a balloon flight before integration into an ACT-prototype balloon flight.

## THICK SILICON DETECTORS

**Requirements.** For the D1 silicon detector assembly, we have baselined 3888 100 cm$^2$ active area, 2-mm thick double-sided silicon strip detectors. For the ISAL and IMDC studies we planned for these detectors to be cooled to -30° C for optimal spectral performance. The recent measured performance of larger area detectors from 150-mm diameter wafers shows that sufficient energy resolution may be achievable at temperatures between 0° and 20° C.

**Status.** These detectors are similar to detectors developed by the Naval Research Laboratory (Figs. J3, J4), and are commercially available. They use intrinsic, high-resistivity silicon, fabricated using standard CMOS processing, and are quite rugged compared to lithium-drifted silicon detectors.

**Challenges.** A primary technical challenge for the SiDs is the development of larger-area, thicker detectors. Thicknesses of 6–8 mm would be ideal for reducing the number of detectors required for the D1 assembly, and thereby reducing overall cost. We are currently baselining intrinsic silicon detectors, since lithium-drifted detectors require cooling, operation in a vacuum environment, and are much more sensitive to environmental conditions. The current limitation on thick intrinsic detectors is ~3 mm. However, NRL has recently shown that a wafer bonding technique may enable thicker detectors [87]. With this technique, two thin (0.5–1mm) single-sided strip detectors would be bonded to a thick (4–6 mm) high-resistivity wafer using a low-temperature bonding process.

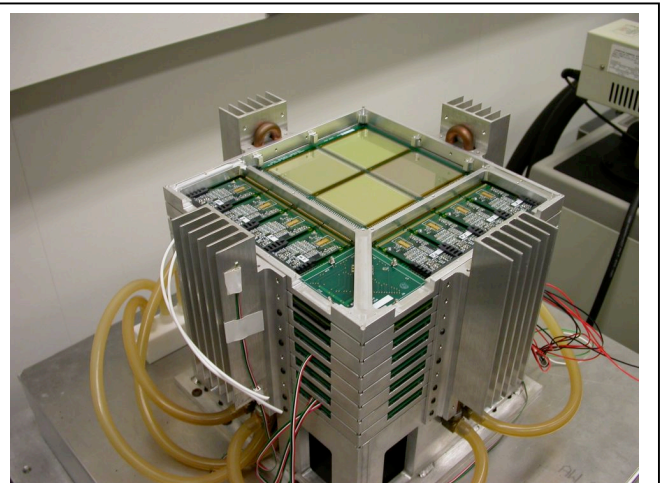

**Fig. J3.** *An 8-layer Si detector prototype array, consisting of 32 2-mm thick detectors, of area 6.3 cm × 6.3 cm each [60].*

A second technology development would be the use of this wafer bonding technology to bond silicon drift detector wafers and/or active pixel sensors to a thick high-resistivity wafer. Based on the significant progress made on silicon drift detectors (SDDs) and active pixel sensors (APSs), these could enable thick detectors with very low noise performance (hence improving the efficiency of the multiple Compton mode) and might also enable electron tracking down to recoil energies of ~250 keV within a detector with an APS with 30–40 micron pixels.

**Recommendations.** We recommend basic detector development support for thicker silicon detectors, including the potential of developing the capability for electron tracking down to useful energies for ACT (~0.2 MeV). In addition, we recommend basic enabling technology developments in terms of production of large numbers of these detectors.

**Demonstration.** These detectors will be extensively tested in the laboratory for their spectral and positioning performance, as well as uniformity and the potential for electron tracking. Tests will be performed at accelerator facilities for radiation hardness. In addition, these detectors should be tested independently on a balloon flight before integration into an ACT-prototype balloon flight.

## LIQUID XE DETECTORS

**Status.** LXe time projection chambers (Fig. J5) have demonstrated large, uniform detector volumes with excellent stopping power and 3D position resolution. They are intrinsically radiation-hard and, due to large drift distances, require by far the smallest number of electronics channels of all 3D position-sensitive concepts studied here. Together with the moderate cryogenic requirements, this results in low cost and low power per unit mass and scalability to very large areas. LXe detectors also provide very fast timing with the potential for ToF measurements in compact geometries.

**Challenges.** The challenge with LXe detectors is primarily that their spectral resolution suffers dramatically relative to semiconductor detectors. Recent advances combining the charge signals with the prompt scintillation signals, however, have led to significant improvements in spectral resolution, from ~8.8% to ~3% FWHM at 1 MeV [5]. For this study, it was assumed that this resolution could be pushed even further to ~1.0% at 1 MeV. While this resolution remains marginal for the ACT requirements, demonstration of this resolution would significantly increase the potential of this detector material for ACT.

**Recommendations.** We recommend support for basic performance demonstrations in the laboratory of improved spectral response of LXe detectors through the combination of charge and scintillation signals, and demonstration of the fast timing for ToF. Further recommendations of this technology in the context of ACT would depend on the outcome of these demonstrations.

**Demonstration.** These detectors should be tested in the laboratory for the realistic spectral performance, fast-timing performance, as well as uniformity.

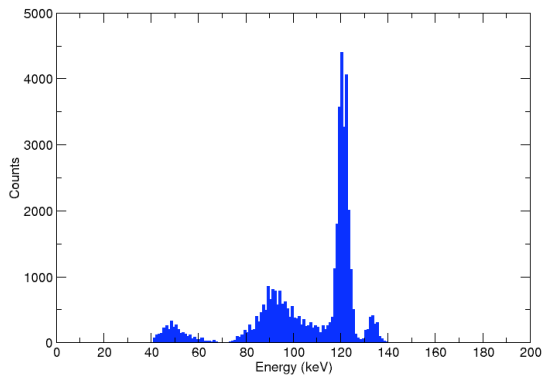

**Fig. J4.** *$^{57}$Co (122-keV) spectra of a full 2×2 daisy-chained array of the Si detectors shown in Fig. J3. Resolutions of ~5 keV FWHM are achieved at room temperature, better resolutions are achievable at colder temperature [60].*

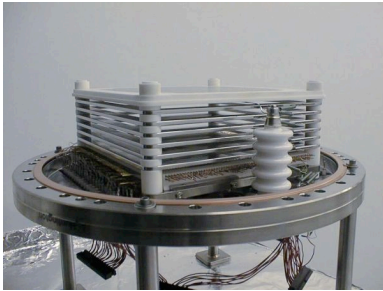

**Fig. J5.** *The LXeGRIT prototype liquid Xe time projection chamber (TPC) which demonstrated uniform ~1 mm³ position resolution in a large 2.5 l volume at 8.8% FWHM spectral resolution at 1 MeV [3–5].*

## FURTHER DETECTOR ADVANCES

### Electron Tracking

Electron tracking is a potentially powerful method of background rejection for a Compton telescope. Non-tracking Compton telescopes record only the energy and position of the recoil electron off which the γ-ray scatters. If the initial direction of the recoil electron can be constrained as well, more complete kinematical information is available for the interaction, allowing the event to be more accurately reconstructed. There are two applications of electron tracking that help to reduce background: (1) If the recoil electron from the initial γ-ray interaction can be tracked with relatively high accuracy, the classical Compton event circle is reduced to an arc [47, 2]. The length of this arc corresponds to the error in measuring the electron direction, which will typically be dominated by Molière electron scattering in the detector material. In this way the solid angle (i.e. the section of the Compton cone) from which the photon could have originated is decreased, which in turn reduces the contamination from diffuse background and incorrectly reconstructed events. (2) If the recoil electrons from multiple Compton interactions can be tracked with even modest accuracy, this provides valuable redundant information to identify the correct event sequence. This increases the efficiency of accurate event reconstruction and aids in the identification of background events, as well as reduces contamination from incorrectly reconstructed events.

**Challenges.** Electron tracking requires a relatively low-density detector and/or fine position resolution, but has only recently been successfully demonstrated in the critical nuclear line range (0.5–

2 MeV) with silicon CDDs [15] (see box) and gas TPCs with micro-well readouts [108]. Other potential technologies for achieving this tracking goal are thin Si strip detectors and gas micro-well detectors. The main challenges facing these technologies are (1) maintaining reasonable stopping efficiencies (and spectral resolutions) such that the overall sensitivity requirements are met, and (2) additional complexity introduced by the large number of channels required for the high spatial resolution.

---

**Controlled Drift Detectors [CDD]**

The ACT collaboration is continually following new detector developments that could improve ACT performance. The silicon CDD device is an interesting example which demonstrated only this past summer electron-tracking to energies below 500 keV, with excellent spectral resolution (few hundred eV at 6 keV) and excellent position resolution (150 μm pixels) [15]. A full study of this technology in terms of ACT mission constraints and overall performance was not possible in time for this report, but it represents one potential technology worth studying further in the context of ACT.

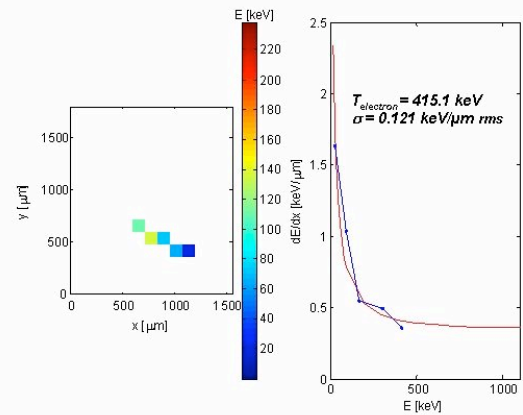

**Fig..** *Track of a 415-keV electron through a Si CDD detector [15]. (A.Castoldi, INFN Italy)*

---

### Fast Timing

As demonstrated by COMPTEL, time-of-flight (ToF) is a potentially powerful technique to reduce background by unambiguously determining the order of photon interactions in the instrument. COMPTEL suffered a significant loss of detection efficiency due to the large separation of detector planes in COMPTEL, driven by the requirement to perform ToF measurements with its relatively slow D2 NaI scintillators. Detectors with sub-nanosecond timing would overcome this restriction and enable compact instrument designs capable of ToF-based event sequencing. With a unique determination of the sequence of events the background rejection capability as well as detection efficiency could be enhanced.

**Challenge.** The most promising technologies for achieving very fast timing are the Xe-based detectors and the novel scintillators such as lanthanum bromide and lanthanum chloride. The main challenge facing these technologies is achieving sufficient spectral resolution such that the overall sensitivity requirement is achieved by adding the fast timing, and maintaining that spectral resolution for large volumes.

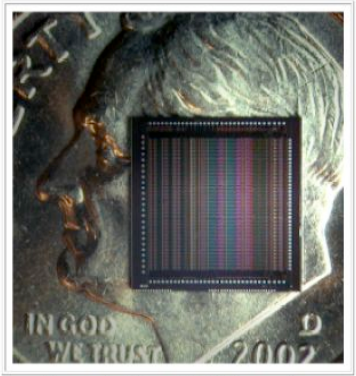

**Fig. J6.** *RENA-2 chip developed by Nova R&D, with 36 channels, 5mW/channel, 100-e rms noise (no input capacitance), and 10-ns timing. (Photo courtesy J. Matteson, UCSD.)*

**READOUT ELECTRONICS**

**Requirements.** A significant challenge for ACT is the design and development of low-power application specific integrated circuits (ASICs) to read out the hundreds of thousands of channels associated with the silicon and germanium strip detectors (ex., Fig. J6). The ASICs for both the silicon and germanium detectors must have low power, large dynamic range, and low noise to maintain the energy resolution associated with these semiconductor detectors, and must provide this performance with the rather large capacitance associated with the strip detectors. Good integral linearity, <0.1%, is desired to allow multiple-site interaction energies to be summed to yield accurate γ-ray energies. In addition, the ASIC must also provide the time resolution (~10 ns) to enable the depth resolution to ensure excellent 3-D positions for γ-ray interactions. They must also allow system-wide coincident time resolutions (<150 ns) required for selecting Compton scatter events in a high rate, large volume detector assembly.

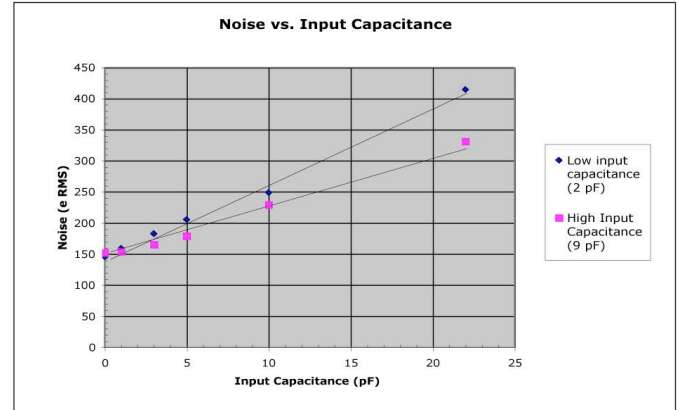

**Fig. J7.** *RENA-2 Input-referred rms noise versus input (detector) capacitance.*

**Status.** The 36-input RENA-2 ASIC [124, 125] meets several of these requirements, which gives us confidence that the overall ASIC requirements for ACT can be met with additional development of this chip and others as well. Nova R&D and UCSD worked closely on the RENA-2 design to assure that it meets the ACT requirements for low noise, and excellent linearity and event timing. Event timing is provided by a unique "fast time stamp" feature, within the chip itself, that provides ~1 ns timing precision for every channel's trigger time. The noise floor is 140 electrons rms (Fig. J7), or 1 keV FWHM in Ge, and an integral linearity of 0.02% has been demonstrated from 0.1–5 MeV with test pulses, as shown in Fig J8. Initial tests of detector-to-detector coincidence time resolution with two CZT detectors showed a resolution of 20 ns FWHM at 511 keV with CZT (Fig. J9). This ASIC was not designed for low power; it has a power consumption of 5 mW/channel. Lower power versions are planned.

**Challenges.** Power is a universal challenge given the large number of channels required. ASIC design is key to achieving the goal of reducing power to below 1 mW/channel. However, 1 mW/channel will be a challenge if timing on the order of 10 ns is necessary for depth measurements in conjunction with excellent spectral resolution. It is possible with today's technology to produce ASICs operating on the μs time scale to consume no more than 1 mW/channel, but as speed is increased power demands also increase.

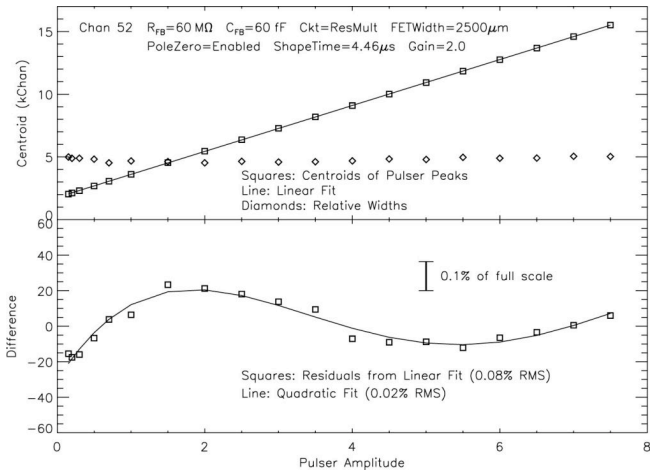

**Fig. J8**. *Integral linearity measurement of one RENA-2 channel. Data taken with test board pulser. Upper panel: full data. Lower panel: deviations from linear fit, which have rms = 0.08% of full scale. This is reduced to 0.02% rms when a polynomial fit, shown, is applied. 1 MeV in CZT corresponds roughly to a pulser amplitude of 1.4.*

**Recommendations.** Development for the ASICs will be based on well-defined requirements for both the SiDs and GeDs, and should include a competitive design and development process involving at least two ASIC designs for both detector types.

**Demonstration.** The noise, dynamic range, linearity, timing, rate, and power performance of the ASICs will be measured in the laboratory when connected to prototype detectors and then detector arrays in a flight-like configuration. In addition to performance tests, the radiation hardness of the ASICs will be tested separately for both γ-ray irradiation and charged particle irradiation (protons and heavier particles) to the radiation levels expected during the extended ACT mission (~ 10 kRad).

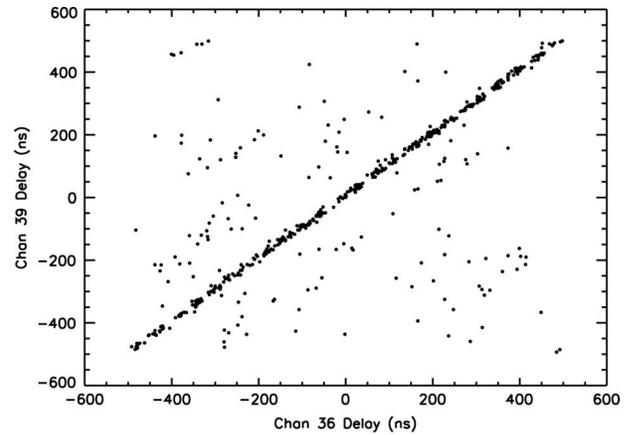

**Fig. J9.** *RENA-2 511 keV pair coincidence, derived from the RENA-2's fast time stamps for two CZT detectors. Coincident events occur on the diagonal, which has a width of 20 ns FWHM; <10 ns resolution is expected when systematics are corrected.*

## CRYOGENIC SYSTEMS

**Requirements.** The ACT baseline instrument requires two separate cryogenic systems for the two detector arrays: a system that goes to 80K for the Ge (D2) detectors, and a system at -30° C for the Si (D1) detectors. Both the ISAL and the IMDC teams recommended that these be separate cryogenic systems given the very different cooling requirements of the two detector types.

**Status.** Flightworthy cryocoolers are now being used, e.g., on the RHESSI mission and NICMOS/HST. The RHESSI unit has demonstrated ruggedness and efficiency, but the demands of ACT will be 20× greater. Thus, for an instrument the size of ACT, the power and the mass associated with the cooling apparatus is an issue. The power consumed can be on the order of a kilowatt, with a correspondingly large set of solar panels and complicated thermal designs further increasing the S/C mass.

**Challenges.** For the Ge detector array at 80 K, assuming an internal power dissipation of ~0.1 mW/channel for the FET within the cryostat at an intermediate temperature of 125 K (ASICs outside at ambient temperatures), the heat lift requirements for the entire array is 61 W. The ISAL team identified the Turbo Brayton mechanical cooler used on HST NICMOS, possibly combined with cooling loops, as the best candidate technology to meet these requirements. This crycooler offers the advantages of essentially zero vibrations, and it is staightforward to scale up. The ISAL team felt that 0.1 mW/chn for the Ge FETs was achievable, allowing them to be located within the cryostat. However, cooled FETs is *not* a requirement for the Ge array—the performance assumed for the Ge array in this study has been achieved with room-temperature FETs (located outside the cryostat) that do *not* require the power overburden of cryogenic cooling. Cooled FETs have the potential of offering even better spectroscopic performance.

For the Si detector array at -30° C, assuming an internal power dissipation of 1 mW/chn at the detector temperature, the total heat lift requirements for the entire array are 253 W. The ISAL and IMDC teams identified the commercial Sunpower cryocoolers currently flying on the RHESSI mission as a good match for this array.

**Recommendations.** We recommend support of the enabling technology development for both potential cryogenic systems (Turbo-Brayton and Sterling cycle) on ACT. In addition, we recommend a detailed technical study of the cryogenic systems, including design, optimization, and redundancy.

**Demonstration.** Cryogenic systems should be demonstrated extensively in the laboratory in terms of mechanical vibration, overall efficiency, and lifetime. In addition, consideration should be made of flying these technologies on a long-duration balloon version of the ACT balloon prototype.

## PASSIVE MATERIALS

**Requirements.** Passive materials (i.e. any materials in the instrument in which interactions are not detectable) play an especially troublesome role in Compton γ-ray telescopes. While passive materials cannot be avoided altogether, it is crucial to minimize the mass of passive materials in the instrument, especially high-Z materials, and materials directly surrounding and supporting the detectors. Passive materials have two harmful effects on sensitivity. First, all passive materials contribute to background γ-rays in space by becoming radioactive when exposed to cosmic rays. (Some are especially problematic if their activation results in decay γ-ray lines at or near lines of astrophysical interest.) Second, excess passive materials within the detector array can also reduce the instrument efficiency if γ-rays scatter in this material before being fully absorbed in the detectors.

**Status.** Both the ISAL and IMDC teams were presented with this challenge of minimizing the passive material on the ACT instrument and S/C bus. The IMDC team developed a mission concept where the mechanical structure for the S/C bus, propulsion system, solar arrays, and communications were composed of lightweight, low-Z carbon fiber composite. This was not seen as presenting any exceptional constraints on the S/C architecture for ACT.

**Challenges.** The ISAL team developed novel approaches for the structure and packaging of both detector arrays, also based on carbon fiber composite. The ISAL team outlined a convenient packaging scheme that allows simple assembly of and access to the arrays. The most radical suggestion of the ISAL team was to design the cryostats as clean barriers, but not vacuum chambers. Most of the structural mass in conventional cryostats is for supporting the pressure loads when the cryostat is under atmospheric pressure on the ground. But there is no need for this vacuum capability in space. By designing the cryostat not to support vacuum, substantial mass savings in and around the detectors can be achieved. The biggest drawback is that all of the instrumental testing on ground would have to be performed in a larger vacuum chamber. Such an approach would need to be studied and tested in more detail before being chosen for ACT.

**Recommendations.** We recommend support of engineering studies and tests of low-Z, low mass support structures and novel cryostat designs for ACT.

**Demonstration.** Novel structural materials would have to be verified in the laboratory in terms of their expected outgassing in a vacuum. In addition, accelerator runs would be needed to characterize induced radioactivity in any materials where the composition is not 100% known. We also recommend a detailed engineering study of the potential for the novel cryostat designs recommended in the ISAL runs.

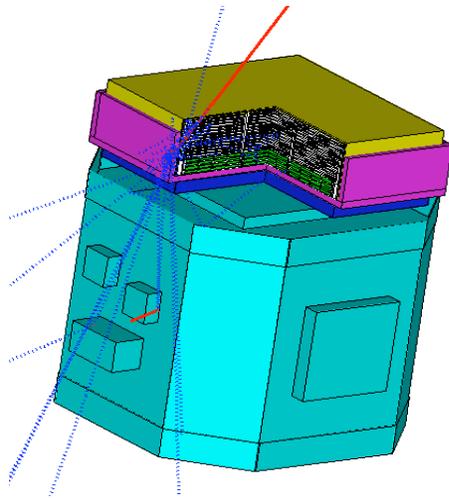

**Fig. J10.** *The ACT baseline simulation mass model, developed within the ACT simulations toolset utilizing the GEANT3 architecture.*

## SIMULATION TECHNOLOGIES

**Requirements.** Monte Carlo simulations will play a critical role in the design and operation of the ACT mission. Indeed, many of the resources for this concept study have been spent on the development, distribution, and application of the ACT simulations tools (Fig. J10, Appendix P). Simulations are required to (1) model the instrument response to γ-ray sources, (2) predict the instrumental background induced by photon, charged particle, and neutron background components, (3) determine the overall sensitivity to astrophysical sources, (4) optimize the instrumental design, (5) study the effects of alternate orbits, (6) develop the on-board and ground processing software, and (7) determine the instrument response matrices. The simulation toolset is an integral part of the ACT technology suite.

**Status.** The current ACT toolset is the most extensively validated simulation system used for MeV γ-ray instruments in the space environment. It was validated against data from several past and current missions.

While the ACT mission concept study was heavily focused on providing realistic estimates of instrument performance based on simulations and models, the limited resources available severely constrained the number and scope of trade studies that could be accomplished. The simulation results obtained thus far are fruitful and promising, but they also indicate that many more simulation studies will be needed to design and operate a real ACT mission.

**Challenges.** The current toolset provides unique and significant capabilities. However, it has been assembled on significantly constrained resources, and is far from the state required for developing the ACT mission. For example, a complete simulation of a single instrument configuration in a particular orbital environment requires weeks of computing resources and hundreds of gigabytes of storage space. The optimization of an ACT mission design will require hundreds or thousands of individual simulation trade studies, making the current toolset untenable.

While the background simulations have been extensively validated against data from several past and current missions, several areas remain where validation is not sufficient for high-fidelity Compton telescope applications. Foremost among these areas are validation of cross sections (particularly for neutron interactions) resulting in activation in both active and passive materials, and the need to validate simulations against a similar instrument operating in a similar environment.

We believe that the background environment models developed for this study are reasonable approximations to the true orbital environment in most cases. However, there are significant deficiencies in our knowledge, particularly in areas related to secondary atmospheric emission (e.g., γ-rays, protons, and neutrons that result from cosmic-ray interactions in the atmosphere).

The most important technologies for improving the scientific return of an ACT mission are those that promise to reduce backgrounds. While we have studied some of the more apparent approaches in terms of instrumental configurations, passive materials, detector performance, and orbit, these studies are by no means comprehensive, and further work is required.

Furthermore, there may be other analysis techniques or detector technologies that could improve or even revolutionize the ability to reduce background. For instance, we studied several approaches to event reconstruction that provide significant improvement in sensitivity. Alternative data analysis techniques may provide similar gains, and further work is strongly recommended.

Limited resources for this study prohibited detailed studies of imaging techniques for Compton telescopes. Given the survey character of ACT and the wealth of information provided by the new detector technologies, a dedicated effort needs to develop efficient imaging algorithms for Compton telescopes beyond the current state-of-the-art, as developed for COMPTEL and medical applications. Only with these new analysis and imaging tools will we be able to fully apply ACT capabilities, and provide reliable maps of various diffuse line and continuum sources that lie at the heart of this mission.

**Recommendations.** We recommend support of short-term upgrades of the existing toolset to significantly reduce CPU and storage requirements of the system and make it accessible to a larger community of instrument developers. In parallel, we recommend a longer-term program to rebuild the Monte Carlo segment toolset from the ground up in the framework of GEANT4, including extensive verification of the new G4 framework with respect to the current toolset and experimental data. Because Compton telescopes are uniquely affected by the angular distribution of atmospheric emission as well as its spectrum and intensity, it is recommended that higher fidelity models be developed and validated against existing measurements. Finally, we strongly recommend that alternative data analysis and imaging techniques be developed

**Demonstration.** Validation of the simulation and activation code against measurements must be performed for those materials that are likely part of the instrument design. It is anticipated that this effort may require new accelerator experiments targeted at specific materials and/or configurations. Both existing and new simulation tools should be validated against data from COMPTEL and balloon payload Compton telescopes, as well as γ-ray missions employing detector materials under considerations, such as RHESSI/SPI (Ge), HXT (Si), IBIS/BAT (CZT), etc. Although COMPTEL is not an exact match for ACT (different materials, configuration, techniques, etc.), it is the only long-duration space instrument that probed the subset of background issues unique to Compton telescopes. Imaging techniques should be tested on simulated maps with realistic background estimates, and, within the constraints of these telescopes, may be tested against data from COMPTEL and balloon payloads.

**On-Board Processing**

The ACT event rate will be two orders of magnitude larger than that of COMPTEL and the amount of information telemetered per event will be much greater because of the fine position resolution and the superior energy resolution that ACT will possess. Source and background simulations performed for this study indicate that the raw event rate of the baseline ACT instrument, set mainly by background, is sufficiently low to fit within anticipated telemetry constraints (Section L). However, given the preliminary nature of the instrument concept and the understanding of the background (and its orbital/solar cycle variations), it is important to study techniques for on-board data reduction that would reduce the telemetry volume. Such measures have been successfully designed and implemented for the GLAST mission, which uses a multi-level hardware/software trigger scheme to reduce the (mainly background) event rate by a factor of several thousand prior to telemetry. We envision that similar measures could be employed for ACT without requiring on-board event reconstruction (computationally-intensive). In normal operating mode, a series of low-level criteria would be applied to raw data so as to reduce background without significantly affecting source efficiency.

This mode would be automatically overridden for short-lived transient events of interest such as γ-ray bursts, and replaced by a special mode designed to capture and buffer as much raw data as possible for a short time, including photoelectric events down to the detector thresholds (~10 keV).

PROTOTYPE BALLOON DEMONSTRATION

To validate the performance of candidate ACT instruments, we recommend a subscale prototype instrument be flown on a high altitude balloon flight (e.g., Fig. J11). Since all ACT are modular, it is straightforward to develop subscale instruments, e.g., for the baseline the prototype will consist of one tower module. Each silicon and germanium layer will have the detectors and electronics as implemented in the full instrument. The silicon layers will be operated near room temperature, whereas the germanium layers will be in a cryostat. The prototype tower will be surrounded by a (plastic) particle anti-coincidence veto shield and a (BGO) γ-ray shield.

This approach using a sub-tower prototype with custom shield was used successfully for the GLAST mission. The prototype will be flown on a conventional (~24-hour) balloon flight. The cryogenic cooling required by the germanium detectors during such a flight could use a flight model cryocooler or be provided by a liquid nitrogen dewar. The full data stream will be stored on-board on disk drives and/or on memory cards, with a subset of the data being telemetered to the ground for quick-look analysis to confirm the nominal operation of the instrument. A successful balloon flight will demonstrate the readiness of the technology. The data from the balloon flight will also serve as "ground truth" for the elaborate simulation package developed to predict the sensitivity of the flight instrument.

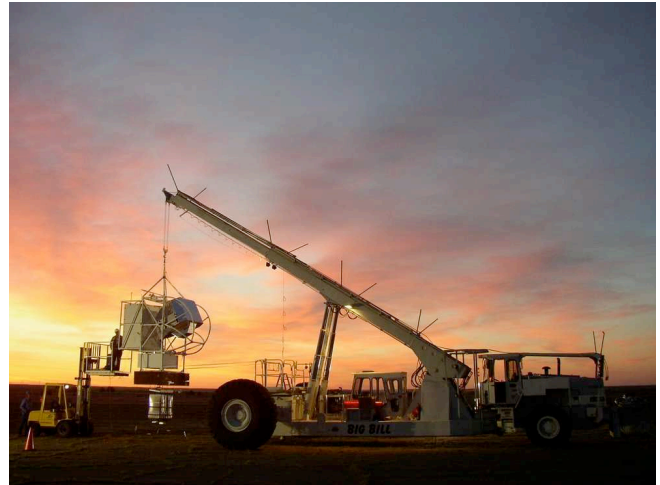

**Fig. J11.** *The NCT balloon payload on the launch vehicle in Ft. Sumner, NM, June 2005. Further balloon tests of individual ACT technologies, and eventually a subscale prototype instrument, are strongly recommended for validation of the technologies and simulation toolset in a space environment [12].*

## K. Deployment

### Orbit

The preferred space environment for ACT is that of a near-equatorial, low-altitude (LEO), circular orbit. This conclusion is driven by the necessity to minimize the instrumental background, and provide nearly continuous instrumental operation, and thereby provide the optimum time-integrated sensitivity. The preferred orbit selection was studied through simulations of a Si-Ge telescope (the initial baseline instrument, not the optimized version) exposed to the model backgrounds expected for different orbits (see Appendix P). As illustrated in Fig. K1, the ACT sensitivity is strongly affected by the intense flux of trapped protons experienced at LEO inclinations above 10°. This background mainly arises from proton activation in the S/C and instrument (particularly in the South Atlantic Anomaly)—resulting in secondary $\gamma$-rays from activated radioactive nuclei. Significant gains in sensitivity are achieved by placing ACT at inclinations less than ~10°—hence our choice of 8° as the near-optimum LEO environment. This orbit was assumed to derive mission requirements in the ISAL and IMDC studies. The IMDC study showed that a Delta IV 4240 could place ACT in a <10° inclination orbit from Kennedy Space Center (KSC).

A high-altitude orbit (HEO), essentially beyond the Earth's radiation belts, was also considered as a means to reduce backgrounds from particles trapped in the magnetosphere, as well as from earth albedo radiation. The lack of albedo radiation in the HEO environment could eliminate the need for a BGO shield and significantly increase the effective instrument FoV. However, the greatly enhanced flux of low-energy cosmic rays in this environment results in decreased on-axis sensitivity compared to a low-inclination LEO environment. Additional simulation trade study is required to fully compare the two options, particularly in taking into account the gains afforded by a larger field of view. Also, if a high orbit were preferred, more extensive ISAL and IMDC studies for this option would be needed.

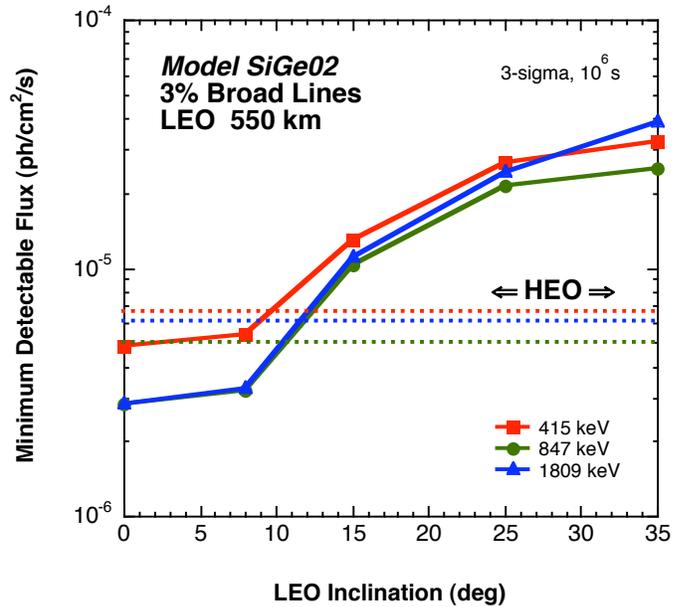

**Fig. K1.** *Simulated broad-line sensitivity (normal incidence) of the initial Si-Ge baseline design for different LEO orbital inclinations. Also shown are the same line sensitivities at three energies for a high-earth orbit (HEO) outside the earth's trapped-particle magnetosphere. (Note this was not the optimized Si-Ge geometry, nor the optimal analysis methods, but we expect trends to be similar among all ACT designs.)*

### Transportation

For the baseline mission to a low-inclination orbit, the IMDC study recommended a Delta IV 4240 (Fig. K2) launch from KSC with an orbit change to a <8° inclination orbit. No specific problems are foreseen. For the high-altitude alternative, a standard mission approach such as that used for *Chandra* or INTEGRAL can be planned.

### Assembly or Deployment

The entire ACT instrument will be integrated and tested with the S/C and its other subsystems at the spacecraft integration facilities before integration with the launch vehicle. After launch, only standard deployment of S/C systems is required (GaAs solar panels, thermal radiators, 1.5-m diameter TDRSS Ku-band high-gain antenna) There is no on-orbit assembly.

### Propulsion

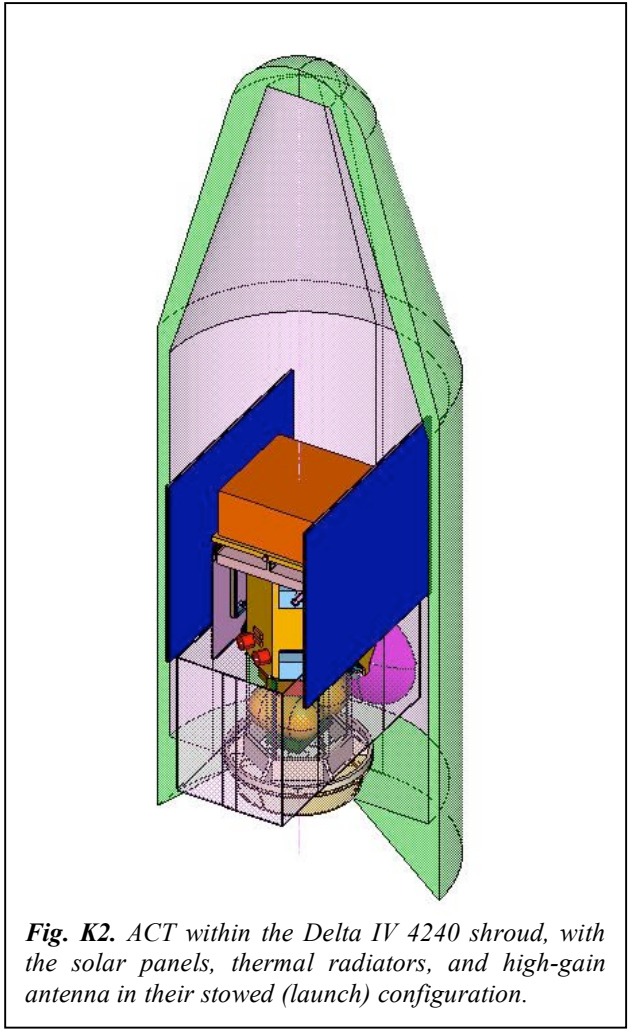

**Fig. K2.** *ACT within the Delta IV 4240 shroud, with the solar panels, thermal radiators, and high-gain antenna in their stowed (launch) configuration.*

For orbital maintenance and controlled re-entry, the IMDC team baselined a propulsion system using proven hydrazine monopropellant thrusters. The system was sized for drag makeup (75 m/sec) and deorbit (143.5 m/sec), with two 100 lb·f thrusters. Four additional 5 lb·f thrusters and four 1 lb·f thrusters provide attitude control during maneuvers. Propulsion tanks (462 kg propellant) are sized to include enough propellant for orbital maintenance and controlled deorbit burns to target a safe landing in the Pacific Ocean.

# L. OPERATIONS

ACT has a very large FoV (>25% of the sky) and for normal operations the center of the FoV will be zenith pointed, or modified zenith pointed with rocking like GLAST (to equalize exposure over the sky). Re-orientation capability will be provided so that the instrument could be pointed closer to the orbit poles when required, for example to monitor transient sources, or to introduce an additional capability to optimize the sensitivity and sky exposure, if required for the primary survey.

The Compton imaging technique, large effective area, and large FoV result in very high event rates and associated data rates. While some on-board processing will be implemented to filter Compton data before transmission to the ground, most of the event processing must be done on the ground requiring high data rates (no higher than 60 Mbps). The IMDC study showed that this could be accomplished through periodic transmissions via TDRSS, and this is the planned approach. An alternative would be to store the data on-board and transmit the stored data directly to a near-equatorial receiving station on every other orbit.

The ACT mission is a moderate sized S/C (~ 4000 kg). It is anticipated that controlled re-entry will be required, and a re-entry implementation using timed burns of a hydrazine thruster system similar to that planned for the GLAST mission can be used.

## COMMUNICATIONS

Original conservative estimates of the instrument data rate made prior to ISAL (and prior to our detailed simulations study) assumed 25000 events/s in ACT, with 10–20 individual interactions per event and 120 bit/interaction, for a total instrument data rate of 60 Mbps. With instrument HK overhead (15 kbps) and CCSDS overhead (15%) this becomes 69 Mbps; this data rate was the basis for the IMDC communications concept design.

Based on our more detailed simulations, we can now say that for the Si-Ge baseline with BGO shield concept this estimate was indeed very conservative. Our simulations indicate that actual event rates, as well as the average number of interactions per event, for a Si-Ge hybrid are much lower. All "2+" interaction events (including random coincidences, not including noise-triggered channels) have an event rate of 3203/s and consist on average of 3.6 interactions. With 120 bits per interaction, plus an additional per-event overhead of 150 bits, this would correspond to a data rate of 1.9 Mbps. What is missing from this are single-site events that get turned into two-site events by the addition of a random noise-induced trigger in one of the ~400k channels of the instrument. With a noise threshold set at 5σ, reading out 400k channels results in one noise-induced additional trigger in every 4 trials, increasing the total "2+" interaction rate to 12265/s and the resulting data rate for a conservative 4 interactions/event to 7.7 Mbps. As two triggers are required to define a valid x-y position, this is a gross overestimate, which is still almost an order of magnitude below the continuous data rate assumed in ISAL. Our instrument study indicates that for sources other than GRBs or strong solar flares, for most instrument concepts sensitivities are best if only events consisting of 3 and more interactions are used. Restricting telemetry to those events would further reduce data rates without impact on most science topics.

ACT's data rate will likely be similar for other BGO-shielded semiconductor concepts relying on few interactions and a complete absorption of the photon's energy. However, concepts such as a many-Si-layer instrument relying on multiple interactions without complete absorption or electron-tracking thin Si or gaseous Xe would have many more interactions per valid event. Non-shielded concepts would face a significantly higher *raw* background event rate—mainly due to albedo photons and secondary radiation from the S/C. Consequently, any such concept would have a data rate much closer to the 60 Mbps assumed at IMDC.

For most observations, obtaining the data on ground is not time-critical. Downlinks every (few) orbit(s) are sufficient from the scientist's point of view, and the required on-board storage (on the order of tens of Gigabyte per orbit) and telemetry (Ku-band downlink via TDRSS) were considered unproblematic by IMDC.

There is one subclass of observations where a delay in data receipt on ground is problematic from a scientist's point of view: ACT should be able to instantly alert ground observers (and other S/C) to interesting astrophysical transient events such as GRBs. This requires very limited bandwidth—on-board analysis can determine a position and rough spectral properties, and only this information need be transmitted instantly. This requirement can be accommodated by S-band transmission via the TDRSS demand access channel.

## GROUND SEGMENT

### Spacecraft Operations

It is anticipated that ACT S/C operations will be managed from Goddard Space flight Center (GSFC). ACT has a very large FoV (> 25% of the sky) and for normal operations will be zenith pointed. While the instrumental design is closer to COMPTEL/CGRO, the S/C operations and data rates are closer to GLAST, and the operational models for this mission will largely apply to ACT.

### Instrument Operations

It is expected that the instrument operations for ACT will be conducted from the Principal Investigator institution, or possibly from GSFC. With the wide FoV instrument, daily instrument operations during the first 5-year zenith-pointed survey mode will not be driven by specific scientific investigator requirements.

Initial operations of ACT will place heavy requirements on fine-tuning the performance of ACT. Specific issues will be optimizing the detector readout system, detector thresholds, operating temperatures, anticoincidence thresholds, and timing. The performance of ACT is also dependent on the processing of the multiple γ-ray interactions occurring simultaneously throughout the ACT detector. The on-board software that processes these data must be adapted to changing configurations to optimize the data return.

The instrument operations facility must also be prepared to implement changes in the instrument configurations to automatically or responsively handle unique observational requirements such as the very high rates associated with γ-ray bursts, solar flares, galactic supernovae, or nearby novae. The instrument operations team must also quickly respond to unexpected anomalies such as the unexpected change in performance or failure of a detector or detector array or associated electronics readout, and the possible requirement for periodic annealing of the solid-state detectors.

### Data Processing

During normal operation mode, an average of 1500 events per second which fulfill the trigger requirements are downlinked and have to be processed. This processing happens in three steps: (1) Calibration of the individual interactions (ADC counts to energies and interaction depths, etc.), (2) Event reconstruction (ordering of the hits in their interaction sequence and evaluation of the event quality) and (3) Scientific analysis. Before each of those steps different levels of simple quick-look analyses can be done to assure the performance of the instrument at all levels.

The initial levels of data processing (1 & 2) occur on an event-by-event basis and thus can be easily parallelized. Even with the most sophisticated currently available event reconstruction algorithms the amount of data can easily be handled by a state-of-the-art (2005) Beowulf cluster (32 CPUs @ 2.4 GHz). The quick-look analysis, which will span data sets of up to several hours, can be done in parallel on several high-end (2005) CPUs.

The final scientific data analysis (3) spans data sets of a few seconds (bursts, flares) to the whole mission lifetime (e.g. all-sky maps). Accordingly the system requirements range from single CPU systems to massively parallel multi-CPU systems. It is not expected that the *relative* data processing challenges for this mission will be larger than they were for ACT's predecessor COMPTEL in the 1990's.

For all these data processing steps preliminary software exists and has been applied to lab measurements and to all the simulations presented in this study [121] (Appendix P).

### Instrument Response Database

Knowledge of the response of an instrument like ACT to well-defined inputs provides the fundamental means of analyzing data. The fidelity of reduced end-product data (fluxes, locations, times, etc.) depends directly upon the fidelity of the underlying instrument response. For instruments like ACT that measure many quantities, the instrument response is a complex function of many variables, such as photon energy, direction, polarization, interaction location, instrument-operating parameters, etc. This function must be reliably captured in a database that can be efficiently incorporated into scientific data processing and analysis systems.

The only practical means of capturing the complex response of ACT is through extensive computer simulations. The particle interaction and transport simulation tools (i.e. MGGPOD for the present study, see Section S) used for instrument response modeling should ideally be the same as those used for instrument design optimization studies. It is of course imperative that simulation codes be extensively cross-validated against experimental measurements.

The above simulation tools alone, however, represent only a small part of the analysis infrastructure needed. Significant research is needed to develop an optimized detector response architecture suitable for the large complexities of a given ACT concept. It is impractical even with modern computing resources to capture and store *all* of the details of any ACT's instrument response function.

Instead, at least for each general detector concept (non-tracking non-ToF, tracking non-ToF, non-tracking ToF, and detector concepts requiring reconstruction of incompletely absorbed multiply-interacting photons) a separate instrument response data space must be designed to preserve the most crucial parameters within a reasonably-sized matrix. Such a response matrix can be made up of several independent sub-matrices addressing different aspects of photon interactions. Of course the parameters spanning the data space must be chosen as orthogonal as possible.

For an ACT telescope measuring a host of parameters per photon, reconstruction and validation of the individual event is a step largely separated from high-level analysis resulting in images, spectra, etc. Correspondingly, the event-reconstruction and imaging responses are separable.

Given currently available computing resources, the only approach feasible to calculate sky images from ACT's data without radical (and detrimental) simplifications of the measured data is to evaluate each measured photon's contribution in sequence ("list mode", see e.g. [122]) instead of performing operations on the full measurement parameter space. In this approach the imaging response for each photon being analyzed is generated "on-the-fly" from a compact imaging response (or simplified analytic function).

While this approach shows promise, further research is needed to investigate whether it is appropriate for a large ACT mission, or whether alternate techniques are needed for specific scientific objectives. For instance, the "photon-by-photon" approach presents a clear advantage for most all-sky imaging objectives—but calculating full-mission all-sky images over wide energy bands would surpass the capabilities of *today's* computers. Today, one would have to revert to a simplified binned data space and imaging techniques similar to those employed by e.g. COMPTEL. Similarly, simplified analysis techniques using binned data spaces might be best suited to rapid on-board GRB localization.

**Science Processing**

In order to transit from the recorded event data, which are primarily described in an instrumental frame of reference, to the 'outside' world of astronomy, one must relate the measurements to auxiliary satellite parameters that describe the exact details and circumstances of the measurement. Since ACT will mostly operate in a continuously scanning mode and the attitude of the instrument is slightly changed for every recorded γ-ray photon one must develop and maintain a continuous model for the satellite attitude allowing the projection of the estimated photon incidence directions or of images made in the instrumental frame of reference to the celestial sphere. Alternatively, each photon's information must be retained in a celestial rather than instrument coordinate system. The most important tasks of science processing will then be the following:

(i) derivation of maps of the γ-ray intensity on the sky

(ii) describe these maps in terms of unresolved sources (point sources), resolved emission regions of limited size (clouds, galactic structures etc.), and a large scale γ-ray brightness forming a more or less uniform background

(iii) analysis of the temporal and spectral properties of the point sources, extended structures, and background

(iv) analysis of special targets like solar flares, γ-ray bursts, terrestrial γ-ray flashes, and possibly investigation of solar system objects.

Task (i) will demand a continuously updated timeline of the exposure integral maps (effective area × exposure time for a pixel on the sky) for a number of spectral bands, either continuum or γ-ray line related, and for a convenient tessellation of the celestial sphere. Since the time of exposure may depend on many factors (systematic variations of the instrument, characteristics of the sources under investigation, preferences of the scientist doing the analysis) it is advisable to generate the exposure maps for the smallest useful pixels on the sky and the shortest practical bins in time. The recorded events and the exposure maps can then be combined to generate maps of 'apparent' γ-ray intensity. However, in any realistic space experiment in the γ-ray range the knowledge of the instrument background is essential to convert the recorded 'apparent' intensity into the true 'astronomical' intensity of the sky. One must therefore prepare via simulation as well as via continuous analysis of the orbital background signatures a database of appropriately integrated background maps. Tasks (i)–(iv) can benefit from a long tradition of analysis procedures that were developed for past and current γ-ray missions. The very much improved sensitivity of ACT will certainly raise new challenges in the science processing: the analysis of crowded source fields, nuclear spectroscopy with more lines, or disentangling faster temporal signatures will be possible with these data.

The overall success of a mission like ACT will of course be judged by the outreach and connections that can be made between a γ-ray mission and the other fields of astronomy and astrophysics. It is therefore crucial to design the science processing system of ACT in a way that enables the 'educated' scientific community to use the data and to perform specialized analysis tasks in a fairly independent manner. The experience from many missions will influence the ACT system and should then lead to a long-term useful data archive.

**Data Archiving**

ACT data will be hosted at the HEASARC archive. HEASARC will be responsible for the archive and will provide access to the data, distribute the software and ensure that data and software follow standards (for data and software) common to other high-energy missions. HEASARC will also interface between the ACT data and software and any future development in the Virtual Observatory project that ensures a larger distribution of data across all spectral wavelengths.

Data (both science and calibration) will be stored in a FITS format with standard layout well known within the high-energy community. The software will follow the standards developed within the FTOOLS.

The telemetered data will be transformed in standard FITS event format for the science data and will use plain FITS tables for the auxiliary and housekeeping information. This first step of the FITS file (Level 0) will be designed to maintain all the information from the telemetry. The usage of standard FITS format will give immediate access to software capable of manipulating the files via the generic FTOOLS. Subsequent processing operating on the Level 0 file will output calibrated and screened data. The processing of the large FoV of ACT will generate different types of databases, each tuned to a specific science objective. The concept is to archive the Level 0 data and the derived high-level data in different databases related either via observation ID or time viewing. Users will have access to both data types.

The data access will be via a specific ACT interface, the generic HEASARC browse interface that will allow for complex query in the data retrieval, or via FTP.

Calibration files derived before and during the mission will be also in FITS format using either a format defined for previous missions or ACT-specific formats for new types of calibration data. The calibration data will be managed within a dedicated calibration database, ensuring access to the latest version of the calibration data and providing a rigorous history of any changes.

The data reduction software will be part of the ACT package and will employ standard libraries to access data files and standard software interfaces. The ACT software package will be part of the FTOOLS package, also known as HEAsoft, which is well supported across the most popular operating systems, and building the ACT software under the FTOOLs umbrella will ensure a long lifetime for data usage.

Data rights, if any, will be honored using encryption software available at the time of the ACT launch.

# M. OPERATIONS ASSURANCE

## SYSTEM RESILIENCE

### Detectors, Electronics, Power Systems

The baseline NACT instrument concept consisting of a large active volume of silicon and germanium strip detectors is an inherently redundant system. The component detectors have proven lifetimes of several years, but some failures ranging from individual strips to detector modules are likely.

Fortunately, no single detector failure (or even a failure of groups of detectors) can cause the entire instrument to fail. Rather, detector failures result in "graceful" degradation of instrument performance that can be largely tolerated and still achieve required science performance. It is envisioned that supporting systems will be designed to capitalize on this inherent detector system redundancy. Detectors will be built as modular units; these modules combined into stacks; and the stacks combined to form columns or towers.

Power distribution, trigger electronics, and data readout electronics can be built around this modular hierarchy to provide further resilience. An array of electronics built around field programmable gate arrays (FPGA) will accumulate trigger signals from the low-level ASIC detector electronics starting at the basic detector module level (e.g., detecting coincident X/Y strip triggers), then moving to the detector stack level (e.g., detecting coincident triggers in more than one layer), and then at the column level (e.g., detecting coincident triggers from two or more layers in both Si and Ge detectors). Finally, column triggers are correlated between different columns and with signals from the BGO and ACD via a central trigger processor.

In addition to providing a great deal of resilience to failures, this approach provides for efficient, fast operation by avoiding excessive data transfer and communications bottlenecks. All data processing requiring fast timing is performed as soon as possible in the processing chain, and close to the detectors. Each level of this trigger electronics hierarchy is ideally supported by a similar scheme of data readout electronics and power distribution, so that individual elements can fail without catastrophic instrument failure.

Unique systems such as the central trigger processor, central instrument control computer, spacecraft power interface, spacecraft data interface, etc., will be made redundant since they are critical to instrument operation and overall mission success.

### Cryogenics

A detailed study of the cryogenic system reliability and redundancy was not performed during the ISAL and IMDC runs, and needs to be performed for ACT.

### ACD and BGO

As discussed above, ACT's ACD design is envisioned very similar to that of GLAST, and the BGO shield design, loosely based on the INTEGRAL/SPI shield concept, follows the same basic philosophy.

The GLAST ACD consists of 25 scintillator tiles on the top surface and 16 tiles on each side for a total of 89 tiles or modules. The scintillation light from each tile is collected by two independent sets of wavelength shifting fibers viewed by two separate PMTs. The 176 electronics channels of the ACD system provide a two-fold readout redundancy. The gain and threshold of each PMT are individually adjustable providing a high level of adaptability in the case of degradation. (In contrast to GLAST, for ACT's ACD we do not expect high-energy back-splash or self-veto to be a problem.)

ACT's BGO shield will consist of ~100 BGO modules. The scintillation light from each will be collected by two separate PMTs or PIN diodes; the two-fold readout system, as well as the shield's modularity, provide redundancy. Individually adjustable gains and thresholds again provide a high level of adaptability.

For GLAST, each ACD module is individually wrapped in a light-tight cover, and the entire ACD is further wrapped in a micrometeoroid thermal blanket. In the event of a collision with a millimeteoroid the light shield of one or more of the ACD modules is likely to be compromised resulting in the loss of that module.

The performance of the ACT telescope in the event of an ACD or BGO module failure depends on the level of modularity of the ACD or BGO design. Failure of one or more ACD or BGO modules will result in progressively more background.

Whether any failure of the ACD can be tolerated depends on the detector, electronics, trigger, etc. While the passage of energetic charged particles through the detector volume should be discernible in the vast majority of cases during Compton event reconstruction, a significantly higher trigger rate of the instrument would likely be unavoidable and result in higher instrument dead times. A small loss of instrument effective area might also result from having to employ schemes such as an interaction in the outermost few mm of the detector being used as veto.

In principle, the same is true for the BGO modules. Since their primary intent, however, is to shield against photons, not charged particles, loss of a BGO module would result in an increase of background components that are much harder to discriminate against during high-level data analysis. However, loss of a single—or even a few—BGO modules would only result in a slight decrease of overall instrument sensitivity.

## MAINTENANCE AND SERVICING

(not applicable)

# N. Safety

## Launch and Near Earth Operations

ACT will be an unmanned launch from KSC and, based on the IMDC study, no launch or near-Earth safety issues are foreseen.

## Planetary Protection

(not applicable)

## End of Mission Safety Issues

The ACT mission is a moderate sized S/C (~4000 kg). For the baseline instrument, it is anticipated that controlled re-entry will be required, and an implementation of this similar to that planned for the GLAST mission is planned.

# O. References

[1] Ahn, K., et al., *Phys. Rev. D* **71**, 121301 (2005).
[2] Akyüz, A., et al., *New Astronomy* **9**, 127 (2004).
[3] Aprile, E., et al., *SPIE* **4140**, 20 (2000).
[4] Aprile, E., et al., *NIM* **A461**, 256 (2001).
[5] Aprile, E., et al., *SPIE* **4851**, 1196 (2003).
[6] Beacom, J.F., et al., *PRL* **94**, 171301 (2005).
[7] Benetti, S., et al, *MNRAS* **348**, 261(2004).
[8] Black, J.K., et al., *SPIE* **4140**, 313 (2000).
[9] Boehm, F., et al., *Phys. Rev. D* **64**, 112001 (2001).
[10] Boggs, S.E., and Jean, P., *A&A* (*Suppl. Ser.*) **145**, 311 (2000).
[11] Boggs, S. E., et al., *SPIE* **4851**, 1221 (2002).
[12] Boggs, S.E., et al., *New Astr. Rev.* **48**, 251 (2004).
[13] Branch, D., et al., *PASP* **107**(1995).
[14] Cassé, M., et al., *ApJ* **602**, L17(2004).
[15] Castoldi, A. et al., *NIM*, to be published (2006).
[16] Churazov E. et al., *MNRAS* **357**, 1377 (2005).
[17] Clayton, D.D., et al., *ApJ* **155**, 75 (1969).
[18] Clayton, D.D, and Silk, J., *ApJ* **158**, L43(1969).
[19] Clayton, D.D., *Nature* **244**, 137 (1973).
[20] Clayton, D.D., and Hoyle, F., *ApJ* **187**, L101 (1974).
[21] Clayton, D.D., *ApJ* **244**, L97 (1981).
[22] Coburn, W., et al., *SPIE* **4784**, 54 (2002).
[23] Coburn, W. and Boggs, S.E., *Nature* **423**, 415 (2003).
[24] Crider, A., et al., *ApJ* **479**, L39 (1997).
[25] Deines-Jones, P., et al., *NIM* **A477**, 55 (2002).
[26] Deines-Jones, P., et al., *NIM* A478, 130 (2002).
[27] Dermer, C.D. and Gehrels, N., *ApJ* **447**, 103 (1995).
[28] Dominguez, I., et al., *Memorie della Societa Astronomica Italiana* **71**, 449 (2000).
[29] Fleishman, G.D., astro-ph/*0502245*.
[30] Fossati, G., *MNRAS* **299**, 433 (1998).
[31] Georgii, R, et al., *A&A* **394**, 517 (2002).
[32] Ghirlanda, G., et al., *ApJ* **613**, L13 (2004).
[33] Ghisellini, G., et al., astro-ph/0504306.
[34] Guessoum, N., and Jean, P., *A&A*, **396**, 157 (2002).
[35] Guessoum, N., et al., *A&A* **436**, 171 (2005).
[36] Hamuy, M., and Pinto, P.A., *AJ* **117**, 1185 (1999).
[37] Hardin, D., et al., *A&A* **362**, 419 (2000).
[38] Harding, A.K., Baring, M.G., and Gonthier, P.L., *ApJ* **476**, 246 (1997).
[39] Harris, M.J., et al., *A&A* **433**, L49 (2005).
[40] Hernanz, M., et al., *ApJ* **526**, L97 (1999).
[41] Hernanz, M., and José, J., *New Astr. Rev.* **48**, 35 (2004).
[42] Hillebrandt, W., and Niemeyer, J.C., *ARA&A* **38**, 191 (2000).
[43] Hoeflich, P., and Khokhlov, A., *ApJ* **457**, 500 (1996).
[44] Hoeflich, P., et al., *ApJ* **492**, 228 (1998).

[45] Jean, P., et al., *A&A*, in press (2005).
[46] Kanbach, G., et al., *AIP Conf. Proc.* **587**: *Gamma 2001*, 887 (2001).
[47] Kanbach, G., et al., *NIM* **A541**, 310 (2005).
[48] Kawai, N., et al., *GCN* **3937** (2005).
[49] Kinzer, R.L., et al., *ApJ* **559**, 282 (2001).
[50] Kippen, R. M., et al., *IEEE Trans. Nucl. Sci.* **47**, 2050 (2000).
[51] Knödlseder, J., *A&A* **441**, 513 (2005).
[52] Knop, R.A., et al., *ApJ* **598**, 102 (2003).
[53] Kozlovsky, B., et al., *ApJ* **316**, 801 (1987).
[54] Kozlovsky, B., et al., *ApJ* **604**, 892 (2004).
[55] Kroeger, R.A., et al., *AIP Conf. Proc* **510**, 794, (2000).
[56] Kuiper, L., Hermsen, W., and Mendez, M., *ApJ* **613**, 1173 (2004).
[57] Kulkarni, S., astro-ph/0510256
[58] Kurfess, J.D., et al., *AIP Conf. Proc* **510**, 789, (2000).
[59] Kurfess, J.D., et al., *NIM* **A505**, 256 (2003).
[60] Kurfess, J.D., et al., *New Astr. Rev.* **48**, 293 (2004).
[61] Lei, F., et al., *Space Sci. Rev.* **82**, 309 (1997).
[62] Leising, M., and Clayton, D.D., *ApJ* **323**, 159 (1981).
[63] Leising, M., et al., *BAAS* **31**, 703 (1999).
[64] Lichti, G.G., et al., *A&A* **292**, 569 (1994).
[65] Lin, R.P., et al., *Solar Physics* **210**, 33 (2002).
[66] Luke, P.N., et al., *IEEE NSS* **39**, 590 (1992).
[67] Luke, P.N., et al., *IEEE NSS*, presented (1998).
[68] MacFadyen, A.I., and Woosley, S. E., *ApJ* **524**, 262 (1999).
[69] Maeda, K., and Nomoto, K., *ApJ* **598**, 1163 (2003).
[70] McConnell, M. L., et al., *AIP Conf. Proc.* **587**: *Gamma 2001*, 909 (2001).
[71] McNaron-Brown, K., et al., *ApJ* **451**, 575 (1995).
[72] Medvedev, M.V., astro-ph/0510472.
[73] Mészáros, P., *ARA&A* **40**, 137 (2002).
[74] Michelson, P.F., *SPIE* **4851**, 1144 (2003).
[75] Milne, P.A., et al., *ApJS* **124**, 503 (1999).
[76] Milne, P.A., et al., *New Astronomy Review* **46**, 617 (2002).
[77] Milne, P.A., et al., *ApJ* **613**, 1101 (2004).
[78] Morris, D.J., et al., *AIP Conf. Proc.* **410**: *Proceedings of the Fourth Compton Symposium*, 1084 (1997).
[79] Nishikawa, K.I., et al., *ApJ* **622**, 927 (2005).
[80] Nomoto, K., et al., *ApJ* **286**, 644 (1984).
[81] Oberlack, U., et al., *SPIE* **4141**, 20 (2000).
[82] O'Neill, T.J., et al., *AIP Conf. Proc.* **587**: *Gamma 2001*, 882 (2001).
[83] Pelling, R. M., et al., *SPIE* **4784**, 21 (2002).
[84] Permutter, S., et al., *ApJ* **483**, 565 (1997).
[85] Phillips, M.M., *ApJ* **413**, L105 (1993).
[86] Phlips, B.F., et al., *NSS/MIC Conf. Rec.*, **N12-12** (2001).
[87] Phlips, B., et al., IEEE NSS/MIC (2005).

[88] Plüschke, S., et al, *ESA* **SP-459**: *Exploring the Gamma-Ray Universe*, 91 (2001).

[89] Preece, R.D., et al., *ApJS* **126**, 19 (2000).

[90] Purcell, W.R., et al., *ApJ* **491**, 725 (1997).

[91] Ramaty, R., and Mandzhavidze, N., *IAU Symp.* **195**: *Highly Energetic Physical Processes and Mechanisms for Emission from Astrophysical Plasmas*, 123 (2000).

[92] Riess, A.G., et al., *AJ* **116**, 1009 (1998).

[93] Rudaz, S., and Stecker, F.W., *ApJ* 325, 16 (1988).

[94] Rudy, R.J., et al., *ApJ* **565**, 413 (2002).

[95] Ruiz-Lapuente, P., et al., *ApJ* **549**, 483 (2001).

[96] Ryan, J.M., et al., *SPIE* **4851**, 885 (2003).

[97] Sambruna, R.M., et al., *ApJ* **474**, 639 (1997).

[98] Savaglio, S., et al., *ApJ* **585**, 638 (2003).

[99] Schönfelder, V., et al., *ApJS* **86**, 657 (1993).

[100] Shah, K.S., et al., *IEEE TNS* **50**, 2410 (2003).

[101] Share, G.H., *ApJ* **595**, L89 (2003).

[102] Share, G.H., et al., *ApJ* **615**, L169 (2004).

[103] Smith, D.M., et al., *ApJ* **595**, L81 (2003).

[104] Smith, D.M., et al., in preparation.

[105] Starrfield, S., et al., *MNRAS* **296**, 502 (1998).

[106] Strigari, L.E., et al., *JCAP* **0504**, 017 (2005).

[107] Sunyaev, R., and Titarchuk, L., *A&A* **143**, 374 (1985).

[108] Tanimori, T., et al., *New Astron. Rev.* **48**, 263 (2004).

[109] van den Bergh, S., *PASP* **102**, 1318 (1990).

[110] Vedrenne, G., et al., A&A 411, L63 (2003).

[111] von Ballmoos P., et al., *A&A* **221**, 396 (1989).

[112] Vreeswijk, P.M., et al., *A&A* **419**, 927 (2004).

[113] Wang, W., et al., astro-ph/0510461.

[114] Watanabe, K., et al., *ApJ* **516**, 285 (1999).

[115] Watanabe, K., et al., *AIP Conf.*, 471 (2000).

[116] Weidenspointner, G., et al., *AIP Conf.*, 467 (2000).

[117] Wood-Vasey, W.M., astro-ph/050564.

[118] Zeh, A., et al., *ApJ* **609**, 952 (2004).

[119] Zhang, P., and Beacom, J.F., *ApJ* **614**, 37 (2004).

[120] Zoglauer, A., et al., *SPIE* **458**1, 1302 (2003).

[121] Zoglauer, A., et al., *New Astr. Rev.*, (in press) (2006).

[122] Zoglauer, A., et al., *ESA* **SP-552**: *The INTEGRAL Universe*, 917 (2004).

[123] http://cryowwwebber.gsfc.nasa.gov

[124] Tümer, T.O., et al., *NSS-MIC Conf. Proc., IEEE* **7**, 4369 (2004).

[125] Matteson, J.L., et al., *NSS-MIC Conf. Proc., in press* (2005).

## P. Numerical Simulations

### Introduction

A major aspect of the ACT concept study was the development and application of computer simulations and models for estimating realistic instrument performance parameters. Space-based instruments operating in the energy range of nuclear lines are subject to complex backgrounds generated by cosmic-ray and neutron interactions, diffuse γ-rays, and random coincidences. The count rate from backgrounds typically far exceeds that from astrophysical γ-ray sources. Maximizing the signal and minimizing the background depends critically on complex event selection and reconstruction algorithms, which in turn depend on the geometry of the instrument. Detailed computer simulations allow us to efficiently explore this vast parameter space and are thus vital for optimizing instrument designs and predicting performance. For the ACT simulation effort, we combined previously existing tools into a complete, powerful package for γ-ray astronomy. These tools were applied to several different ACT instrument concepts operating in different environments as described throughout the report.

### Simulation Tools and their Heritage

As illustrated in Fig. P1, any simulation system used for detailed instrument performance estimation must necessarily include many different components or "tools." For simulating γ-ray instruments in a space environment four basic types of tools are required: (1) models to describe source and background environment in terms of particle distributions, (2) models to describe the detailed geometry and mass of the instrument and spacecraft, (3) Monte Carlo simulation tools to predict the response of the instrument to particle sources and backgrounds, and (4) tools to process and analyze the resulting data in a manner emulating anticipated operational capabilities. There was no single simulation solution that included all these required tools. The greatest challenge of the ACT concept study simulation effort was to integrate and augment existing tools into a cohesive integrated system. The major components of the resulting toolset are described below.

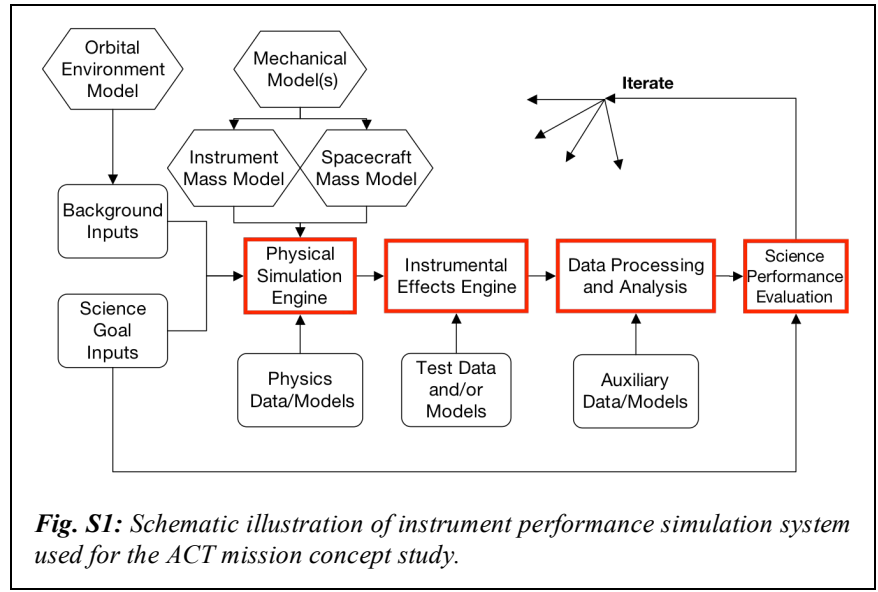

**Fig. S1:** *Schematic illustration of instrument performance simulation system used for the ACT mission concept study.*

### Source and Environment Models

Accurate space environment models are crucial inputs for reliable prediction of instrument performance. For the ACT study we generated a tool capable of producing particle distributions expected from a comprehensive set of backgrounds using available modeling resources and published results. The outputs are designed to be directly compatible with the chosen Monte Carlo simulation framework. The primary resource used for background models is the CREME96 package (Tylka et al. 1997) from the U.S. Naval Research Laboratory. This code is widely used to predict radiation doses for determining satellite electronics design constraints and has been shown to be accurate at predicting galactic cosmic-ray, anomalous cosmic-ray, and solar flare components of the near-Earth environment. The CREME96 code also includes a well-tested geomagnetic transmission calculation algorithm, and uses the established AP8 models for predicting magnetospherically trapped protons. The atmospheric neutron environment model (i.e., secondary neutrons that result from cosmic-ray interactions with the earth's atmosphere) is based on empirical data from COMPTEL reported by Morris et al. (1995) and references therein. For electron/positron cosmic rays, diffuse photons, and albedo photons the analytical models compiled in Mizuno et al. (2004) are used, convolved with geomagnetic transmission functions for specific orbits supplied by CREME96. Cosmic-ray electrons are extended to energies below 7 GeV for high Earth orbital environment based on data from Ferreira and Potgieter (2002). A compilation of all the backgrounds used in the study for different orbital environments is shown in Fig. P2.

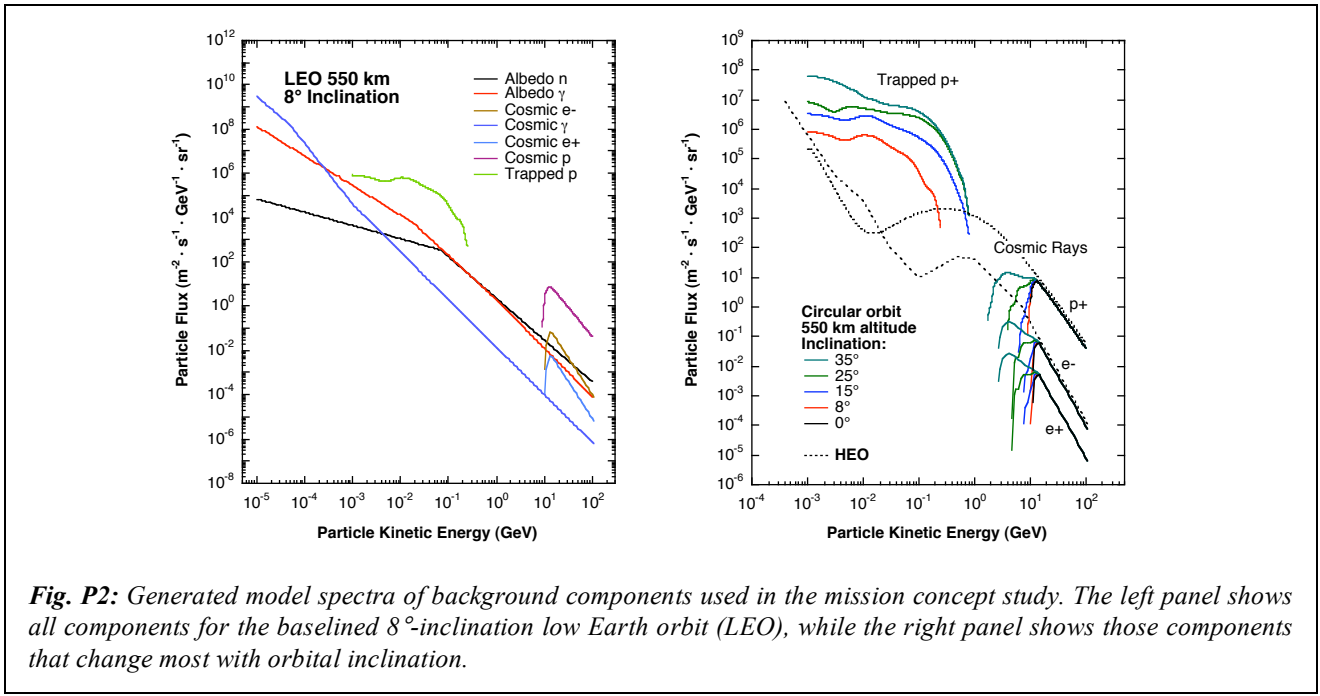

***Fig. P2:*** *Generated model spectra of background components used in the mission concept study. The left panel shows all components for the baselined 8°-inclination low Earth orbit (LEO), while the right panel shows those components that change most with orbital inclination.*

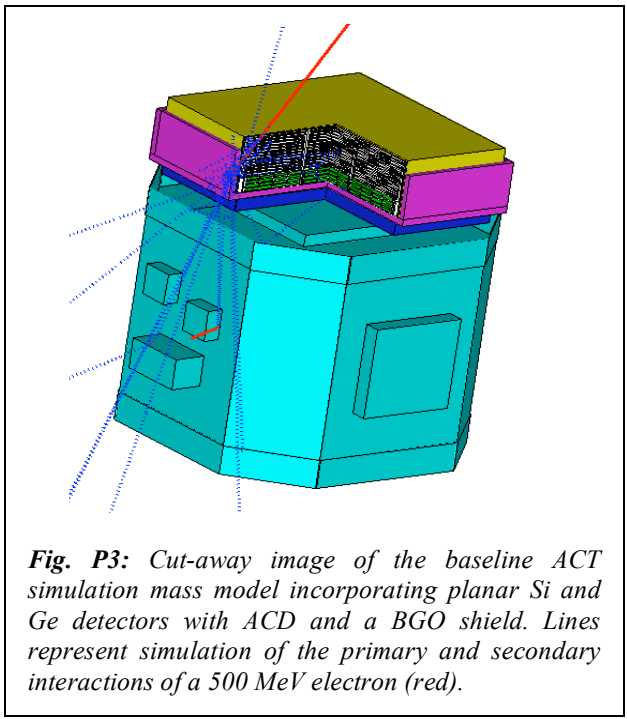

***Fig. P3:*** *Cut-away image of the baseline ACT simulation mass model incorporating planar Si and Ge detectors with ACD and a BGO shield. Lines represent simulation of the primary and secondary interactions of a 500 MeV electron (red).*

## Geometry and Mass Models

The process of modeling detailed instrument geometry and mass distributions for use in realistic Monte Carlo simulations is often long and tedious. For modeling different detector concepts, we need to be able to easily modify detector materials and geometries—including structural materials—while the problem of background lines from activation in the S/C and structural materials force us to include a fair level of detail even in rough-estimate mass models. To confront these challenges we developed a "universal" ACT mass model generation tool that conveniently supports different detector and structural materials and combines them with a reasonably detailed spacecraft model (based roughly on the GLAST spacecraft). Use of this tool also facilitates the comparison of performance simulations for different detector concepts, including "hybrids," because the same modeling constraints are included in all concept models. Fig. P3 shows a mass model, generated by the ACT mass model tool, for a hybrid instrument consisting of Si and Ge layers surrounded by a plastic scintillator anticoincidence shield.

## Monte Carlo Radiation and Particle Transport

The Monte Carlo package at the heart of the ACT simulation toolset is called MGGPOD (Weidenspointner et al. 2005). This package is a suite of Monte Carlo codes built around the GEANT3 framework to simulate the physical processes relevant for the production of instrumental backgrounds at γ-ray energies. These processes include the build-up and delayed decay of radioactive isotopes as well as the prompt de-excitation of excited nuclei—both of which give rise to a plethora of instrumental γ-ray background lines in addition to continuum backgrounds. The MGGPOD package has been successfully applied to modeling the instrumental backgrounds of the Wind-TGRS (Weidenspointner et al. 2005), INTEGRAL-SPI (Weidenspointner et al. 2003), and RHESSI (Wunderer et al. 2004) instruments. To illustrate the performance of MGGPOD, we depict a comparison of simulated and measured instrument backgrounds for Wind-TGRS in Fig. P4. For inclusion in the ACT simulation package, additional input particle geometries, an interface to new environment and mass models, and more detailed event output were added to the package while the core simulation code remained unchanged. Enhanced neutron cross sections—based on the JENDL data—were incorporated into the GEANT3/GCALOR package for the relevant isotopes of Ge, Si, and Xe. An accurate description of the prompt de-excitation of excited Ge, Si, and Xe nuclei with resulting emitted particles was added. The MGGPOD suite also includes the GLECS/GLEPS (Kippen 2004, McConnell et al., 2006) packages for simulating the detailed effects of atomic binding and polarization on photon scattering processes that are important for Compton telescopes.

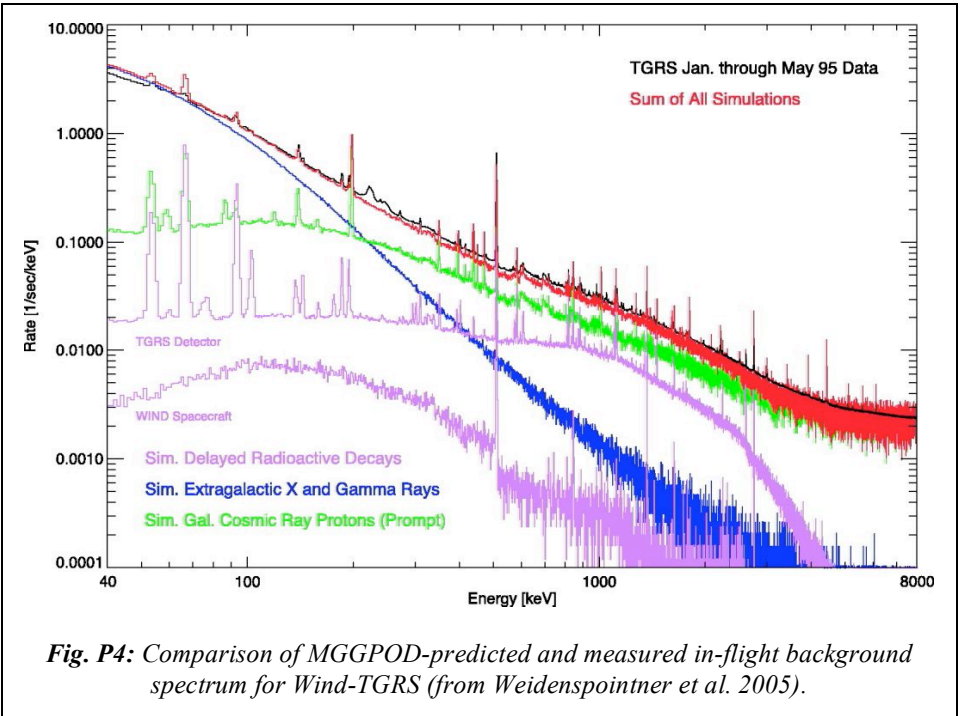

**Fig. P4:** *Comparison of MGGPOD-predicted and measured in-flight background spectrum for Wind-TGRS (from Weidenspointner et al. 2005).*

**Data Processing and Analysis**

The MEGAlib package (Zoglauer 2005, Zoglauer et al., 2006) was originally developed for

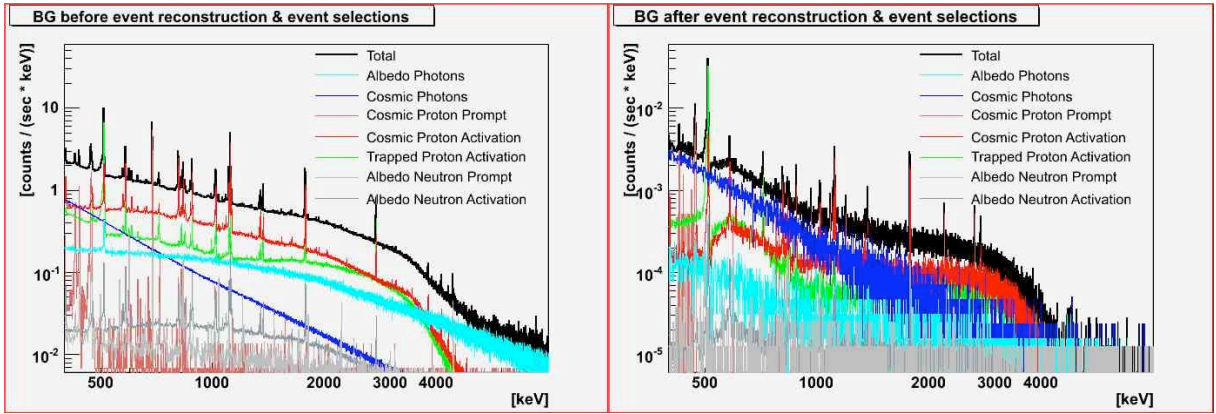

**Fig. P5:** *Background in the Si-Ge baseline instrument before and after event reconstruction and event selections. Note scales are different by three orders of magnitude.*

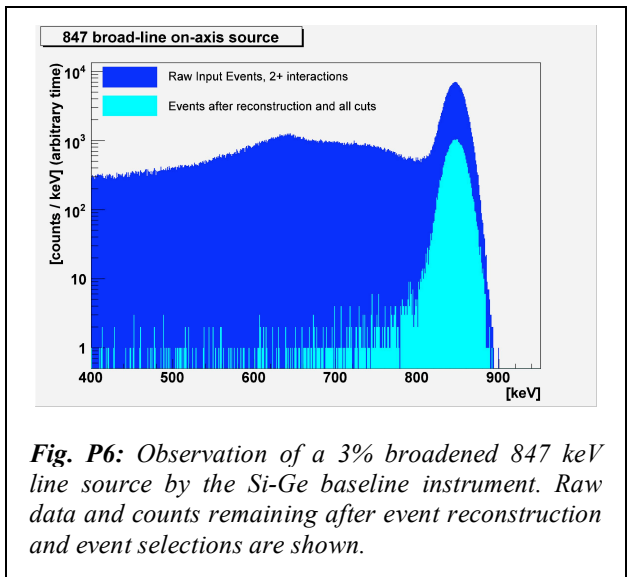

*Fig. P6: Observation of a 3% broadened 847 keV line source by the Si-Ge baseline instrument. Raw data and counts remaining after event reconstruction and event selections are shown.*

analysis of simulation and calibration data related to the MEGA laboratory/balloon prototype, a German-developed Compton telescope consisting of a thin Si tracker and a CsI calorimeter (Kanbach et al. 2003). The package comprises the complete data analysis chain for Compton telescopes—from discretizing simulation data and calibrating real measurements, to the reconstruction and selection of events, up to high-level data analysis, i.e. image reconstruction, background estimation, and polarization analysis. For the ACT study, the package was enhanced to include the reconstruction of incompletely absorbed events with four or more interactions (relevant in particular to a thick-Si instrument), time-of-flight

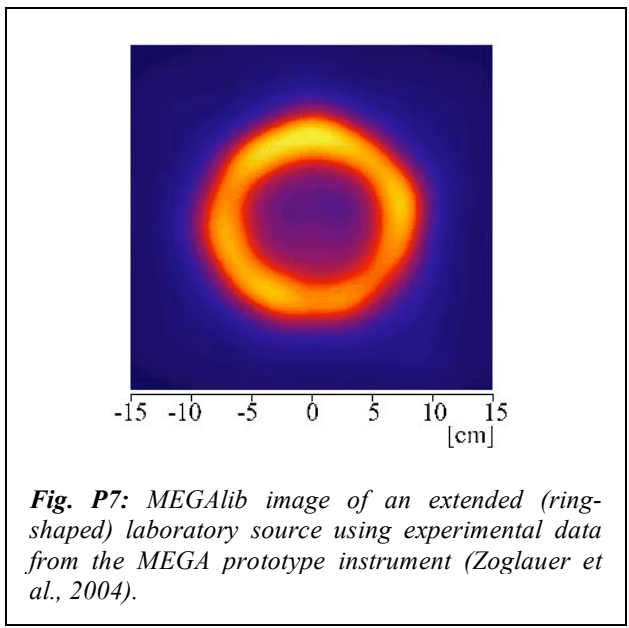

*Fig. P7: MEGAlib image of an extended (ring-shaped) laboratory source using experimental data from the MEGA prototype instrument (Zoglauer et al., 2004).*

information, background due to random coincidences, and more.

The most critical part in the data analysis is the reconstruction of photon energy and direction (and polarization) from raw measured quantities (interaction positions and energies). Event reconstruction is the key to differentiating source events from background. Different approaches are implemented or under development in MEGAlib. The most promising technique is based on Bayesian statistics. It uses a seven-dimensional pre-calculated instrument response matrix to determine the most probable direction of motion, and computes a quality factor for each event describing the probability that the given event sequence is correct and originated from a completely absorbed photon. Rather than simpler, analytical approaches to event discrimination, this approach requires neither projections of the event space onto a single dimension nor use of figures-of-merit that can render relative quality ratings of different events problematic. This Bayesian approach provides a powerful means of reducing background events, which typically have lower statistical quality.

Figs. P5 and P6 illustrate the power of this approach. In Fig. P5 we show the background rates induced in the instrument by the different components of the radiation environment as function of energy—first for all raw events consisting of 2 or more interactions in the instrument, then for only those events that pass event reconstruction and all event selections. Note that background levels are reduced by roughly a factor of 1000. Fig. P6 shows data from a point source of 3% broadened 847 keV line emission—again "raw" data are compared to the outcome of event reconstruction and event selections. The photopeak intensity is reduced by a factor of 6.8, with a simultaneous suppression of the Compton continuum by over 2 orders of magnitude.

The capabilities of MEGAlib to correctly reconstruct incident directions of photons and to ultimately produce images even of diffuse emission regions are illustrated in Fig. P7 by the reconstructed image of an extended source observed by MEGA in the laboratory.

## Coincidence Timing

The baseline science instrument consists of Si and Ge strip detectors surrounded by BGO and plastic anticoincidence shields. Photon interactions of interest consist of two or more simultaneous interactions in the detectors. Given an instrument consisting of several hundred kilograms of active detector mass read out by hundreds of thousands of channels, random coincidences between unrelated events are a concern for two reasons: the effects on the overall telemetry rates, and the production of an additional background component. We have estimated the magnitude of these potential problems with our simulations of the baseline instrument.

For the Ge strip detectors, the achieved depth resolution of 0.4 mm corresponds to signal timing of ~10 ns, and Si timing on the same order should be achievable. To be conservative, we can start with an assumed coincidence window of 150 ns between the detectors, and a comparable window for the anticoincidence system. Combining simulated observations of all background components and generating random coincidence interactions from them results in a total of ~3200 cts/s in the instrument (~9.5% random coincident events) from events consisting of 2 or more interactions (~1480 cts/s, 5% randoms for events consisting of 3 or more interactions). Random events are dominated by the addition of low-energy deposits—for the "3 or more interaction" events, less than 4% have a total energy of more than 300 keV.

These numbers demonstrate that with a 150-ns coincidence window, random coincidences will not significantly affect the ACT telemetry rate. In addition, given further event reconstruction on the ground, the expected contribution should not be a dominant background component. However, computational and time limitations made it impossible to explicitly include this component in any of the sensitivity estimates shown in this report.

Another type of random event occurs independent of coincidence timing—it results from randomly high signal levels in individual channels during readout. For a noise threshold of 5σ, one such event occurs roughly every 4 times the 400,000 channels are read out, randomly adding a single triggering strip to 25% of the single-hit events. Assuming no matching orthogonal strip were required for a second hit, this would turn the event into one consisting of two interactions.

For the first type of random coincident events,

our simulations demonstrate that a coincidence window of 150 ns keeps this component from becoming a problem. Given that a coincidence window between Si and Ge of a few tens of ns should be achievable, this type of background will be a minor contributor to the full background encountered by this instrument.

The contribution of random doubles due to noisy channels can be drastically reduced both by requiring a signal from at least two orthogonal strips in any given detector (i.e. at least one complete interaction) and by setting readout thresholds to 7σ or 8σ rather than the fairly low 5σ assumed in these estimates.

## Status

The first-generation ACT simulation pipeline is in place and has been tested end-to-end at multiple user sites. We have verified our environment model by comparison with the models used for RHESSI and TGRS background modeling, and through comparison with published literature. The simulation toolset has been applied to many different ACT instrument concepts, configurations, and environmental conditions—representing tens of thousands of hours of computing time. Improving and enhancing the event reconstruction code and event selection criteria is an ongoing process that constituted a large portion of the ACT Vision Mission Study task. The set of event reconstruction tools necessary and the optimum event selection criteria depend on the ACT detector concept and instrument geometry under consideration. The preliminary simulation results presented in this report represent our current best effort, but future optimization work is likely to result in additional performance gains.

## ACT Core Simulation Team

Despite limited funding and resources we were able to accomplish extraordinary results, thanks in part to the dedication of a core team of researchers dedicated to the development, test, and application of simulation systems. This team is listed at the beginning of the report.

## References

Ferreira, S. E. S. and Potgieter, M. S. 2002. "The modulation of 4- to 16-MeV electrons in the outer heliosphere: Implications of different local interstellar spectra" JGR **107(A8)**, 1221.

Kanbach, G., et al. 2003. "Concept study for the next generation medium-energy gamma-ray astronomy mission: MEGA." *Proc. SPIE* **4851**, 1209.

Kippen, R. M. 2004. "The GEANT Low Energy Compton Scattering (GLECS) Package for use in Simulating Advanced Compton Telescopes." *New Astron. Rev.* **48(1-4)**, 221 (http://public.lanl.gov/mkippen/actsim/).

McConnell, M., et al. 2006, ApJ, in preparation.

Mizuno, T., et al. 2004. "Cosmic-Ray Background Flux Model Based on a Gamma-Ray Large Area Space Telescope Balloon Flight Engineering Model." ApJ **614**, 1113.

Morris, D., et al. 1995. "Neutron Measurements in near-Earth Orbit with COMPTEL." JGR **100**, 12243.

Tylka, A. J., et al. 1997. "CREME96: A Revision of the Cosmic Ray Effects on Micro-Electronics Code." *IEEE Trans. Nucl. Sci.* **44**, 2150 (https://creme96.nrl.navy.mil/).

Weidenspointner, G., et al. 2003. "First identification and modelling of SPI background lines." A&A **411**, L113.

Weidenspointner, G., et al. 2005. "MGGPOD: a Monte Carlo Suite for Modeling Instrumental Line and Continuum Backgrounds in Gamma-Ray Astronomy." ApJS **156**, 69. http://sigma-2.cesr.fr/spi/MGGPOD/

Wunderer, C., et al. 2004. "Modelling of the Detector Background Spectrum for the Low-Earth Orbit GE Spectrometer RHESSI with MGGPOD." In *The INTEGRAL Universe (ESA SP-552)*, Eds. V. Schönfelder, G. Lichti & C. Winkler, p.913.

Zoglauer, A. 2005. PhD Thesis Techn. U. Munich.

Zoglauer, A., Andritschke, R. and Kanbach, G. 2004. "Data Analysis for the MEGA Prototype." *New Astron. Rev.* **48(1-4)**, 231.

Zoglauer, A., et al. 2006. New Astr. Rev., submitted.

# Q. INSTRUMENT CONCEPTS

## INTRODUCTION

One aspect of the ACT Vision Mission Concept Study was to determine the "most promising technologies". Since different detector technologies and the resulting possible ACT instrument designs have very different advantages and disadvantages, requirements and merits, a means of comparing one instrument's performance to another had to be found.

In terms of a "science performance parameter", the on-axis sensitivity to a 3%-broadened 847 keV line, in combination with the instrument's energy resolution as discussed in Section G and illustrated in Fig. G8, was the obvious choice given the prime science objective of distinguishing SN Ia models. Of course the sensitivity evaluation includes simulations of all relevant components of the space radiation environment. Random coincidence events were not studied in detail for each instrument concept due to computational limitations see discussion in Sections G & P.

In order to compare very different instrument concepts on as level a playing field as possible, a means of limiting the size, pixelization, etc. of the instruments had also to be determined. The natural choice here was to use constraints on instrument mass, instrument power, and instrument size from the IMDC (and ISAL) runs.

The ACT toolset was used for all simulations. This ensured the same assumptions about the radiation environment, similar levels of detail in the instrument modeling as well as identical S/C bus architecture in the models, the same physics implemented in the Monte Carlo code, and use of the same event reconstruction and event selection toolset.

## COMMON CONSTRAINTS

We based our common set of constraints on a Delta 4240 launcher into the low-inclination orbit studied during IMDC, the launch envelope of a 4-m or 5-m Delta fairing, and instrument power of 2000 W without contingencies.

**Mass**

IMDC determined that ~ 4700 kg could be lifted to our orbit of choice using a Delta 4240 or 4450. Subtracting fuel, S/C mass, and contingency (30%) margins as determined during the IMDC study, the ACT instrument including the mass of any cryocooler needed can have a mass up to 1850 kg. (Based on ISAL information, any cryocooler was estimated to have 0.1 kg of mass per watt of power consumed by the unit.)

**Power**

The IMDC investigated an instrument power requirement of 2500 W including margins (2000 W without margins). Subtracting generic power requirements for main electronics box, heaters, and LV power, 1730 W are available for cryocooling and front-end electronics.

ISAL determined 1mW/channel ASIC power requirements for high-resolution detectors such as Si or Ge. Cryocooler power requirements for Si, Ge, and liquid Xe instruments were estimated based on ISAL information, scaling from a combination of channel numbers and to-be-cooled detector mass.

**Size**

The size of any instrument appeared to be the least problematic. The IMDC run had considered a fairly compact Si-Ge unit, but no instrument studied would exceed the size limits set by a Delta 4-m or 5-m shroud before violating either the mass or the power limits explained above.

## R. THIN SILICON - CdZnTe
University of California at Riverside

### General Description

This concept is based on using multiple layers of thin silicon strip detectors both as the Compton converter and the recoil electron tracker, providing both the electron's energy and direction. Position sensitive CdZnTe detectors on five sides of the silicon modules absorb the Compton-scattered photon, to measure its energy and direction. The complete Compton event can be reconstructed for the incident γ-ray's unique direction and energy. This results in a much more confined point-spread-function to eliminate background events that originate from the full event annulus determined without the knowledge of the recoil electron direction.

The use of a low-Z (14) silicon converter/tracker assures that Compton interactions dominate over ACT's full 0.2–10 MeV energy range and minimizes the effect of Doppler broadening on the angular resolution. The detailed energy loss and multiple-Coulomb scattering record of the electron track through successive silicon layers gives the direction-of-motion needed for proper event reconstruction. The high-Z (~49) CdZnTe calorimeter is designed to fully absorb the scattered photon so that the measured incident energy is unaffected by Doppler broadening in the silicon. The converter/tracker surrounded by a calorimeter naturally optimizes the detection probability within the FoV by favoring forward-scattered electron events with higher energies for better detection and a significantly larger solid angle for scattered photons at large Compton scatter angles. The calorimeter also serves as an effective active shield for the albedo γ-ray background.

### Status of Hardware Development

Silicon strip detectors were originally developed for tracking charged particles in high-energy accelerator experiments in the early 1990s. Double-sided detectors with orthogonal strips on opposite sides provide the location as well as the energy loss of the charged particle traversing the detector. They are currently available with strip pitches as small as 20 μm, thicknesses from 50–1500 μm, and wafer sizes up to 6 in. An aluminum metallization layer above each strip provides AC coupling. The 150 μm detectors used for this study are fully depleted at a bias voltage of 20 volts and have bias resistors fabricated on the wafer. The low voltage and room temperature operation eliminates the need for guard rings and cooling. We use detectors with a pitch of 0.76 mm and dimensions of 10 cm × 10 cm. These parameters can be adjusted for optimum sensitivity and number of channels (i.e., power). This was not done in this thin Si-CdZnTe study.

Minimum ionizing particles losing ~26 keV/100 μm in silicon produce 277 electron-hole pairs for each keV of energy loss. These statistics with a small Fano Factor for stopping electrons yield an excellent inherent energy resolution. The front-end electronics (ASIC), individual strip capacitance (i.e., thickness and width) and number of silicon layers traversed by the electron ultimately determine the energy resolution for stopping electrons. A number of self-triggered front-end ASIC readout chips are available. As an example, the TAA-1 chip (IDE AS) has 128 channels, 0.29 mW/ch and (160 + 6.1/pF) electron noise. For the detector configuration in this study, we have assumed a single-layer resolution of 5 keV (1-σ).

Cadmium-Zinc-Telluride detectors with 3-dimension position sensitivity offer major advantages for advanced Compton telescopes operating at room temperature. Arrays of CdZnTe detectors with dimensions of 2 cm × 2 cm × 2 cm and resolutions of 3 keV (1-σ) and 1 mm are assumed for this study. Resolutions of ~2 keV (1-σ) and 1 mm with pixilated 1.5 cm × 1.5 cm × 1 cm detectors have been achieved[1].

## Simulated Performance

**Description of Mass Model**

In our final simulations the D1 silicon converter/tracker consisted of 16 (4 × 4) modules, each with a sensitive area of 1600 cm² for a total active area of 25,600 cm². Each D1 module had 80 layers of 150 μm thick silicon for 2.8 g/cm² of converter material. The layer spacing was 0.5 cm. Each D1 module consisted of 4.5 kg of silicon and 3.5 kg of support structure. The trays separating the layers were assumed made from low-density Rohacell™ rigid foam with carbon composite and thin circuit board side panels for the ASICs to minimize the passive material traversed by the scattered gammas. The total D1 mass was 128 kg. Double-thickness CdZnTe arrays for D2 and D3 surrounded the 16-module D1 on five sides. The total mass for D2 and D3 was 1000 kg and the full assembly mass was 1789 kg. A plastic scintillator anticoincidence shield surrounded the telescope on all six sides. For the configuration in the final simulation D1 and D2/D3 required $2.62 \times 10^6$ and $0.102 \times 10^6$ channels of readout, respectively, and 2.72 kW (at 1 mW/ch). While this far exceeds the nominal science power budget, lower power ASICs and silicon detector pitch and thickness trade-offs can significantly reduce this number.

|  | Assumed in Mass Model | | | | Achieved in Laboratory | | | |
|---|---|---|---|---|---|---|---|---|
|  | ΔE (1σ) (keV) | Δx̄ (mm) (x/y; z) | Thresh (keV) | Δt (s) | ΔE (1σ) (keV) | Δx̄ (mm) (x/y; z) | Thresh (keV) | Δt (s) |
| D1 | 5.0 | 0.78; 0.15 | 50 | — | 4.6 | 0.78; 0.3 | 100 | — |
| D2 | 3.0 | 1.0; 1.0 | 100 | — | 2.0 | 1.0; 1.0 | 30 | — |
| D3 | 3.0 | 1.0; 1.0 | 100 | — | 2.0 | 1.0; 1.0 | 30 | — |
| AC | — | — | 100 | — | — | — | 100 | — |

**Table thinSi-CZT-1:** *Detector performances achieved and assumed for the instrument simulation for the thinSi-CZT ACT concept. Note: The silicon detector resolution achieved in the laboratory was for a 10 cm × 1 cm × 300 μm detector with 0.78 mm pitch.*

**Simulation Results**

|  | 511 keV narrow | 847 keV narrow | 847 keV 3% broad | 1809 keV narrow |
|---|---|---|---|---|
| ΔE (keV, 1σ) | - | 10.9 | - | - |
| ARM (FWHM°) | - | 2.8 | - | - |
| A_eff (cm²) | 228 | 387 | - | 21.2 |
| FOV (HWHM°) | - | 45 | - | - |
| 3σ sens (on-axis) | $4.4 \times 10^{-6}$ | $1.5 \times 10^{-6}$ | - | $9.3 \times 10^{-6}$ |

**Table thinSi-CZT-2:** *Simulated instrument performance for the thinSi-CZT ACT concept. Note that sensitivities are in photons cm⁻² s⁻¹ for 10⁶ s exposure. Only albedo photons and cosmic photons are included in background simulations; effective areas and sensitivities are for tracked events. (Chance coincidences are not included, though we estimate their effect to be small.)*

## References

[1] Z. He, private communication.

# S. GERMANIUM ACT WITH BGO SHIELD
### University of California at Berkeley

## GENERAL DESCRIPTION

ACT's main motivation is the quest to understand the origin of the elements, pursued via observations of γ-ray line emission from radioactive nuclei in the cosmos. In order to take maximum advantage of the information encoded in line profiles and shifts, the logical approach is to build a Nuclear Advanced Compton Telescope using detectors with the best energy resolution available—Germanium.

Compared to most other detector concepts studied here, Ge has the advantage of a reasonably high stopping power as well as fairly fast timing and fine detector voxelization, all at modest channel counts if Ge-strip detectors are used. A Ge instrument aims to fully contain each photon; optimum background rejection is possible if this happens in 3 or more interactions.

The Ge-ACT described here achieved the best 847 keV broad-line sensitivity of all instrument concepts under consideration—$1.1 \cdot 10^{-6}$ ph/cm$^2$s using Bayes reconstruction methods—as well as best narrow-line sensitivities, best energy resolution, and one of the largest fields-of-view.

## STATUS OF HARDWARE DEVELOPMENT

The Nuclear Compton Telescope[1,2] (NCT) is a Germanium Compton Balloon instrument. It utilizes the same type of Ge-strip detectors envisioned for a Ge-ACT. Ultimately a twelve-detector long-duration-balloon instrument, NCT had its first technology-demonstration 5-hour balloon flight with two detectors in June 2005 from Ft. Sumner, NM. Extensive laboratory calibrations demonstrated NCT's capability to image sources even from interactions in only one Ge detector. Each NCT detector is 1.6 cm thick and has 37×37 strips (2 mm pitch); 2-mm guard rings are instrumented and could be used as anticoincidence. Relative timing of anode and cathode signals to 10 ns enables depth resolution to 0.4 mm in each detector [3]. The current balloon electronics is based on conventional surface-mount technology, chosen to keep initial development cost at a minimum—switching to ASIC readout for the balloon's upgrade to more detectors is under consideration.

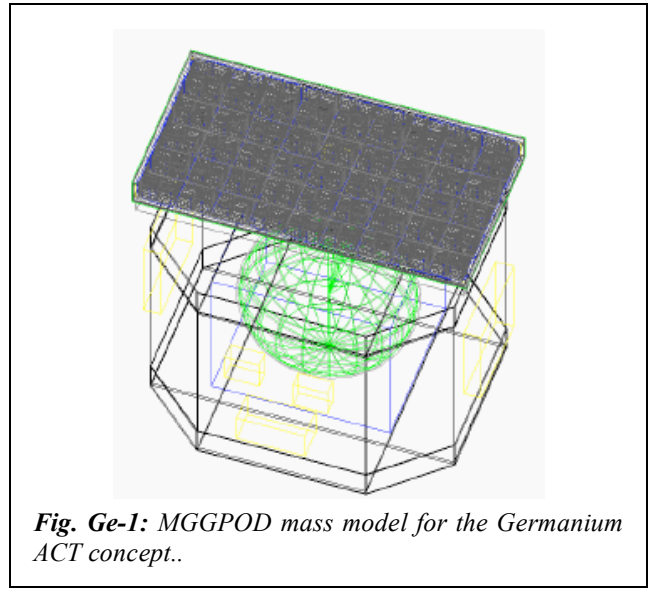

**Fig. Ge-1:** *MGGPOD mass model for the Germanium ACT concept..*

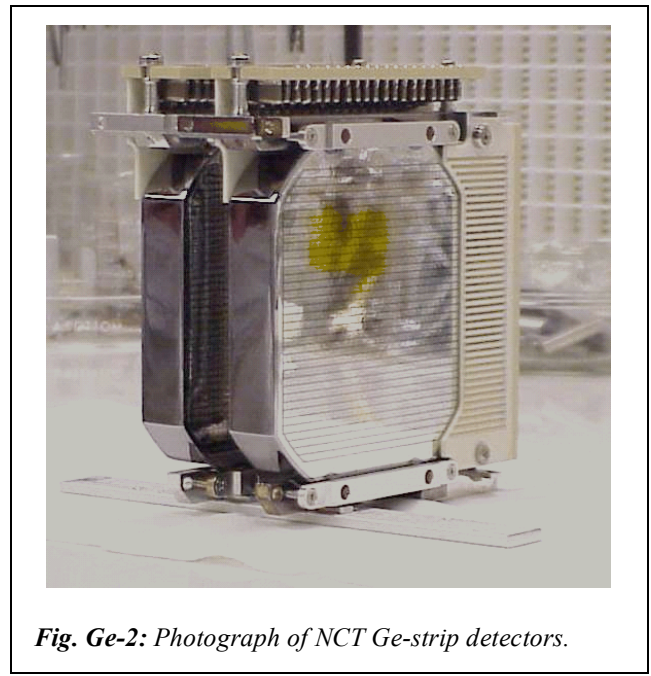

**Fig. Ge-2:** *Photograph of NCT Ge-strip detectors.*

## SIMULATED PERFORMANCE

### Description of Mass Model

The instrument consists of 6 layers of 10×18 Ge detectors, each 9.2×9.2 cm$^2$ (9.0×9.0 cm$^2$ active) and 1.6 cm thick, with NCT's strip pitch of 2 mm. Strip interpolation will enable 0.5 mm position resolution [4,5] or better. The detector is surrounded at sides and bottom by 3 cm BGO to reduce background from albedo photons as well as the impact of secondary photons from the S/C. At top and sides, a plastic scintillator rejects charged particles.

| | Assumed in Mass Model | | | | Achieved in Laboratory | | | |
|---|---|---|---|---|---|---|---|---|
| | $\Delta E$ (1σ) (keV) | $\Delta \vec{x}$ (mm) (x/y; z) | Thresh (keV) | $\Delta t$ (s) | $\Delta E$ (1σ) (keV) | $\Delta \vec{x}$ (mm) (x/y; z) | Thresh (keV) | $\Delta t$ (s) |
| D1 | 0.42 @ 0 keV 0.47 @ 60 keV 0.52 @ 200 keV 0.66 @ 666 keV 0.76 @ 1 MeV | x: 0.5; y: 0.5; z: 0.18 (z: 1σ) | Noise: 8 Trigger: 15 | ~ 10 ns | same for best det. / strips | x: 2.0, y: 2.0; z: 0.18 (z: 1σ) | Noise: 8 Trigger: 40 | ~ 10 ns |

*Table Ge-1: Detector performances achieved and assumed for the instrument simulation for the Ge ACT concept. Note: Guard rings assumed 1 mm thick (NCT 2 mm) and not instrumented as anticoincidence (i.e. act as passive material even though NCT detectors demonstrate the capability). A failure rate of Ge strips of 0.5% has been assumed in the simulations.*

The instrument has 97200 Ge channels; the whole Ge detector assembly weighs 851 kg, the whole instrument 1827 kg (including 711.5 kg for BGO shield assembly and 200 kg electronics and cyrocooler). Of the Ge assembly mass, only 8.5% is passive material—a benefit of the compactness of the Ge detectors that reduces the number of incompletely absorbed events the system has to recognize.

The Ge detector requires 1.65 kW including cooling; adding BGO readout brings this to 1.7 kW. This instrument design is constrained primarily by power. The thickness of the BGO shield is limited by mass rather than power budgets, with the 3 cm used here a reasonable baseline.

## Simulation Results

| | 415 keV narrow | 511 keV narrow | 847 keV narrow | 847 keV 3% broad | 1809 keV narrow | 6103 keV narrow |
|---|---|---|---|---|---|---|
| $\Delta E$ (keV, 1σ) | 1.0 | 1.1 | 1.2 | not applicable | 1.5 | 2.3 |
| ARM (FWHM°) | 3.3 | 2.6 | 1.8 | Cut: 1.3 | 1.4 | 2.2 |
| $A_{eff}$ (cm²) | 819 | 801 | 711 | 859 | 479 | 32 |
| FoV (HWHM°) | | | | 50 | | |
| 3σ sens (on-axis) | 7.0e-7 | 1.7e-6 | 2.5e-7 | 1.1e-6 | 2.1e-7 | 3.9e-7 |

*Table Ge-2: Simulated instrument performance for the Ge ACT concept. Note that sensitivities are in photons cm⁻² s⁻¹ for 10⁶ s exposure, obtained using Bayesian event reconstruction. Dominant background component for 847 keV line: cosmic photons. (Chance coincidences are not included, though we estimate their effect to be small.)*

## REFERENCES

[1] Boggs, S. E., et al., *SPIE* **4851**, 1221 (2002).

[2] Boggs, S.E., et al., *New Astr. Rev.* **48**, 251 (2004).

[3] Amrose, S., et al., *NIM* **A505**, 170 (2003).

[4] Coburn, W., et al., *SPIE* **4784**, 54 (2002).

[5] Burkes, M., et al., *IEEE NSS* **2**, 1114 (2004).

## General Description

One concept investigated for the Advanced Compton Telescope is an instrument that uses only thick, position-sensitive silicon detectors for the primary detector. These detectors are the basis for the D1 silicon detector array used in the ACT baseline concept discussed in Section G. The instrument studied is shown in Fig. 1 and consists of a 3×5 array of towers. Each tower consists of 64 layers of detectors, with each layer based on a 4×4 array of double-sided strip detectors 10cm × 10cm × 3mm thick. (Referred to as NRL04 in Fig. G8.)

Advantages of this approach include: (1) near room operating temperature for the solid-state detectors (eliminating the 80 K required for the germanium detectors); (2) reduced Doppler broadening that provides better angular resolution for the direction of the incident γ-ray; (3) larger FoV that improves the sky coverage on each orbit and also improves the sensitivity for a zenith pointed instrument; (4) excellent energy resolution from the silicon detectors. (Note that for multiple Compton interactions that will be associated with each incident γ-ray for the low-Z detector, the energy of the incident γ-ray can be determined independently at each Compton scatter site. Averaging these reduces the Doppler broadening effect for events that are not fully absorbed. A narrow line energy resolution of 4–5 keV is expected at 847 keV). The use of near room-temperature detectors enables low-power ASICs to be located very close to the detectors which is advantageous for reducing input capacitance and cross-talk and thereby providing excellent energy resolution.

## Status of Hardware Development

NRL has been developing position-sensitive germanium and silicon detectors for over ten years. We have demonstrated Compton imaging in both technologies [Wulf et al. 2003, Wulf et al. 2004] and have specifically focused on silicon Compton imagers due to the near room operating temperature, lower Doppler broadening and potential for developing thicker silicon detectors ideally suited to Compton imaging. We are currently finishing the assembly of an 8-layer device with a 2×2 array of 63mm × 63mm × 2mm thick silicon detectors in each layer [Kurfess et al. 2004]. Table thickSi-1 lists the laboratory detector performance and that assumed for the ACT mission.

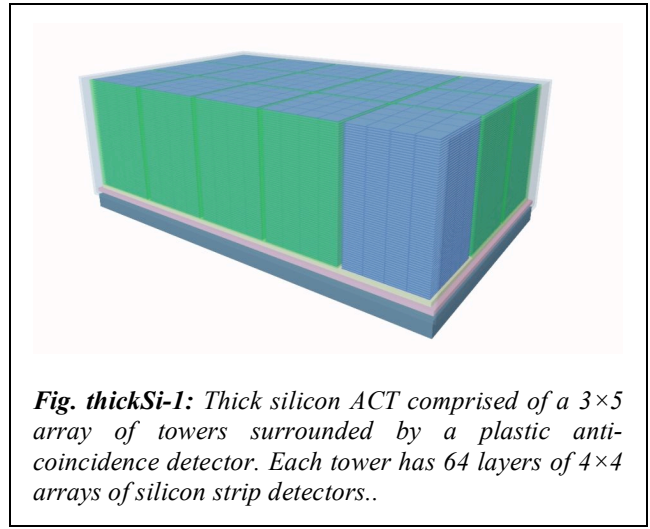

**Fig. thickSi-1:** *Thick silicon ACT comprised of a 3×5 array of towers surrounded by a plastic anti-coincidence detector. Each tower has 64 layers of 4×4 arrays of silicon strip detectors..*

## Simulated Performance

### Description of Mass Model

The Thick silicon ACT instrument is shown in Fig. thickSi-1. It consists of a 3×5 array of towers surrounded by an ACD detector array. Each tower has 64 layers of 10cm × 10cm × 3mm thick double sided-silicon strip detectors. There are 80 strips per side providing an x-y spatial resolution of less than 1.2 mm. Total instrument power is 1655 watts. This assumes the ISAL baseline of 1 mW/ch for front-end electronics and 615 watts for the cryogenic cooler (assuming operation at –30° C). The power required for the cryogenic cooler would probably be substantially less based on the excellent room temperature performance of the latest detectors. Total instrument mass is 1690 kg. The overall instrument meets all requirements identified in the ISAL and IMDC studies.

| | Assumed in Mass Model (4×4 daisy) | | | | Achieved in Laboratory (single detector) | | | |
|---|---|---|---|---|---|---|---|---|
| | $\Delta E$ (1σ) (keV) | $\Delta \vec{x}$ (mm) (x/y; z) | Thresh (keV) | $\Delta t$ (s) | $\Delta E$ (1σ) (keV) | $\Delta \vec{x}$ (mm) (x/y; z) | Thresh (keV) | $\Delta t$ (s) |
| D1 | 1.7 | 1.2/0.4 | 15 | 10 | 1 | 0.9/0.5 | 10 | 10 |

**Table thickSi-1:** *Detector performances achieved and assumed for simulation for the thick-Si ACT concept.*

**Simulation Results**

The on-axis sensitivities for selected lines are shown in Table 2. The overall ACT sensitivity for a zenith-pointed ACT must also include the sensitivity as a function of the off-axis angle. The preliminary sensitivities to the 847 keV line as a function of off-axis angle are shown in Fig. thickSi-2 for several instrument configurations in low-Earth and high-Earth orbit. Although we have concentrated on the performance of the thick silicon ACT in a low-Earth equatorial orbit, we have also done preliminary simulations for a high altitude orbit outside the earth's magnetosphere. This would enable a much larger FoV instrument (greater than $3\pi$ steradian), while also eliminating the Earth $\gamma$-ray and neutron albedo background contributions, but increasing the cosmic-ray background. As can be seen in Fig. thickSi-2, the combination of good sensitivity and broad FoV make this an attractive option to explore.

It may also be possible to achieve low-energy electron tracking using the technology development discussed in Section J. This would, for example, reduce the direction of the incident $\gamma$-ray from a cone to a segment of a cone, enable a larger fraction of the incident celestial $\gamma$-rays to be accepted, and also enable rejection of a significant portion of atmospheric albedo $\gamma$-rays. These would all contribute to a significant improvement in the sensitivity of a silicon ACT.

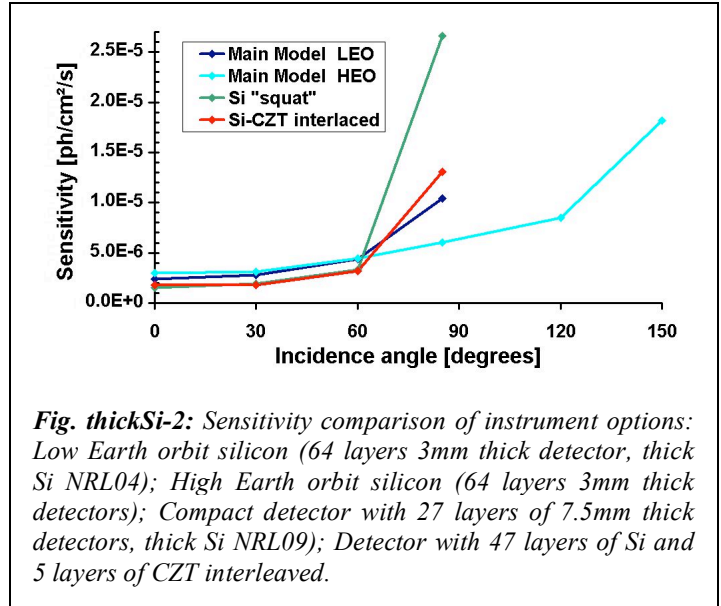

**Fig. thickSi-2:** *Sensitivity comparison of instrument options: Low Earth orbit silicon (64 layers 3mm thick detector, thick Si NRL04); High Earth orbit silicon (64 layers 3mm thick detectors); Compact detector with 27 layers of 7.5mm thick detectors, thick Si NRL09); Detector with 47 layers of Si and 5 layers of CZT interleaved.*

Disadvantages of an all-silicon instrument are the overall small fraction of events that are fully absorbed (thereby requiring the three-Compton reconstruction for determination of the incident energy and the resulting degradation in energy resolution) and the higher probability of energy losses in passive material or below the low-energy threshold of the input discriminators. These latter two issues may be significantly reduced with the modifications discussed below.

A significant constraint on the performance of an all silicon instruments is the probability of an interaction in passive material. This is especially a problem if such an interaction occurs before the 3rd interaction in the sequence. Therefore, minimizing the passive material within the envelope of the active silicon detectors is a primary goal. An alternative is to interleave room temperature, high-Z detectors (e.g. CZT) to limit the number of interactions that take place before full energy absorption (this is a variant of the baseline Si-Ge instrument that would not require cooling to 80 K). We have also studied this option and have plotted some preliminary sensitivity data in Table 3 and also show the sensitivity as a function of off-axis angle in Fig. 2. Use of interleaved CZT layers improves the overall sensitivity relative to an all-silicon instrument. In addition, we have also done simulations for a compact all silicon instrument using 27 layers of 7.5mm thick detectors and for the baseline instrument but in a high-altitude orbit. A brief description of the four configurations we have simulated, along with on-axis sensitivities for a broad 847 keV line and the sensitivity as a function of off-axis angle, are shown in Table 3 and Fig. thickSi-2. All of these instrument configurations are consistent with the mass and power constraints imposed in the ISAL and IMDC studies. Another option would be to include a BGO shield for a LEO mission (as in the baseline instrument).

| | 511 keV narrow | 847 keV narrow | 847 keV, 3% broad | 1809 keV narrow |
|---|---|---|---|---|
| $\Delta E$ (keV, 1$\sigma$) | 3.1 | 3.4 | 16.2 | 6.7 |
| ARM (FWHM°) | 1.6 | 1.2 | 1.4 | 0.8 |
| $A_{eff}$ (cm$^2$) | 420 | 460 | 860 | 433 |
| 3$\sigma$ sens (on-axis) ($\gamma$/cm$^2$-s in $10^6$ s) | $5.2 \times 10^{-6}$ | $1.9 \times 10^{-6}$ | $2.4 \times 10^{-6}$ | $2.0 \times 10^{-6}$ |

**Table thickSi-2:** *Simulated instrument performance for the thick Si ACT concept. Note that sensitivities are in photons cm$^{-2}$ s$^{-1}$ for $10^6$ s exposure. (Chance coincidences are not included, though we estimate their effect to be small.)*

We have investigated silicon Compton telescopes that use detectors currently available, and that would provide a significant improvement to the sensitivity for nuclear astrophysical investigations. A promising technology that could dramatically improve the sensitivity of ACT would be the capability for electron tracking within silicon detectors. This has recently been demonstrated in silicon Controlled Drift Detectors for electron energies below 500 keV [5]. These detectors could provide several advantages, all of which would provide significant improvement to the sensitivity.

Those specifically associated with electron tracking include: (1) reduction of the incident γ-ray cone to an arc segment, (2) reduction of background associated with Earth albedo, and (3) improved event reconstruction. In addition, these detectors have excellent energy resolution and low thresholds (less than 5 keV) at room temperature, and these will also increase the number of incident γ-rays that can be properly reconstructed. We expect that these combined effects will significantly improve the sensitivities shown in Fig. thickSi-2, and potentially beat the $7\times10^{-7}$ γ/cm$^2$-s goal for broad $^{56}$Co lines from Type Ia supernovae identified in our concept.

REFERENCES

1. Wulf, E.A., et al., IEEE Trans. Nucl. Sci. 50, p. 1182–1189 (2003)
2. Wulf, E.A., et al., IEEE Trans. Nucl. Sci. 51, p. 1997–2003 (2004)
3. Kurfess, J.D., et al., New Astronomy Reviews, 48, p. 293–298 (2004)
4. Strüder, L. et al., Proc. SPIE, Vol. 5165, p. 10–18 (2004)
5. Castoldi, A. et al., ' Multi-linear Drift Detectors for X-ray and Compton Imaging', 10th European Symposium on Semiconductor Detectors, Wildbad-Kreuth, June 12–16, 2005 (NIM)

| Table 3. Alternative Configurations Investigated | | | | | |
|---|---|---|---|---|---|
| | Detector Configuration | | | | |
| Concept | # Layers | Layer Separation | Frontal Are (active Si) | Height | On-axis 847 keV Broad line Sensitivity γ/cm$^2$-s in $10^6$ s |
| Baseline | 64-3mm thick silicon | 10mm | 2.2 m$^2$ | 64cm | $2.4 \times 10^{-6}$ |
| Compact | 27-7.5mm thick silicon | 11.5mm | 2.2 m$^2$ | 31cm | $1.6 \times 10^{-6}$ |
| Si-CZT | 47-3mm thick Silicon 5-5mm thick CZT interleaved | 7mm | 2.2 m$^2$ | 38cm | $1.8 \times 10^{-6}$ |
| High-Earth Orbit | 64-3mm thick silicon | 10mm | 2.2 m$^2$ | 64 cm | $3.0 \times 10^{-6}$ |

*…Witness to the Fires of Creation*

# U. Liquid Xenon Time Projection Chambers with Enhanced Spectroscopy and Time-of-Flight: LXeACT

Rice University, Columbia University

## General Description

The LXeACT concept is based on a 6 × 6 array of D1-D2 towers built with Liquid Xenon Time Projection Chambers (LXeTPC), vertically separated by 10 cm. A third LXe volume on the bottom, operated as a scintillation detector, is used as an anti-coincidence shield (D3). Each LXe volume has an area of 22×22 cm$^2$, viewed by a total of 64 square 1" photomultiplier tubes (PMTs). The total area including the shroud spans 144 × 144 cm$^2$. D1, D2, and D3 have a thickness of 3 cm, 7 cm, and 5 cm, respectively. In the TPCs D1 and D2, simultaneous detection of ionization and scintillation signals produced by radiation in LXe provide precise measurements of energy and 3D position on an event-by-event basis. Within its large homogeneous volume, instrumented with a 2D charge readout and triggered by the prompt scintillation light (175 nm), a LXeTPC efficiently records the history of each γ-ray, by measuring arrival time, energy deposited, and coordinates of each interaction above a threshold of 30 keV.

Detected by VUV sensitive PMTs with a very fast time response, the fast (~30 ns decay) Xe scintillation light allows not only outstanding timing but also a time-of-flight (ToF) measurement. ToF is a powerful tool for both background suppression and efficiency enhancement, allowing unique ordering of interaction sequences even for events with only two interactions, while maintaining and improving ordering with kinematic Compton sequence reconstruction. The combination of millimeter position resolution and ToF is a unique feature of the LXeTPC approach to an ACT, provided by no competing technology. The double energy measurement from light and charge greatly improves energy resolution over the charge measurement alone [1]. It also enables discrimination of γ-rays and nuclear recoils, hence eliminating prompt hadronic interactions in the sensitive volume (e.g., elastic neutron scattering) as a background component. The LXeTPC approach to Compton imaging thus provides a detector with superior background rejection capability based not only on γ-ray tracking within a large homogeneous volume but also on ToF between two TPC modules. This detector concept is highly efficient while minimizing the number of electronics channels and power, resulting in a highly cost-effective concept that is both modular and scalable to areas significantly larger than simulated for this study.

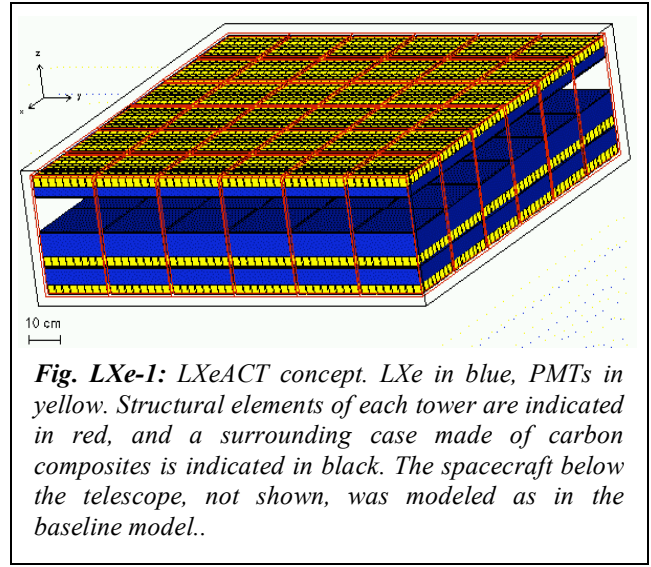

**Fig. LXe-1:** *LXeACT concept. LXe in blue, PMTs in yellow. Structural elements of each tower are indicated in red, and a surrounding case made of carbon composites is indicated in black. The spacecraft below the telescope, not shown, was modeled as in the baseline model..*

## Status of Hardware Development

The LXeTPC concept for γ-ray astronomy has been developed at Columbia University for a number of years [2]. A large detector module, LXeGRIT, of 20×20×7 cm$^3$ sensitive volume was built, and extensively tested both in the laboratory and in two balloon flight campaigns from Ft. Sumner, NM, in 1999 and 2000 [3]. Extensive data analysis and imaging software has been developed for this prototype. The custom-designed data acquisition system comprised 128 low-noise charge readout channels and 4 light readout channels [4]. The light trigger efficiency was characterized in detail [5]. Recent developments in PMT technology have enabled us to greatly improve light collection. As a consequence, a large improvement in energy resolution has been achieved by combining light and charge, which have been shown to be strongly anti-correlated [6]. With compact metal channel PMTs operated at LXe temperature, a factor of 3 improvement over LXeGRIT resolution has already been accomplished. Studies continue at Columbia and Rice, with the goal of further improvement in light collection and noise reduction to enable finer spectroscopy with LXe. The fast timing characteristics of the new PMTs allow us to use ToF in a compact ACT configuration. With test setups, a timing of 250 ps FWHM at 511 keV has already been measured [7], and a ToF measurement with two large-area LXe scintillator detectors for medical imaging is currently being analyzed. Alternative solid state devices such as Avalanche Photodiodes (APD) and Silicon Photomultipliers (SiPM) have also been tested in LXe and more work is underway to adapt these new photosensors for the light readout of a LXeTPC [8]. Other technology developments, currently driven by the application of a LXeTPC for Dark Matter detection [9] include an efficient cryocooler optimized for LXe [10], an effective purification scheme with continuous circulation of the LXe vapor through a commercial getter, and the application of high-performance plastics (PEEK, Cirlex, PTFE, Kapton) as structural materials inside the liquid, without loss of purity.

| | Assumed in Mass Model | | | | Achieved in Laboratory | | | |
|---|---|---|---|---|---|---|---|---|
| | ΔE (1σ at 1 MeV, in keV ) | Δx̄ (mm) (x/y; z) | Thresh (keV) | Δt (s) | ΔE (1σ at 1 MeV, in keV) | Δx̄ (mm) (x/y; z) | Thresh (keV) | Δt (s) (1σ, 511 keV) |
| D1; D2 | 4.76 | 0.44; 0.1 | 30 | 100 ps | 13 | 1.0; 0.3 | 80 | 106 ps |
| D3 ACS | n/a | n/a | 30 | 100 ps | n/a | n/a | 30 | 106 ps |

**Table LXe-1:** *Detector performances achieved and assumed for the instrument simulation for the LXe ACT concept. Note: Spatial resolution in x/y is assumed to be ¼ of the readout pitch.*

## SIMULATED PERFORMANCE AND RESULTS

The mass model was limited in size to 1.4×1.4 m², as assumed in the ISAL study, yielding a sensitive area of 1.74 m². At this size, the LXeACT concept was not limited by any of the given constraints: the detector has a mass of 1260 kg (30% passive), 18,000 charge readout channels, 4,608 light readout channels, and requires 311 W of power including cooling. If scaled to larger areas, with the chosen shield and conservative assumptions on the amount of passive materials required, the first limitation within the assumptions of this study would be mass. Due to the low number of electronics channels and the moderate cooling requirements, this concept may be scaled up greatly before encountering power limits. The light readout as simulated here is based on available PMT technology (Hamamatsu R8520), providing ~1/3 of the passive mass. Future technology developments in solid-state photosensors, such as APD or SiPM, or hybrid photomultiplier tubes have the potential of greatly improving light collection while minimizing the amount of passive materials. A design more suitable to a pressurized liquid detector, as opposed to the constraints of the baseline model, optimized for a solid state design, could further decrease the amount of passive mass. The kinematic Compton sequence software employed in this study requires improvement for this detector type, as only ~50% of the sequences without ToF were properly reconstructed in this study, hampering sensitivity. A higher reconstruction efficiency would not only proportionally increase effective area but also reduce background.

| | 511 keV narrow | 847 keV narrow | 847 keV 3% broad | 1809 keV narrow |
|---|---|---|---|---|
| ΔE (keV, 1σ) | 3.7 | 4.5 | n/a (3%) | 6.1 |
| ARM (FWHMº) | 6.6 | 3.6 | 3.6 | 1.7 |
| A$_{eff}$ (cm²) | 521 | 606 | 616 | 529 |
| FOV (HWHMº) | 60 | 65 | 65 | 70 |
| 3σ sens (on-axis) | 3.3 10⁻⁶ | 1.3 10⁻⁶ | 2.0 10⁻⁶ | 7.4 10⁻⁷ |

**Table LXe-2:** *Simulated instrument performance for the LXe ACT concept. Note that sensitivities are in photons cm⁻² s⁻¹ for 10⁶ s exposure. Dominant background components for 847 keV broad line are cosmic diffuse γ-rays, activation from trapped protons, and albedo γ-rays. (Chance coincidences are not included, though we estimate their effect to be small.)*

## REFERENCES

[1] E. Aprile et al., 2005, NIM A, submitted
[2] E. Aprile , M. Suzuki, 1989, TNS 36, 311
[3] E. Aprile et al., 2002, SPIE 4851, 140, astro-ph/0212005;
 A. Curioni et al., 2002, SPIE 4851, 151, astro-ph/0211584
[4] E. Aprile et al., 2001, TNS 48, 1299;
 E. Aprile et al., 1998, NIM A, 412, 425
[5] U. Oberlack et al., 2001, TNS 48, 1041
[6] E. Conti et al, 2003, Phys. Rev. B 68, 54201
 E. Aprile et al., 2005, in prep.
[7] E. Aprile et al., 2005, TNS, accepted
[8] E. Aprile et al., 2005, NIM A 551, 356;
 E. Aprile et al., 2005, NIM A, accepted
[9] E. Aprile et al., 2002, Xenon01 Workshop, astro-ph/0207670;
 E. Aprile et al., 2005, New Ast. Rev. 49, 289
[10] T. Haruyama et al., 2004, Adv. Cryogenic Eng. 710, 1459

# V. GASEOUS XENON – LABR₃

## V. GASEOUS XENON – LaBr₃

NASA/Goddard Space Flight Center & University of New Hampshire

## GENERAL DESCRIPTION

This concept is based on the use of advanced gas and scintillator detector technologies to achieve optimum electron tracking. In a low-density gaseous medium the recoil electron from the Compton interaction may be tracked, even at low energies, more accurately than in any other detector material. The complete kinematic information obtained in this manner may be used to reject background and improve imaging performance.

This is a two-detector approach. The scattering detector (D1) is made of large-volume time projection chambers (TPC) read out by gas micro-well detectors (MWD). A MWD is a type of gas proportional counter based on micro-patterned electrodes. Each sensing element consists of a charge-amplifying well with the cathode around the rim and the anode at the bottom. Ionization electrons created by the deposition of energy in the gas volume drift into the well, where an ionization avalanche occurs which produces the signal. The cathodes and anodes of the wells are each connected in "crossed-strip" fashion to allow the readout of a large area with fine spatial resolution. The third spatial dimension can be obtained by timing the drift of the ionization electrons. We assume Xe gas at 3 atm pressure for the detection medium, which allows the use of the UV scintillation light from the recoil electron as a trigger for coincidence operation with the calorimeter as well as for absolute event timing. For optimum track resolution it is necessary to control the diffusion of the ionization charges as they drift to the MWD plane. The addition of an electro-negative ion such as $CS_2$ has been found to cause the formation of negative ions from the ionization electrons, which subsequently drift with much lower diffusion than electrons, albeit much more slowly.

The calorimeter detector (D2) is made of individual crystals of $LaBr_3$, read out as pixellated arrays. $LaBr_3$ is a promising new scintillator material with far better light output, energy resolution, and timing properties than traditional scintillators such as NaI or CsI. For a Compton telescope its most attractive property is its very fast light decay time (~26 ns), which allows for a very short coincidence window and thus a very small contamination from random coincidences.

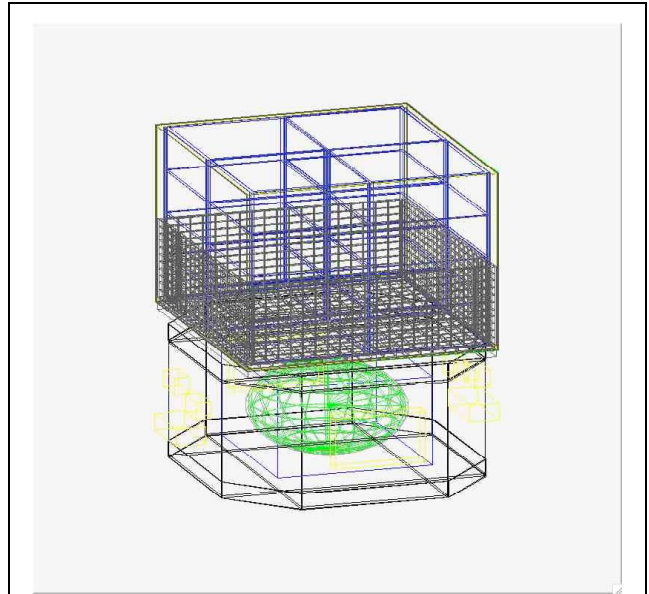

**Fig. gXe 1:** *MGGPOD mass model for the gaseous Xe/LaBr₃ ACT concept..*

## STATUS OF HARDWARE DEVELOPMENT

Goddard Space Flight Center has been producing small MWDs for the past several years on rugged polymide substrates using UV laser ablation. Very good detector performance has been observed[1-3], including a FWHM energy resolution of 18% for 5.9 keV X-rays in 95% Xe/5% $CO_2$ gas, and a FWHM position resolution of ~ 200 μm for 400 μm well pitch. Mikro Systems Inc. has recently applied their proprietary micro-casting and laminating technologies to large-area MWD fabrication, and 5×5 cm$^2$ detectors are undergoing testing. The use of $CS_2$ to reduce gas diffusion has been demonstrated in gas detectors[4]. UV avalanche photodiodes suitable for recording Xe scintillation light exist[5]. $LaBr_3$ crystals ~1 cm in size have been shown to have an energy resolution of ~3% at 662 keV and a coincident timing resolution of 260 ps[6] for 511-keV photons.

## SIMULATED PERFORMANCE

### Description of Mass Model

For the Concept Study simulations mass model, D1 consists of 2 × 2 × 4 modules, each 80 cm × 80 cm × 50 cm, filled with 97% Xe/3% $CS_2$ gas at 3 atm pressure. A pitch of 200 μm is assumed. The walls of the modules are polyamide held by a carbon composite frame; the passive mass fraction is about 20%. D2 is made of 5 mm × 5 mm $LaBr_3$ crystals, 8 cm thick on the bottom and 4 cm thick on the sides. A depth resolution of 1 cm is assumed. The telescope is surrounded by the standard plastic anticoincidence shield on all six sides. Assuming 1 mW per channel for D2 gives a total D2 power of 644 W, which leaves 8.5 mW per channel for the D1 readout if the limit of 1730 W is not to be exceeded. This has yet to be demonstrated. No cooling is required. The mass model is ultimately limited by the mass constraint of 1850 kg.

| | Assumed in Mass Model | | | | Achieved in Laboratory | | | |
|---|---|---|---|---|---|---|---|---|
| | ΔE (1σ) (keV) | $\Delta \vec{x}$ (mm) (x/y; z) | Thresh (keV) | Δt (s) | ΔE (1σ) (keV) | $\Delta \vec{x}$ (mm) (x/y; z) | Thresh (keV) | Δt (s) |
| D1 | 0.46 @ 6 | 0.2; 0.2 | 0.022 | — | 0.46 @ 6 | 0.2; - | ~1 | — |
| D2 | 7.2 @ 662 | 5; 10 | 20 | $5 \times 10^{-10}$ | 6.8 @ 662 | 5; - | ~8 | $2.6 \times 10^{-10}$ |

**Table gXe 1:** *Detector performances achieved and assumed for the instrument simulation for the Xe/LaBr₃ ACT concept. Note: ΔE is assumed to scale as sqrt(E) . D1 threshold is for single e⁻ in one well. Drift timing is assumed to give a D1 z-resolution the same as the (x,y) resolution.*

### Simulation Results

| | 511 keV narrow | 847 keV narrow | 847 keV 3% broad | 1809 keV narrow |
|---|---|---|---|---|
| ΔE (keV, 1σ) | 10.4 | 12.3 | 21 | 23.5 |
| ARM (FWHM°) | 4.6 | 2.8 | 3.2 | 2.0 |
| $A_{eff}$ (cm²) | 503 | 210 | 295 | 75 |
| FoV (HWHM°) | | ~60 | ~60 | |
| 3σ sens (on-axis) | | | $3.6 \times 10^{-6}$ | |

**Table gXe 2:** *Simulated instrument performance for the Xe/LaBr₃ ACT concept. Note that sensitivities are in photons cm⁻² s⁻¹ for 10⁶ s exposure. Only albedo photons and cosmic photons are included in background simulations; albedo photons cause the most background. All effective areas and sensitivities include a cut of 25° on the deviation of the recoil electron from the true direction. The broad line sensitivity is 67% worse without the electron direction cut. (Chance coincidences are not included.)*

## REFERENCES
[1] Black, J. K., et al. 2000, Proc. SPIE, 4140, 313
[2] Deines-Jones, P., et al. 2002, NIM A, 477, 55
[3] Deines-Jones, P., et al. 2002, NIM A, 478, 130
[4] Martoff, C. J., et al. 2000, NIM A, 440, 355
[5] Lopes, J. A. M., et al. 2001, IEEE Trans. Nucl. Sci., 48, 312
[6] Shah, K. S., et al. 2003, IEEE Trans. Nucl. Sci., 50, 2410

# W. Fast Scintillator ACT

University of New Hampshire

## General Description

This concept is based on the latest fast scintillator technology with the goal of preserving time-of-flight (ToF) discrimination and a very short coincidence window as the main means of rejecting background. These techniques were proven effective in space by the COMPTEL telescope. This concept is, in effect, a "modern COMPTEL" that takes full advantage of the lessons learned and uses simple, well-understood technology. Due to the experience with COMPTEL, the background predictions for this concept will be quite reliable.

Like COMPTEL, this concept employs a scattering detector, D1, and a calorimeter detector, D2. COMPTEL took advantage of the good intrinsic time resolution of scintillators to measure the relative timing of signals in D1 and D2; any event that did not match the expected ToF profile of an initial scatter in D1 followed by absorption in D2 could be rejected as background. In addition, such high time resolution allowed the use of a very short coincidence window, preventing random coincidences of unrelated interactions from becoming a significant source of background. The main background contamination was photons generated by particle interactions in material too close to D1 to be distinguished by the detectors' time resolution. Drawing on the knowledge gained from COMPTEL and to take advantage of well-understood technology, this ACT concept also emphasizes high time resolution, a short coincidence window, and low-Z passive material in the vicinity of D1. Very high spatial and energy resolution are NOT emphasized, allowing the design to remain relatively simple and low power. D1 is made of small, individual cells of low-Z scintillator, either plastic or liquid, with a fast scintillation decay time. D2 is made of similar small cells of a new, fast (26 ns decay time), high-Z, high-light-output scintillator called lanthanum bromide ($LaBr_3$). Small crystals of $LaBr_3$ have already been shown to have an energy resolution rivaling that of solid-state detectors like CdZnTe, permitting an energy resolution for the telescope of a few percent near 1 MeV. Passive structural material in the D1 assembly is made of Be. The readout and power requirements of this design

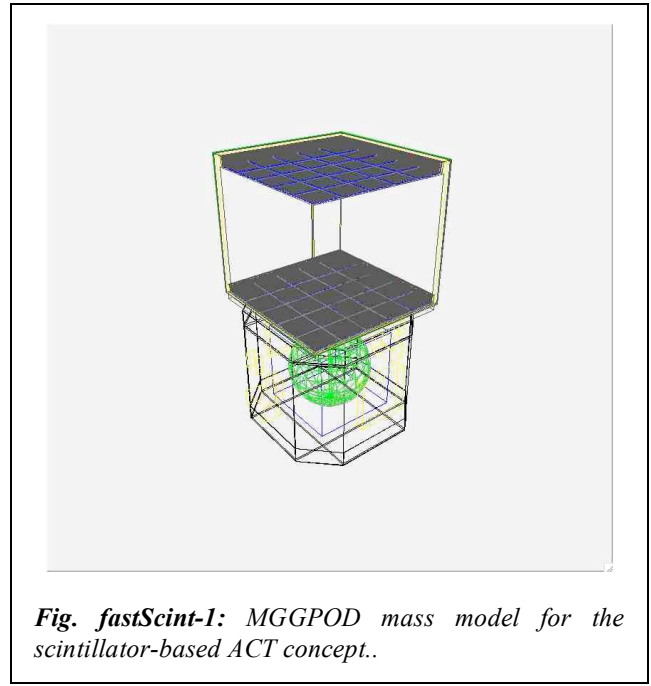

**Fig. fastScint-1:** *MGGPOD mass model for the scintillator-based ACT concept..*

would be modest compared to other ACT concepts.

## Status of Hardware Development

This concept draws on the entire body of experience from the COMPTEL telescope on the CGRO mission: construction, calibration, and nine years of on-orbit operations[1]. Plastic scintillator technology is fully mature and standard in any laboratory. Liquid scintillator, such as that used on COMPTEL, has the advantage of higher light output. In addition, deuterated plastic and liquid scintillator is available from companies such as Saint Gobain and ELJEN Technology; this would have the advantage of eliminating internal 2.2 MeV background from neutron capture. $LaBr_3$ crystals ~1 cm in size have been shown to have an energy resolution of ~ 3% at 662 keV and a coincident timing resolution of 260 ps[2]. Manufacturers have produced crystals up to ~2.5 cm in size that can be read out with either PMTs or avalanche photodiodes.

## SIMULATED PERFORMANCE

### Description of Mass Model

For the Concept Study mass model, D1 consists of 5 layers of 2 cm × 2 cm × 2 cm cubes of plastic scintillator. Each layer has 60 × 60 cubes, for a total of 18000 D1 elements. Each cube has 0.5 mm of Si on the bottom to represent a readout diode. The energy resolution and threshold values are those of COMPTEL's D1 detector[1]. No sub-cube position resolution is assumed. The structural material near D1 is Be and carbon composite, with a passive mass fraction of 60%. There is a 150 cm separation between the bottom of D1 and the top of D2. D2 consists of 4 layers, also 60 × 60 elements, of identical 2 cm LaBr$_3$ cubes, with the same Si readout. This gives 14400 channels for D2. The structural material is carbon composite, with a passive mass fraction of 26%. A time resolution of 500 ps for both D1 and D2 is conservatively assumed. To allow for faster electronics, we assume 2 mW per readout channel, giving a total power of 65 W. No cooling is required. The telescope is surrounded by the standard plastic anticoincidence shield on the top and sides. The mass model is ultimately limited by the mass constraint of 1850 kg.

| | Assumed in Mass Model | | | | Achieved in Laboratory | | | |
|---|---|---|---|---|---|---|---|---|
| | ΔE (1σ) (keV) | Δ$\vec{x}$ (mm) (x/y; z) | Thresh (keV) | Δt (s) | ΔE (1σ) (keV) | Δ$\vec{x}$ (mm) (x/y; z) | Thresh (keV) | Δt (s) |
| D1 | 45 @ 662 | 20; 20 | 50 | $5 \times 10^{-10}$ | 45 @ 662 | 5; 5 | 50 | ~$1 \times 10^{-10}$ |
| D2 | 7.2 @ 662 | 20; 20 | 20 | $5 \times 10^{-10}$ | 6.8 @ 662 | 5; - | ~8 | $2.6 \times 10^{-10}$ |

**Table fastScint-1:** *Detector performances achieved and assumed for the instrument simulation for the fast scintillator ACT concept. Note: D1 ΔE and threshold are those of the D1 detector of COMPTEL. "Achieved" position resolution merely reflects the size of the crystals used, since no sub-crystal resolution is needed.*

### Simulation Results

| | 511 keV narrow | 847 keV narrow | 847 keV 3% broad | 1809 keV narrow |
|---|---|---|---|---|
| ΔE (keV, 1σ) | 12.6 | 13.7 | 18.2 | 23.8 |
| ARM (FWHM°) | 5.1 | 2.7 | 2.9 | 1.6 |
| A$_{eff}$ (cm$^2$) | 74 | 115 | 123 | 111 |
| FoV (HWHM°) | | | ~45 | |
| 3σ sens (on-axis) | $1.6 \times 10^{-5}$ | $6.1 \times 10^{-6}$ | $6.3 \times 10^{-6}$ | $3.3 \times 10^{-6}$ |

**Table fastScint-2:** *Simulated instrument performance for the Xe/LaBr$_3$ ACT concept. Note that sensitivities are in photons cm$^{-2}$ s$^{-1}$ for $10^6$ s exposure. Only albedo photons and cosmic photons are included in background simulations; albedo photons cause the most background. (Chance coincidences are not included, though we estimate their effect to be small.)*

### REFERENCES

[1] Schönfelder, V., et al. 1993, ApJS, 86, 657

[2] Shah, K. S., et al. 2003, IEEE Trans. Nucl. Sci., 50, 2410